\documentclass[amssymb,useAMS,prd,aps,amsmath,onecolumn,superscriptaddress,nofootinbib,preprintnumbers]{revtex4}

\pdfoutput=1
\usepackage{amsmath}
\usepackage[utf8]{inputenc}
\usepackage[T1]{fontenc}
\usepackage{float}
\usepackage{multirow}
\usepackage{graphicx}
\usepackage{array}
\usepackage{psfrag}
\usepackage{braket}
\usepackage[usenames,dvipsnames]{color}
\usepackage[caption=false]{subfig}
\usepackage{epsfig,amssymb,fancyhdr}   
\usepackage[normalem]{ulem}
\usepackage{mwe}
\usepackage{slashed}
\usepackage{mathtools}
\usepackage[titletoc]{appendix}
\usepackage{datetime}
\usepackage{alltt} 
\usepackage{esvect}
\usepackage{natbib}
\usepackage{empheq}
\newlength\dlf  

\usepackage{epstopdf}

\usepackage{diagbox}

\newcommand{\beq}{\begin{equation}}
\newcommand{\eeq}{\end{equation}}
\newcommand{\ben}{\begin{eqnarray}}
\newcommand{\een}{\end{eqnarray}}
\newcommand{\bi}{\begin{itemize}}
\newcommand{\ei}{\end{itemize}}

\def\epsv{{\vec{\epsilon}}}

\newcommand{\FC}{Faraday conversion }

\newcolumntype{P}[1]{>{\centering\arraybackslash\Large}p{#1}}

\newcolumntype{b}{>{\large}c}
\usepackage{graphics}
\usepackage{psfrag}
\usepackage{graphicx}

\newcommand{\remove}[1]{}

\allowdisplaybreaks[4]

\def\refe@jnl#1{{#1}}
\def\aj{\refe@jnl{Astron.~J.}}
\def\araa{\refe@jnl{Annu.~Rev.~Astron.~Astrophys.}}
\def\apj{\refe@jnl{Astrophys.~J.}}
\def\apjl{\refe@jnl{Astrophys.~J.~Lett.}}
\def\aap{\refe@jnl{Astron.~Astrophys.}}
\def\mnras{\refe@jnl{Mon.~Not.~R.~Astron.~Soc.}}
\def\prd{\refe@jnl{Phys.~Rev.~D}}
\def\fcp{\refe@jnl{Fund.~Cos.~Phys.}}
\def\physrep{\refe@jnl{Phys.~Rep.}}
\def\physlett{\refe@jnl{Phys.~Lett.}}

\def\pe{{p_{e^-}}}

\def\pe1{{p_{e_1}}}

\def\piki{{p_1 \!\! \cdot \! k_1}}
\def\piko{{p_1 \!\! \cdot \! k_2}}

\def\cpo{ c_{\Phi_1}}
\def\cpt{ c_{\Phi_2}}
\def\spo{ s_{\Phi_1}}
\def\spt{ s_{\Phi_2}}

\def\P{$\mathbf{P}$}
\def\Pmat{$\mathbf{P}$--matrix }

\begin{document}

\hfill {\tt IPPP/19/24}

\vspace{1cm}

\title{Polarisation of high energy gamma--rays after scattering}

\author{C\'eline B\oe hm} 
\email{celine.boehm@sydney.edu.au} \thanks{ORCID: \href{http://orcid.org/0000-0002-5074-9998}{http://orcid.org/0000-0002-5074-9998}}
\affiliation{School of Physics, The  University of Sydney, NSW 2006, Australia}
\affiliation{LAPTH, U. de Savoie, CNRS,  BP 110, 74941 Annecy-Le-Vieux, France}
\affiliation{Perimeter Institute, 31 Caroline St N., Waterloo Ontario, Canada N2L 2Y5}
\author{Andr\'es Olivares-Del Campo}
\email{andres.olivares@durham.ac.uk}\thanks{ORCID: \href{https://orcid.org/0000-0001-7015-7650}{https://orcid.org/0000-0001-7015-7650}}
\affiliation{Institute for Particle Physics Phenomenology, Durham University, South Road, Durham, DH1 3LE, United Kingdom}
\author{Maura Ramirez-Quezada} 
\email{maura.e.ramirez-quezada@durham.ac.uk}\thanks{ORCID: \href{https://orcid.org/0000-0003-2321-0759}{https://orcid.org/0000-0003-2321-0759}}
\affiliation{Institute for Particle Physics Phenomenology, Durham University, South Road, Durham, DH1 3LE, United Kingdom}
\author{Ye-Ling Zhou} 
\email{ye-ling.zhou@soton.ac.uk} \thanks{ORCID: \href{https://orcid.org/0000-0002-3664-9472}{https://orcid.org/0000-0002-3664-9472}}
\affiliation{School of Physics and Astronomy, University of Southampton, Southampton SO17 1BJ, United Kingdom}

\preprint{}
\begin{abstract}
The polarisation of sunlight after scattering off the atmosphere was first described by Chandrasekhar using a geometrical description of Rayleigh interactions. Kosowsky later extended Chandrasekhar's formalism by using Quantum Field Theory (QFT) to describe the polarisation of the Cosmological Microwave Background radiation. Here we focus on a case that is rarely discussed in the literature, namely the polarisation of high energy radiation after scattering off particles. After demonstrating why the geometrical and low energy QFT approaches fail in this case, we establish the transport formalism that allows to describe the change of polarisation of high energy photons when they propagate through space or the atmosphere.  We primarily focus on Compton interactions but our  approach is general enough to describe e.g. the scattering of high energy photons off  new particles or through new interactions. Finally we determine the conditions for a circularly polarised $\gamma$--ray signal to keep the same level of circular polarisation as it propagates through its environment. 
\end{abstract}
\maketitle

\section{Introduction\label{sec:I}}

The polarisation of light is a cornerstone of modern astrophysics. Observations of both linear and circular polarisation have been used to understand the nature of astrophysical sources emitting electromagnetic radiation  \cite{De:2014qza,King:2016exc,doi:10.1093/mnras/208.2.409}. The associated formalism was first introduced by Chandrasekhar, who described the polarisation of starlight after scattering off dust particles in the atmosphere using modified Stokes parameters, namely $(I_l, I_r, U, V)$  \cite{chandrasekhar1960radiative}. In the conventional formalism of the Stokes parameters, the $I$--parameter measures the intensity of the polarisation signal, the $Q$-- and $U$--parameters provide information regarding the linear polarisation of that signal and the $V$--parameter indicates whether the observed light is circularly polarised.  

Chandrasekhar did not possess a Quantum Field Theory (QFT) description of particle interactions at the time so he used a geometrical description of the Rayleigh interactions to describe the radiative transfer of the visible light through the atmosphere. 
To perform his calculations, he introduced modified Stokes parameters, referred to as $(I_l, I_r, U, V)$ where $I_l, I_r$ stand for the decomposition of the intensity along the two main axes of the polarisation plane. His results for the radiative transfer are encapsulated in the so-called \P--matrix which describes the change in polarisation (and Stokes parameters) after scattering. In 1994 Kosowsky extended Chandrasekhar's formalism and described the polarisation of the Cosmological Microwave Background (CMB), i.e. mm radiation, as it propagates in an expanding (inhomogeneous) Universe using a QFT approach (see Ref.~\cite{Kosowsky:1994cy}). 
Due to the nature of Rayleigh and Thomson scattering interactions, both Chandrasekhar and Kosowsky concluded that the $V$--parameter was secluded. In other words, a low energy ($E_\gamma <m_e$) circular polarisation signal cannot generate a linearly polarised component nor can it be produced by the scattering of a linear polarisation signal off cosmic material. Yet the intensity of a circularly polarised signal can  change as the light scatters off ambient material. This change can be understood as follows. Photons have  two helicity states. Each of them are associated with one circular polarisation state (refer to as left--handed and right--handed in the following);  if one helicity state dominates over the other one, the observed light will be circularly polarised and the measured $V$--parameter will be non-zero. We refer to this case as  a net circular polarisation. If however the number of photons with $\pm$ helicity state is the same, there is no net circular polarisation and the $V$--parameter is essentially zero. The photon interactions which  change the number of photon polarisation states can thus change the fraction of net circular polarisation. They can also change the properties of the linearly polarised light. The recent  upper limits on  CMB circular polarisation can be found in ref. \cite{Nagy:2017csq}.

The aim of this paper is to describe how the polarisation of high energy ($\sqrt{s} \geq m_e$) electromagnetic signals changes as they propagate through space or in the atmosphere. Electromagnetic interactions at high energy ($\sqrt{s} > m_e$) are described by the Compton scattering cross section in the high energy regime (which is different from the Klein-Nishima regime) and critically includes helicity-flip processes such as $e^-_R \gamma_L \rightarrow e^-_L \gamma_R$.  In this regime, the (classical) radiative transfer approach is no longer appropriate since it assumes that the scattering only changes the direction of the outgoing photons and ignores significant transfer of energy or flip of the helicity configuration of the particles involved. The QFT approach that Kosowsky developed is more suited to describe the nature of the interactions, (though the Physics of interactions at high energy is different from that at low energy) but the propagation of high energy photons in space does not require to take into account the evolution of the Universe and is therefore different from the polarisation of the CMB in an expanding Universe.

 The correct formalism thus has to be a mixed of the two; I.e., one needs to embed the QFT formulation in a radiative transfer framework. Here we  develop such a formalism and show how to recast Chandrasekhar's low energy \P--matrix in terms of the (QFT) scattering matrix amplitude elements, thus addressing an important gap in the  literature. Our formalism is general enough to be applied in a different context,  including for example to describe the evolution of the Stokes parameters after the light scatters off  generic new particles.

The paper is organised as follows. In Section \ref{sec:classical}, we present the classical radiative transfer formalism; we define the Stokes parameters and how they change after scattering. In Section \ref{sec:quantum}, we present the Quantum formalism and the relation between the Stokes parameters before and after scattering with the scattering amplitude. We show that it is consistent with the radiative transfer in the low energy limit in Appendix~ \ref{App:Low_energy}. We generalise the latter in presence of generic interactions in Section~\ref{sec:compton} before discussing the special case of (high energy) Compton interactions. We focus on circular polarisation in Section \ref{sec:CP}. This  discussion is particularly relevant since circularly polarised $\gamma$-ray signals could reveal the nature of the  particles in cosmic accelerators \cite{Boehm:2019yit}. Finally we provide the transport equations of the Stokes parameters in space (or the atmosphere) in Section~\ref{sec:boltzman}. We conclude in Section~\ref{sec:conclusion}.

\section{Classical Formalism} \label{sec:classical}

In this section we review the radiative transfer formalism introduced by Chandrasekhar to determine the polarisation of the visible light after Rayleigh scattering. 

\subsection{Electric field definitions and basis}

\begin{figure}
\includegraphics[width= .6\textwidth, angle = 0]{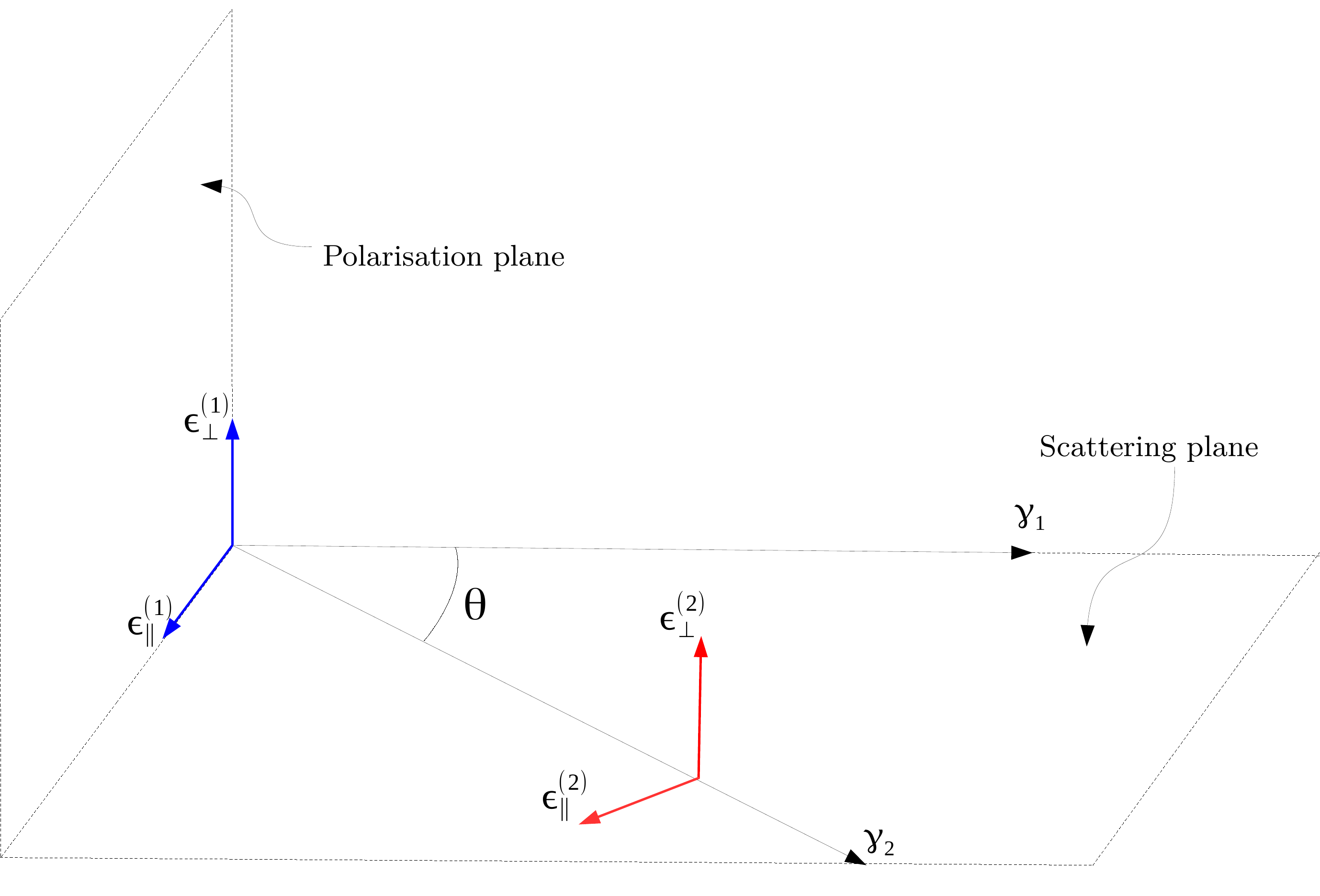}\caption{Illustration of the scattering plane formed by the incoming and outgoing photon  directions. Before the scattering, the  $\epsilon_r$ and $\epsilon_l$ vectors that define the polarisation plane are parallel and perpendicular to the scattering plane, so we denote them by $\epsilon_\perp^{(1)}$ and $\epsilon_{||}^{(1)}$ respectively. After the  scattering, the polarisation plane (defined by $\epsilon_\perp^{(2)}$ and $\epsilon_{||}^{(2)}$) forms an angle $\theta$ with respect to the initial polarisation plane.}\label{fig:scpln}
\end{figure}

The  electric field $\vec{E}$ can be expressed as the linear combination of  two perpendicular polarisation vectors $\vec{\epsilon}_l$ and $\vec{\epsilon}_r$, 
\begin{equation} 
\vec{E}(\mathbf{x},t) = (E_r \ \vec{\epsilon}_r \ + \ E_l \ \vec{\epsilon}_l)  e^{i(wt-\mathbf{k}\cdot\mathbf{x})} \,, \label{eq:E_field_def}
\end{equation} 
where $E_r=a_r e^{i \delta_r}$ and  $E_l=a_l e^{i \delta_l}$ with $a_{l,r}$ being real and $\delta_{l,r}$ being the phases of $E_{l,r}$, respectively.  From this definition, it follows that the two (orthogonal) vectors $\vec{\epsilon}_l$ and
$\vec{\epsilon}_r$, which define the  polarisation plane (see Fig. \ref{fig:scpln}), can be written as \cite{HAGIWARA19861}
\begin{align}
\epsv_l(\mathbf{k})=&\frac{1}{k_0 k_T}(k_x k_z, k_y k_z , -k^2_T) \,,  \nonumber \\
\epsv_r(\mathbf{k})=&\frac{1}{ k_T}(-k_y , k_x , 0) \,\label{eq:pol_vec_def}\,, 
\end{align}
where $\mathbf{k}$ refers to
 the 3-momentum of the propagating light $\mathbf{k} = (k_x, k_y, k_z)$ and $k_T = \sqrt{k_x^2 + k_y^2}$. 
When the two phases are the same $\delta_l =
   \delta_r = \delta$, the electric field is linearly polarised, i.e., it oscillates in a plane,  and can be expressed as 
\begin{equation}
\vec{E}(\mathbf{x},t) = (a_r   \, \epsv_r \, + a_l \, 
\epsv_l \, ) \ e^{i \delta} e^{i (wt-\mathbf{k} \cdot \mathbf{x})} \, . 
\end{equation} 
When the two phases differ by $\delta_l- \delta_r = \pm \pi/2$ and the amplitudes are the same ($a_r=a_l = a$), the electric field rotates around the propagation direction and the light is circularly polarised. The electric field then reads 
\begin{equation}
\vec{E}(\mathbf{x},t) = (\epsv_r \, \pm \, 
i \, \epsv_l ) \ a \ e^{i \delta_r} e^{i (wt-\mathbf{k}\cdot\mathbf{x})}\,.
\end{equation}
For convenience, we will define another
set of perpendicular vectors  $\epsv_\pm$  as a
linear combination of $\epsv_{l,r}$, which can be written as
\begin{align} 
\epsv_\pm(k)=& \frac{1}{\sqrt{2}} (\mp \epsv_l - i \,  \epsv_r)\,\label{eq:circ_pol_vec_def}, 
\end{align}
where $\epsv_+$ and $\epsv_-$ describe photons with positive and negative  helicity along the propagating direction respectively. The electric field in this $\pm$ basis,  reads as 
\begin{equation}
\vec{E}(\mathbf{x},t) = (E_{+} \, \epsv_+ \, + \, E_{-}  \, \epsv_- ) \ e^{i (wt-\mathbf{k}\cdot\mathbf{x})} \,,
\label{eq:circ_pol_E_field_def}
\end{equation}
where $E_\pm$ are given by
\begin{align}
E_+=& -\frac{a_l\,e^{i\delta_l}-
  ia_r\,e^{i\delta_r}}{\sqrt{2}}\,,\nonumber\\
E_-=&\frac{a_l\,e^{i\delta_l}+
  ia_r\,e^{i\delta_r}}{\sqrt{2}}\,.
\end{align}
Without loss of generality, one can always  re-parametrise $E_\pm$ as $E_\pm = a_\pm e^{i \delta_\pm}$ with $a_\pm$ being absolute values in the $\pm$ basis and $\delta_\pm$ the  associated phases respectively.
While both the linearly and circularly polarised light can be described using $\vec{\epsilon}_{l,r}$ and $\vec{\epsilon}_{\pm}$, it is worth noticing that the $(l,r)$ basis (i.e. $\vec{\epsilon}_{l,r}$)  is more convenient to describe the linearly polarised light while the $\pm$ basis (corresponding to $\vec{\epsilon}_{\pm}$)  is more appropriate to describe circularly polarised light. We will  use both in the following, depending on whether the emphasis is on circular or linear polarisation.

\subsection{Stokes parameters \label{Subsec4:Stokes_parameters}}

One can describe the polarisation of light using four Stokes parameters $I$, $Q$, $U$, and $V$, where $I$ represents the intensity, $Q$ and $U$ give some information about the linear polarisation properties and $V$ provides some information regarding the net circular polarisation of the signal.
In the $(l,r)$ basis, the Stokes parameters are defined by  
\begin{eqnarray}  
I &=& |\epsv_l \cdot \vec{E}|^2 + |\epsv_r \cdot \vec{E}|^2 \, =  a_l^2  +  a_r^2 \,, \nonumber\\
Q &=& |\epsv_l \cdot \vec{E}|^2 - |\epsv_r \cdot \vec{E}|^2 \, =  a_l^2  -  a_r^2 \,, \nonumber\\
U &=&  2 \ \text{Re} [(\epsv_r \cdot \vec{E})^{*} \times (\epsv_l \cdot \vec{E}) ] \, = 2 \  a_r \, a_l \cos(\delta_r-\delta_l) \,, \nonumber\\
V &=&  -2 \ \text{Im}[(\epsv_r \cdot \vec{E})^{*} \times (\epsv_l \cdot \vec{E}) ]  \, = 2 \  a_r \, a_l \sin(\delta_r-\delta_l) \, , \label{eq:stokes_linear_parameters}
\end{eqnarray} 
while in the $\pm$ basis (defined by Eq.~(\ref{eq:circ_pol_vec_def})), the Stokes parameters read as 
\begin{eqnarray}  
I &=& |\epsv_+ \cdot \vec{E}|^2 + |\epsv_- \cdot \vec{E}|^2 \, = a_+^2  +  a_-^2 \,,  \nonumber\\
Q &=&  -2 \ \text{Re} [(\epsv_+ \cdot \vec{E})^{*} \times (\epsv_- \cdot \vec{E}) ]\, = -2 \, a_+ \, a_- \cos(\delta_+-\delta_-) \,, \nonumber\\
U &=&   2 \ \text{Im}[(\epsv_+ \cdot \vec{E})^{*} \times (\epsv_- \cdot \vec{E}) ]\, = 2 \,  a_+ \, a_- \, \sin(\delta_+-\delta_-) \,, \nonumber\\
V &=&  |\epsv_+ \cdot \vec{E}|^2 - |\epsv_- \cdot \vec{E}|^2 \, = a_+^2  -  a_-^2 \,. \label{eq:circ_stks_def} 
\end{eqnarray} 
This last equality indicates that there is no net circular polarisation ($V = 0$) when the number of polarisation states are the same ($ a_+^2 =  a_-^2$).

\subsection{$\mathbf{R}$--matrix and $\mathbf{P}$--matrix and the modified ($I_l, I_r, U,V$) Stokes parameters at low energy  \label{ssec:geometrical}} 

Chandrasekhar was able to predict the polarisation of sunlight after scattering off dust particles \cite{chandrasekhar1960radiative} by using a geometrical description of the scattering and by introducing a modified set of Stokes parameters, referred to as  ($I_l, I_r, U,V$), where $I_l, I_r$ are the intensity components ($I =I_l + I_r$) of the signal in the scattering plane, which is defined by the $\epsilon_{l,r}$ vectors shown in Fig.~\ref{fig:scpln}.  These vectors are themselves defined with respect to the ($\epsilon_{||}$,$\epsilon_\perp$) vectors that define the polarisation plane. With these conventions in mind, one can also define $Q=I_l - I_r$, with $I_r = a_r^2$ and $I_l = a_l^2$. The reason why Chandrasekhar could describe the radiative transfer using such a geometrical approach instead of Quantum Field Theory is that at low energy, the energy of the outgoing particles is similar to that of the incoming particles. In other words the scattering changes the direction of the outgoing particles but has a negligible effect on the energy of the scattered particles. Hence it is possible to express the radiative transfer in 2D using the one angle between the incident and outgoing photons (see Fig.~\ref{fig:scpln}). This makes the relation between the outgoing and incoming (modified) Stokes parameters extremely simple.  In 3D, the same geometrical approach requires 4 angles. It is thus somewhat easier to first describe the scattering in 2D and then embed the result in 3D. The description of the radiative transfer in 2D is encapsulated in the so-called $\mathbf{R}$--matrix while the 3D description is encapsulated in the  $\mathbf{P}$--matrix, which can be summarised as

\begin{eqnarray}
\begin{bmatrix}
  I_l \\
  I_r \\
  U \\
  V
\end{bmatrix}^{(2)}
= \mathbf{R} \ \ 
\begin{bmatrix} 
     I_l \\
  I_r \\
  U \\
  V
 \end{bmatrix}^{(1)} 
\label{eq:R_mat_general}, 
\end{eqnarray}
where the superscript $(1)$ and $(2)$  denote the parameters before and after scattering  respectively. Since at low energy and in the scattering plane the photons are scattered with an angle $\theta$ in the $l$ direction (see Fig.~\ref{fig:scpln}), one can deduce that  $I_l^{(2)} =  \cos^2 \theta \ I_l^{(1)} $ and  $I_r^{(2)} =   I_r^{(1)} $. 
Therefore  the $\mathbf{R}$--matrix for Thomson scattering reads as 
\begin{eqnarray}
\mathbf{R}_{\rm Chandrasekhar}
= \begin{bmatrix}
  \cos^2\theta&0&0&0 \\
    0&1&0&0  \\
    0&0 &\cos \theta&0 \\
    0&0&0&\cos\theta 
\end{bmatrix} \,.
\label{eq:R_mat_Chandra}
\end{eqnarray}

\begin{figure}
\subfloat[]{\label{fig:a0}\includegraphics[width=0.65\linewidth]{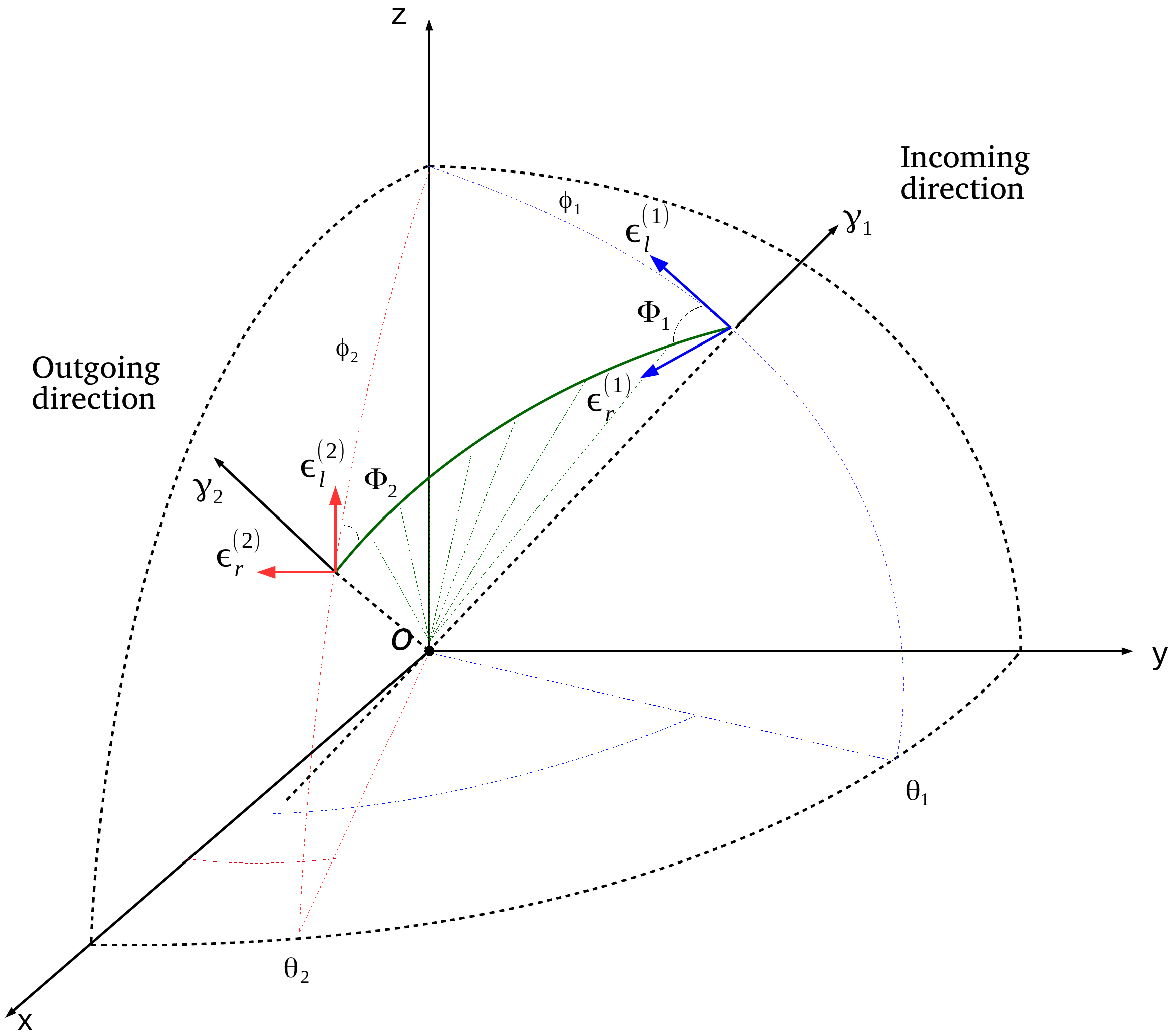}}
\subfloat[]{\label{fig:b0}\includegraphics[width= .25\textwidth, angle = 0]{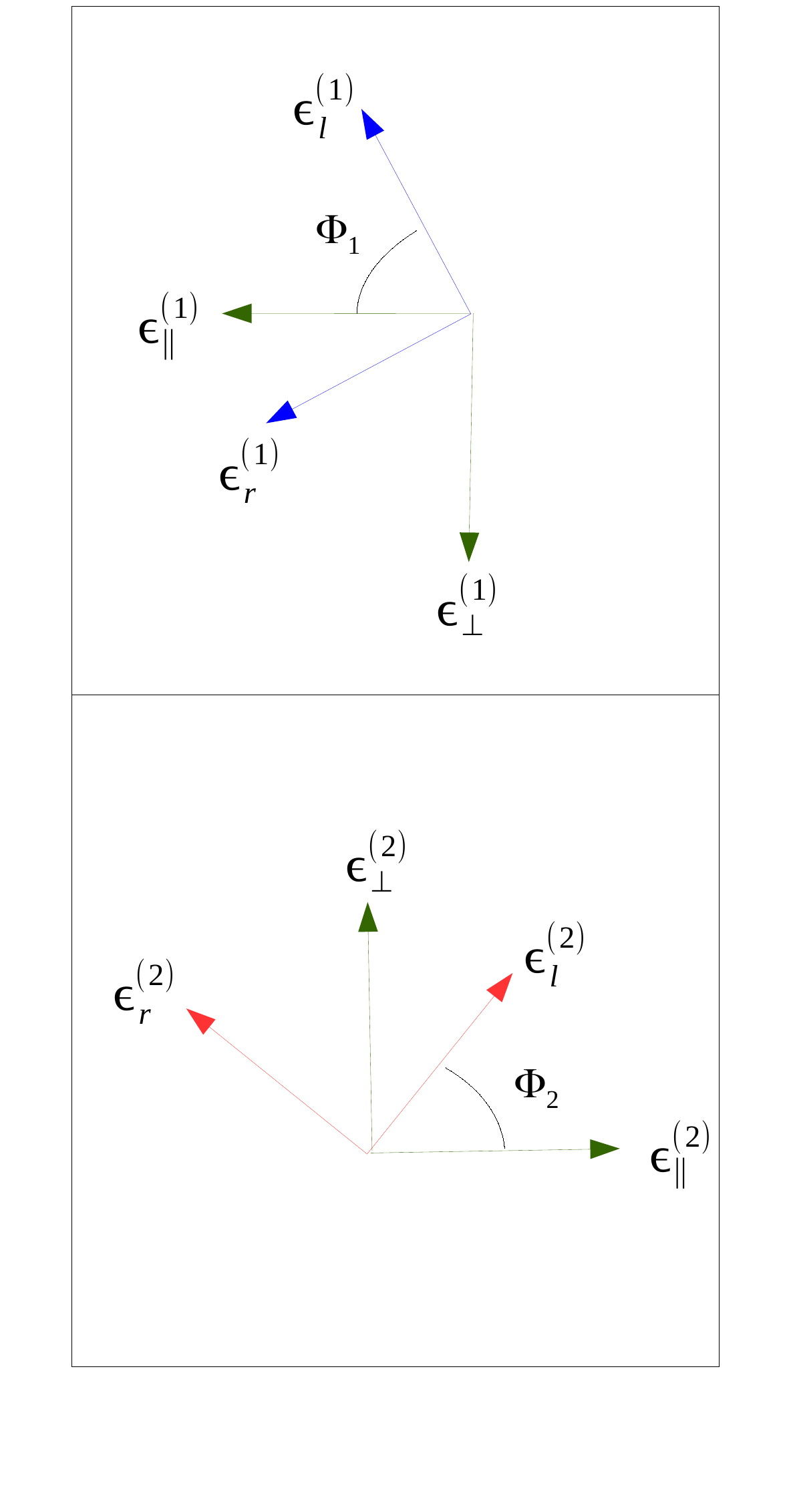}}\caption{In panel (a), we show the scattering plane in an absolute referential frame. The latter is defined by the scattering of an incoming photon off a particle located in $O$ (see green vectors). One can associate a polarisation state for each incoming and outgoing particle, i.e., ($\epsilon_l^{(1)}$, $\epsilon_r^{(1)}$) and ($\epsilon_l^{(2)}$, $\epsilon_r^{(2)}$), noticing that $\epsilon_l$ must be parallel  (and $\epsilon_r$  perpendicular) to the plane formed by the $z$--axis  and the corresponding photon direction. In this figure, $\Phi_{1,2}$ are the angles between the $\epsilon_l^{(1,2)}$ vectors,and the scattering plane respectively. In panel (b) we define, using the green colour, the parallel and perpendicular directions to the scattering plane as $\epsilon_{||,\perp}^{(1,2)}$ . We show the two rotations that are needed to obtain the \P--matrix, namely  $\mathbf{L}(-\Phi_1)$ to rotate the $\epsilon_{l,r}^{(1)}$ to the  $\epsilon_{||,\perp}^{(1)}$ basis and  $\mathbf{L}(\pi-\Phi_2)$ to rotate the $\epsilon_{||,\perp}^{(2)}$ to  $\epsilon_{l,r}^{(2)}$.} \label{Fig:LRL}
\end{figure}

The \P--matrix is then readily obtained by first rotating the plane defined by the incoming polarisation vectors by an angle $ -\Phi_1$  and then rotating the outgoing direction by $\pi-\Phi_2$, as shown in Fig. \ref{fig:b0}. This leads to the following relationship between the \P-- and the $\mathbf{R}$--matrices \cite{chandrasekhar1960radiative}
\begin{equation}
\mathbf{P}_{\rm Chandrasekhar}=\mathbf{L}(\pi-\Phi_2) \ \mathbf{R}_{\rm Chandrasekhar}  \ \mathbf{L}(-\Phi_1) \,, \label{eq:rotation3}
\end{equation}
where, $\mathbf{L}(\Phi)$ is defined as
\begin{eqnarray} 
\mathbf{L} = \left(
  \begin{array}{cccc}
   \cos^2\Phi & \sin^2\Phi  &- \tfrac{1}{2}\sin2\Phi & 0 \\
   \sin^2\Phi & \cos ^2\Phi  & \tfrac{1}{2}\sin 2\Phi & 0 \\
\sin2\Phi & -\sin 2\Phi & \cos 2\Phi &  0 \\
  0 & 0 &  0 &  1 \\
  \end{array}
  \right)\, \, \label{eq:rotation_chandra}
\end{eqnarray}
and leads to the relation displayed in Appendix \ref{Ap:Pmatrix_chandra} for the $\mathbf{P}$--matrix, which eventually (after relating the angles $\Phi_{1,2}$ and $\theta$ to the angles that define the most generic frame i.e., the fixed frame, see Appendix \ref{Ap:kinematics}, namely $\phi_{1,2}$ and $\theta_{1,2}$, see Fig. \ref{Fig:LRL}) leads to

\begin{eqnarray}
\mathbf{P}_{\rm Chandrasekhar}
\!=\! \begin{bmatrix}
   \mathbf{P_{11}} & \mu_2^2  \, s_{12}^2 &\mathbf{P_{13}}& 0 \\
 \mu_1^2  \, s_{12}^2  &  c_{12} &(\mu_1 \, c_{12} \, \, s_{12}) & 0 \\
\mathbf{P_{31}}  & -2(\mu_2 \, c_{12}   \, s_{12})   &  \mathbf{P_{33}}  &  0\\
   0 & 0  &  0  & \mathbf{P_{44}}\\
\end{bmatrix}
\label{eq:Chandrasekhar_Pmatrix}
\end{eqnarray}
with
\begin{align}
   \mathbf{P_{11}} &=   \Big(\mu_1 \,\mu_2\,  c_{12}  + \sqrt{1-\mu_1^2}\ \sqrt{1-\mu_2^2}\,\,\Big)^2 \,,\nonumber\\
    \mathbf{P_{13}} &=\tfrac{1}{2} \, \mu_2^2 \, \mu_1 \, \sin2(\theta_2-\theta_1) + \mu_2 \, \sqrt{1-\mu_1^2}\ \sqrt{1-\mu_2^2} \, s_{12}, \nonumber\\
     \mathbf{P_{31}} &=-\tfrac{1}{2} \, \mu_2 \, \mu_1^2 \, \sin2(\theta_2-\theta_1)-\mu_2 \, \sqrt{1-\mu_1^2}\ \sqrt{1-\mu_2^2} \, s_{12}, \nonumber\\
   \mathbf{P_{33}}&=\sqrt{1-\mu_1^2}\ \sqrt{1-\mu_2^2}\ c_{12}  + \mu_1 \, \mu_2 \ \cos2(\theta_2-\theta_1),\nonumber\\
   \mathbf{P_{44}}&=\sqrt{1-\mu^2_1} \ \sqrt{1-\mu_2^2} \ c_{12} + \mu_1 \, \mu_2\, ,\nonumber\\
         \end{align} 
    and 
$ s_{12} = \sin(\theta_2-\theta_1)$,  $c_{12} = \cos(\theta_2-\theta_1) $, $ \mu_{1,2}= \cos\phi_{1,2}$.

We note that it is actually possible to get both $\mathbf{R}$ and $\mathbf{P}$--matrices without using geometrical arguments. Instead, one can use the definitions of the  electric field in Eq. (\ref{eq:E_field_def}) and  Stokes parameters in Eq.~(\ref{eq:stokes_linear_parameters}). This is demonstrated in Appendix \ref{Ap:Pmatrix_direct} where we substituted the polarisation vectors by their rest frame kinematics. This is important as this means that one should be able to use the  definitions of the square matrix amplitude for Thomson interactions which is given in terms of the polarisation vectors and replace them by their kinematics to obtain the  $\mathbf{P}$--matrix. Indeed the square matrix amplitude for Thomson interactions read as 

\begin{eqnarray}
{|{\cal{M}}|^2}&\propto& 
\sum_{\lambda=r,l} \sum_{\beta=r,l}  |\epsilon^{(1)}_{\lambda}\cdot\,\epsilon^{(2)}_{\beta}|^2  \
\end{eqnarray}
where $\epsilon^{(1)}_{l,r}$ and $\epsilon^{(2)}_{l,r}$ are the polarisation vectors associated with the $l,r$ directions in the scattering plane. 

Since this amplitude essentially conveys the information about the intensity of the signal, one can write it as the following matrix: 

 \begin{eqnarray}
\begin{bmatrix} 
     I_l \\
  I_r \\
 \end{bmatrix}^{(2)} 
= \begin{bmatrix}
    |\epsilon^{(1)}_{l}\cdot\epsilon^{(2)}_{l}|^2  & |\epsilon^{(1)}_{l}\cdot\,\epsilon^{(2)}_{r}|^2  \\
    |\epsilon^{(1)}_{r}\cdot\,\epsilon^{(2)}_{l}|^2  & |\epsilon^{(1)}_{r}\cdot\,\epsilon^{(2)}_{r}|^2   \\
        \end{bmatrix} \,
    \begin{bmatrix} 
     I_l \\
  I_r \\
   \end{bmatrix}^{(1)}. 
\label{eq:Chandrasekhar_Plr}
\end{eqnarray}
which after substitution (assuming electrons at rest) reads as
\begin{eqnarray}
\begin{bmatrix} 
     I_l \\
  I_r \\
 \end{bmatrix}^{(2)} 
= \begin{bmatrix}
    \cos \theta ^2  &0 \\
  0  & 1  \\
        \end{bmatrix} \,
    \begin{bmatrix} 
     I_l \\
  I_r \\
   \end{bmatrix}^{(1)}. 
\label{eq:Chandrasekhar_Pnr}
\end{eqnarray}
Summing over all the elements to get the total intensity $I$, we obtain 
\begin{eqnarray}
{|{\cal{M}}|^2} &\propto& 1 + \cos^2 \theta  \,,
\end{eqnarray}
which is indeed the square matrix element for the Thomson (i.e. low energy) interactions. 
The same exercise in 3D leads to the elements of the $\mathbf{P}$--matrix as given by Chandrasekhar in \cite{chandrasekhar1960radiative}  and eventually gives the following square matrix element  
     \begin{eqnarray}
{|{\cal{M}}|^2}_{\rm{Thomson}} &\equiv& 
\frac{3}{4} \ \ \Big[1 + \mu_1^2\, \mu_2^2 +
(1-\mu_1^2) (1-\mu_2^2) \ c_{12}\nonumber\\
&& + \  2 \mu_1\, \mu_2 \sqrt{(1-\mu_1^2)} \sqrt{(1-\mu_2^2)} \ c_{12}\Big]\,.
\label{chandracomplete} 
\end{eqnarray}
which  again leads to the well-known expression of the Thomson cross section in the electron rest frame, namely 
\begin{eqnarray}
{|{\cal{M}}|^2}_{\rm{Thomson, rest}} &\equiv& 1+ {\cos^2 \phi_2}\,. 
\end{eqnarray}
where $\phi_2$ is the angle between the outgoing photon direction and the initial electron (i.e., $\theta$ in the rest frame as depicted in Fig. \ref{fig:scpln}).
There is therefore a strong connection between the \Pmat that Chandrasekhar obtained and the square matrix amplitude that one can derive using QFT. 


\subsection{$\mathbf{R}$--matrix and \P--matrix and the ($I,Q,U,V$) Stokes parameters at low energy} 
So far we have been using Chandrasekhar's modified Stokes parameters ($I_l, I_r, U,V$). However  it is possible to define  the equivalent of the $\mathbf{R}$--matrix and \Pmat in terms of the $(I,Q,U,V)$ Stokes parameters. 
To avoid a possible confusion, we will denote by $\mathbf{R}'$ and \P $'$ the equivalent $\mathbf{R}$ and \P--matrices obtained using the $(I,Q,U,V)$ Stokes parameters. In other words, 

\begin{eqnarray}
\begin{bmatrix}
  I \\
  Q \\
  U \\
  V
\end{bmatrix}^{(2)}
= \mathbf{R'}  \ \  (\rm{or} \ \ \mathbf{P'}) \ \ 
\begin{bmatrix} 
     I \\
  Q \\
  U \\
  V
 \end{bmatrix}^{(1)} 
\label{eq:Rprime_mat_general}, 
\end{eqnarray}
depending on whether this relation is computed in the scattering plane ($\mathbf{R'} $) or in 3D ($\mathbf{P'}$).  
Using  the definitions of the Stokes parameters in the $\pm$ basis, see Eq.~\eqref{eq:circ_stks_def}, we obtain

\begin{equation}
  \mathbf{R'} = \left(
\begin{array}{cccc}
 3+\cos2\theta &  -2 \sin^2\theta & 0 & 0 \\
  -2 \sin^2\theta &  3+\cos2\theta & 0 & 0 \\
 0 & 0 & 4 \cos\theta & 0 \\
 0 & 0 & 0 & 4 \cos\theta \\
\end{array}
\right)
\end{equation}
and 
\begin{equation}
\mathbf{P'}=\left(
\begin{array}{cccc}
\mathbf{P'_{11}} & \mathbf{P'_{12}} & \mathbf{P'_{13}} & 0 \\
 \mathbf{P'_{21}} & \mathbf{P'_{22}} & \mathbf{P'_{23}} & 0 \\
\mathbf{P'_{31}} &\mathbf{P'_{32}} & \mathbf{P'_{33}} & 0 \\
 0 & 0 & 0 &  \mathbf{P'_{44}} \\
\end{array}
\right)
\end{equation}
with
\begin{align}
    \mathbf{P'_{11}}= & \left(\mu_1^2 \mu_2^2+1\right)c_{12}^2 +2 \sqrt{1-\mu_1^2} \sqrt{1-\mu_2^2}\ \mu_1 \mu_2 c_{12}+(1-\mu_1^2) (1-\mu_2^2)+s_{12}^2 \left(\mu_1^2+\mu_2^2\right) \, ,\nonumber\\
    \mathbf{P'_{12}}= &\left(\mu_1^2 \mu_2^2-1\right) c_{12}+2 \sqrt{1-\mu_1^2} \sqrt{1-\mu_2^2}\ \mu_1 \mu_2 c_{12}+(1-\mu_1^2) (1-\mu_2^2)+s_{12}^2 (\mu_1^2-\mu_2^2) \,,\nonumber\\
    \mathbf{P'_{13}} = & \ - 2\, s_{12} \left(\sqrt{1-\mu_1^2} \sqrt{1-\mu_2^2}\ \mu_2+c_{12} \mu_1 \left(\mu_2^2-1\right)\right) \,,\nonumber\\
    \mathbf{P'_{21}} = &\left(\mu_1^2 \mu_2^2-1\right) c_{12}+2 \sqrt{1-\mu_1^2} \sqrt{1-\mu_2^2}\ \mu_1 \mu_2 c_{12}+(1-\mu_1^2) (1-\mu_2^2)+s_{12}^2 \left(\mu_2^2-\mu_1^2\right) \,,\nonumber\\
    \mathbf{P'_{22}} = &\left(\mu_1^2 \mu_2^2+1\right) c_{12}+2 \sqrt{1-\mu_1^2} \sqrt{1-\mu_2^2}\ \mu_1 \mu_2 c_{12}+(1-\mu_1^2) (1-\mu_2^2)-s_{12}^2 \left(\mu_1^2+\mu_2^2\right) \,,\nonumber\\
    \mathbf{P'_{23}} = & \ 2\, s_{12} (\sqrt{1-\mu_1^2} \sqrt{1-\mu_2^2}\ \mu_1+c_{12} \mu_2 \left(\mu_1^2+1\right)) \,,\nonumber\\
   \mathbf{P'_{31}} = &  -2\, s_{12} \left(\sqrt{1-\mu_1^2} \sqrt{1-\mu_2^2}\  \mu_1+c_{12} \left(\mu_1^2-1\right) \mu_2\right) \,,\nonumber\\
  \mathbf{P'_{32}}  = & -2\, s_{12} \left(\sqrt{1-\mu_1^2} \sqrt{1-\mu_2^2}\ \mu_2+c_{12} \left(\mu_2^2+1\right) \mu_1\right) \,,\nonumber\\
   \mathbf{P'_{33}} = & \ 2\, \left(\sqrt{1-\mu_1^2} \sqrt{1-\mu_2^2}\ c_{12} + \mu_1 \mu_2\cos2(\theta_2-\theta_1) \right) \,,\nonumber\\
    \mathbf{P'_{44}} = & \ 2\, \left( c_{12} \sqrt{1-\mu_1^2} \sqrt{1-\mu_2^2}+ \mu_1 \mu_2 \right)
    \end{align}
 with $s_{12}=\sin(\theta_2-\theta_1)$, $c_{12}=\cos(\theta_2-\theta_1)$, $\mu_1=\cos\phi_1$ and $\mu_2=\cos\phi_2$. The transformation from the $\mathbf{R}'$--matrix to the  \P$'$--matrix is given in Appendix \ref{Ap:Pprimematrix_chandra}. 

\section{Quantum formalism} \label{sec:quantum}

One can generalise the geometrical formalism proposed by Chandrasekhar and understand it in a more fundamental way by using Quantum Field Theory (QFT). 
In the following, we show how to relate the Stokes parameters to the scattering matrix amplitude $\mathcal{M}$ associated with microscopic interactions.

\subsection{Stokes operator \label{ssec:IVA}}

To describe the change in polarisation and relate the Stokes parameters before and after scattering using Quatum principles, we first need to remind the reader of the definition of a photon quantum state. The latter reads as \cite{Landau}
\begin{equation}
    |\gamma^{(\alpha)}  \rangle = a_r \, e^{i\theta_r}| \, \epsilon_{r}^{(\alpha)} \rangle \,  + \, a_l \, e^{i\theta_l}| \, \epsilon_{l}^{(\alpha)}\rangle \,, 
\end{equation}
where $|\epsilon_{r, \,l}^{(\alpha)} \rangle$ are the polarisation states and  $\alpha = 1,2$ denotes the initial and final states, respectively. 
Let the operator for the Stokes parameters be $\hat{S}=(\hat{I},\hat{Q},\hat{U},\hat{V})$, then the associated observables can be constructed using the relationship  
\begin{equation}
S^{(\alpha)} \equiv \langle  S^{(\alpha)}  \rangle =  \langle \gamma^{(\alpha)}\, |\, \hat{S}^{(\alpha)}\, |\, \gamma^{(\alpha)} \rangle \,,\label{eq:exp_S}
\end{equation}
with $\hat{S}^{(\alpha)} $ the Stokes operators for the initial or final state defined as
\begin{equation}
\begin{centering}
\hat{S}^{(\alpha)}=\mathcal{\mathbf{W}} \, |\epsilon_{j}^{(\alpha)} \rangle \, \langle \epsilon_{i}^{(\alpha)}| 
\label{eq:stokes_vector_def} 
\end{centering}
\end{equation}
with $i,j = r,l$ or $\pm$. 
In the $(l,r)$ basis, the Stokes operator takes the form 
\begin{eqnarray}
\begin{bmatrix}
  \hat{I} \\
  \hat{Q} \\
  \hat{U} \\
  \hat{V}
\end{bmatrix}^{(\alpha)}
=
\underbrace{
\begin{bmatrix}
   1 & 0  & 0 & 1\\
   1 & 0  & 0 & -1 \\
  0 & 1 &  1&  0 \\
  0 & -i &  i &  0 \\
\end{bmatrix}}_{\mathcal{\mathbf{W}}}
\begin{bmatrix} 
   |\epsilon_l \rangle\langle \epsilon_l| \\
   |\epsilon_r \rangle\langle \epsilon_l| \\
   |\epsilon_l \rangle\langle \epsilon_r| \\
   |\epsilon_r \rangle\langle \epsilon_r| 
\end{bmatrix}^{(\alpha)} 
\label{eq:stoke_op_linear_def}, 
\end{eqnarray}
which  leads to the definitions in Eq. (\ref{eq:stokes_linear_parameters}). 
In the $\pm$ basis,  the Stokes operators take the form 
\begin{eqnarray}
\begin{bmatrix}
  \hat{I} \\
  \hat{Q} \\
  \hat{U} \\
  \hat{V}
\end{bmatrix}^{(\alpha)}
= 
\underbrace{
\begin{bmatrix}
   1 & 0  & 0 & 1\\
   0 & -1  &  -1 &  0 \\
  0 & i &  -i&  0 \\
  1 & 0 &  0 &  -1 \\
\end{bmatrix}}_{\mathcal{\mathbf{W}}}
\begin{bmatrix}
   |\epsilon_+ \rangle\langle \epsilon_+| \\
   |\epsilon_+ \rangle\langle \epsilon_-| \\
   |\epsilon_- \rangle\langle \epsilon_+| \\
   |\epsilon_- \rangle\langle \epsilon_-| 
\end{bmatrix}^{(\alpha)} \,,
\label{eq:Stoke_definition}
\end{eqnarray}
leading to the definitions in Eq. (\ref{eq:circ_stks_def}).

\subsection{Relation between the Stokes parameters and the scattering matrix amplitude\label{ssec:IVB}} 

Now that we have defined the Stokes operators, we can relate them to scattering matrix amplitudes. In the remainder of the paper we will focus on Compton interactions but we are keeping the formalism general enough so that it can be applied to any light scattering process at any energy, that is $\mathcal{M}(X \, \gamma_i \to X \, \gamma_{i'}) \equiv M_{i'i}$, with $X$ being the particle the photon scatters off (e.g. an electron in the case of Thomson or Compton scatterings).  The outgoing polarisation state ($|\epsilon_{i'}^{(2)}\rangle$) is related to the initial polarisation state ($|\epsilon_i^{(1)} \rangle$) by the matrix element  
$  M_{i'i} \equiv  \langle \epsilon_{i'}^{(2)}|  \epsilon_{i}^{(1)} \rangle$ where $i, i'$ refers to the $(l,r)$ or the $\pm$ basis, leading to 
\begin{equation}
| \epsilon_{j'}^{(2)} \rangle \langle \epsilon_{i'}^{(2)}| = \, M_{i'i} \, M_{j' j}^* \, | \epsilon_{j}^{(1)} \rangle\langle \epsilon_{i}^{(1)}| \, .
\label{eq:transition_states_matrix}
\end{equation}

Using Eq.(\ref{eq:stokes_vector_def}) and Eq.(\ref{eq:transition_states_matrix}), we then obtain 
\begin{equation}
\hat{S}^{(2)}= \mathcal{\mathbf{W}} \, |\epsilon_{j'}^{(2)} \rangle\langle \epsilon_{i'}^{(2)}| = \mathcal{\mathbf{W}} \, M_{i' i} \, M_{j' j}^* | \, \epsilon_{j}^{(1)} \rangle\langle \epsilon_{i}^{(1)}| \, =  \ \mathcal{\mathbf{W}} \, M_{i' i} \, M_{j' j}^* \,  \mathcal{\mathbf{W}}^{-1} \ \hat{S}^{(1)},
\label{eq:S_relatedto_M}
\end{equation}
which we will write in the following as 
\begin{equation}
\hat{S}^{(2)} =  \mathbf{A'}_{i' i j' j}  \ \hat{S}^{(1)} \,  \label{eq:M_def}
\end{equation}
with $i'ij'j$ indices that refer to the different polarisation of the photons in the initial and final states. We can now express the $\mathbf{A'}_{i' i j' j} $  matrix in the $(l,r)$ basis, that is 
\begin{eqnarray}
\begin{bmatrix}
  \hat{I} \\
  \hat{Q} \\
  \hat{U} \\
  \hat{V}
\end{bmatrix}^{(2)}
=
\,\underbrace{\mathcal{\mathbf{W}}
  \begin{bmatrix}
  M_{ll}  M^*_{ll} & M_{ll}M^*_{lr} & M_{lr}M^*_{ll} &  M_{lr} M^*_{lr}\\
M_{ll} M^*_{rl} &M_{ll} M^*_{rr} & M_{lr} M^*_{rl} & M_{lr}M^*_{rr} \\
    M_{rl}M^*_{ll} &M_{rl}M^*_{lr} & M_{rr}  M^*_{ll} &  M_{rr}  M^*_{lr}\\
    M_{rl}  M^*_{rl}&  M_{rl}M^*_{rr}  &  M_{rr}M^*_{rl} & M_{rr} M^*_{rr}\\
\end{bmatrix}\,
 \mathcal{\mathbf{W}}^{-1}}_{\mathbf{A'} }\,
\begin{bmatrix} 
  \hat{I} \\
  \hat{Q} \\
  \hat{U} \\
  \hat{V}
\end{bmatrix}^{(1)}\label{eq:stokes2_stokes1_amplr_def}.
\end{eqnarray}
or, in the $\pm$ basis, 
\begin{eqnarray}
\begin{bmatrix}
  \hat{I} \\
  \hat{Q} \\
  \hat{U} \\
  \hat{V}
\end{bmatrix}^{(2)}
=
\,\underbrace{\mathcal{\mathbf{W}}
  \begin{bmatrix}
  M_{++}  M^*_{++} & M_{+-}M^*_{++} & M_{++}M^*_{+-} &  M_{+-} M^*_{+-}\\
M_{-+} M^*_{++} &M_{--} M^*_{++} & M_{-+} M^*_{+-} & M_{--}M^*_{+-} \\
    M_{++}M^*_{-+} &M_{+-}M^*_{-+} & M_{++}  M^*_{--} &  M_{+-}  M^*_{--}\\
    M_{-+}  M^*_{-+}&  M_{--}M^*_{-+}  &  M_{-+}M^*_{--} & M_{--} M^*_{--}\\
\end{bmatrix}\,
 \mathcal{\mathbf{W}}^{-1}}_{\mathbf{A'} }\,
\begin{bmatrix} 
  \hat{I} \\
  \hat{Q} \\
  \hat{U} \\
  \hat{V}
\end{bmatrix}^{(1)}\label{eq:stokes2_stokes1_amp_def}.
\end{eqnarray}

\subsubsection{$A'$-matrix definition in the $\pm$ basis } 

Replacing the $\mathbf{W}$ matrix by Eq. \eqref{eq:Stoke_definition}, we find that the $\mathbf{A'}$--matrix takes the following form in the $\pm$ basis 
\begin{eqnarray}
\begin{bmatrix}
  \hat{I} \\
  \hat{Q} \\
  \hat{U} \\
  \hat{V}
\end{bmatrix}^{(2)}
=
 \, \begin{bmatrix}
  \mathbf{A'}_{11} &\mathbf{A'}_{12} &\mathbf{A'}_{13} &\mathbf{A'}_{14} \\
    \mathbf{A'}_{21} &\mathbf{A'}_{22} &\mathbf{A'}_{23} &\mathbf{A'}_{24}  \\
    \mathbf{A'}_{31} &\mathbf{A'}_{32} &\mathbf{A'}_{33} &\mathbf{A'}_{34}  \\
    \mathbf{A'}_{41} &\mathbf{A'}_{42} &\mathbf{A'}_{43} &\mathbf{A'}_{44} 
\end{bmatrix}\,
\begin{bmatrix} 
  \hat{I} \\
  \hat{Q} \\
  \hat{U} \\
  \hat{V}
\end{bmatrix}^{(1)},
 \label{eq:A_matrix_general}
\end{eqnarray}
 with
%

%
\begin{empheq}[box=\fbox]{align}  \label{eq:M_matrix_elements}
\mathbf{A'}_{11} &\!=\! \tfrac{1}{2} \!\left(|M_{++}|^2 + |M_{+-}|^2 + |M_{-+}|^2 + |M_{--}|^2 \right) \ 
&  \mathbf{A'}_{31} &\!=\! \text{Im} \!\left(M_{+-}M_{--}^* + M_{++}M_{-+}^* \right) \nonumber\\ 
\mathbf{A'}_{12} &\!=\! - \text{Re} \left(M_{--}M_{-+}^* + M_{+-}M_{++}^* \right)  \ 
& \mathbf{A'}_{32} &\!=\! \text{Im}\! \left(M_{--}M_{++}^* + M_{-+}M_{+-}^* \right)  \nonumber\\
\mathbf{A'}_{13} &\!=\!  \text{Im} \left(M_{--}M_{-+}^* + M_{+-}M_{++}^* \right)  \ 
& \mathbf{A'}_{33} &\!=\! \text{Re} \left(M_{++}M_{--}^* - M_{+-}M_{-+}^* \right)  \nonumber\\
\mathbf{A'}_{14} &\!=\! \tfrac{1}{2} \!\left(|M_{++}|^2 + |M_{-+}|^2 - |M_{+-}|^2 - |M_{--}|^2 \right) \ 
&\mathbf{A'}_{34} &\!=\!  \text{Im}\!\left(M_{++}M_{-+}^* + M_{--}M_{+-}^* \right)  \\
\mathbf{A'}_{21} &\!=\! - \text{Re} \left( M_{++}M_{-+}^* + M_{--}M_{+-}^* \right)  \ 
&\mathbf{A'}_{41} &\!=\! \tfrac{1}{2}\! \left(|M_{++}|^2 + |M_{+-}|^2 - |M_{-+}|^2 - |M_{--}|^2 \right)  \nonumber\\
\mathbf{A'}_{22} &\!=\! \text{Re} \left(M_{--}M_{++}^* + M_{-+}M_{+-}^* \right)  \ 
&\mathbf{A'}_{42} &\!=\! \text{Re} \left(M_{--}M_{-+}^* - M_{+-}M_{++}^* \right)  \nonumber\\
\mathbf{A'}_{23} &\!=\! \text{Im}\! \left(M_{++}M_{--}^* + M_{-+}M_{+-}^* \right)  \ 
&\mathbf{A'}_{43} &\!=\! \text{Im} \!\left(M_{-+}M_{--}^*+ M_{+-}M_{++}^* \right) \nonumber\\
\mathbf{A'}_{24} &\!=\! \text{Re} \left(M_{--}M_{+-}^* - M_{++}M_{-+}^* \right)   \ 
&\mathbf{A'}_{44} &\!=\! \tfrac{1}{2} \!\left(|M_{++}|^2 + |M_{--}|^2 - |M_{+-}|^2 - |M_{-+}|^2 \right) \nonumber\,
 \end{empheq}
%

\subsubsection{$A'$-matrix definition in $(l,r)$ basis}

In the $(l,r)$ basis, the $\mathbf{A'}$--matrix reads as 

\begin{empheq}[box=\fbox]{align}  \label{eq:M_matrix_elements_lr}
\mathbf{A'}_{11} &\!=\! \tfrac{1}{2} \!\left(|M_{ll}|^2 + |M_{lr}|^2 + |M_{rl}|^2 + |M_{rr}|^2 \right) \ 
&  \mathbf{A'}_{31} &\!=\! \text{Re} \!\left(M_{ll}M_{rl}^* + M_{lr}M_{rr}^* \right) \nonumber\\ 
\mathbf{A'}_{12} &\!=\! \tfrac{1}{2} \!\left(|M_{ll}|^2 - |M_{lr}|^2 + |M_{rl}|^2 - |M_{rr}|^2 \right)  \ 
& \mathbf{A'}_{32} &\!=\! \text{Re}\!\left(M_{ll}M_{rl}^* - M_{lr}M_{rr}^* \right)  \nonumber\\
\mathbf{A'}_{13} &\!=\!  \text{Re} \left(M_{ll}M_{lr}^* + M_{rl}M_{rr}^* \right)  \ 
& \mathbf{A'}_{33} &\!=\! \text{Re} \left(M_{ll}M_{rr}^* - M_{lr}M_{rl}^* \right)  \nonumber\\
\mathbf{A'}_{14} &\!=\! -\text{Im}\!\left(M_{ll}M_{lr}^* +M_{rl}M_{rr}^* \right) \ 
&\mathbf{A'}_{34} &\!=\!  -\text{Im}\!\left(M_{ll}M_{rr}^* + M_{rl}M_{lr}^* \right)  \\
\mathbf{A'}_{21} &\!=\!\tfrac{1}{2} \!\left(|M_{ll}|^2 + |M_{lr}|^2 - |M_{rl}|^2 - |M_{rr}|^2 \right)  \ 
&\mathbf{A'}_{41} &\!=\!\text{Im}\!\left(M_{ll}M_{rl}^* + M_{lr}M_{rr}^* \right)   \nonumber\\
\mathbf{A'}_{22} &\!=\! \tfrac{1}{2} \!\left(|M_{ll}|^2 - |M_{lr}|^2 - |M_{rl}|^2 + |M_{rr}|^2 \right)  \ 
&\mathbf{A'}_{42} &\!=\! \text{Im}\!\left(M_{ll}M_{rl}^* + M_{rr}M_{lr}^* \right)  \nonumber\\
\mathbf{A'}_{23} &\!=\! \text{Re}\! \left(M_{ll}M_{lr}^* - M_{rl}M_{rr}^* \right)  \ 
&\mathbf{A'}_{43} &\!=\! \text{Im} \!\left(M_{ll}M_{rr}^*+ M_{lr}M_{rl}^* \right) \nonumber\\
\mathbf{A'}_{24} &\!=\! -\text{Im} \left(M_{ll}M_{lr}^* - M_{rl}M_{rr}^* \right)   \ 
&\mathbf{A'}_{44} &\!=\!  \text{Re}\! \left(M_{rr}M_{ll}^* - M_{rl}M_{lr}^* \right)  \nonumber\, .
 \end{empheq}

\subsubsection{Relationship between the initial and final Chandrasekhar Stokes parameters} 

To compare our results with that of Chandrasekhar, one needs  to perform the following transformation 

\begin{equation}
\begin{bmatrix}
 \hat{I_l} \\
  \hat{I_r} \\
  \hat{U} \\
  \hat{V}
\end{bmatrix} =
 \mathbf{C}
\begin{bmatrix}
 \hat{I} \\
  \hat{Q} \\
  \hat{U} \\
  \hat{V}
\end{bmatrix}  \, , \,\, \text{with}\, \,  
\mathbf{C}=
\begin{bmatrix}
\frac{1}{2} & \frac{1}{2} & 0 & 0 \\
\frac{1}{2} & -\frac{1}{2} & 0 & 0 \\
0 & 0 & 1 & 0 \\
0 & 0 & 0 & 1 
\end{bmatrix}\,. 
\label{eq:stokes_rotation}
\end{equation}
 This transformation is only valid in the $(l,r)$  basis where we can decompose the intensity $I$ in terms of  $I_l$ and $I_r$. 
 Using this transformation, the change in the modified Stokes parameters after scattering reads as 
\begin{equation}
\begin{bmatrix}
\hat{I_l}\\\hat{I_r}\\\hat{U}\\\hat{V}
\end{bmatrix}^{(2)} =
 \underbrace{\mathbf{C} \ \mathbf{A'} \  \mathbf{C}^{-1}}_{\mathbf{A}} 
\begin{bmatrix} 
\hat{I_l}\\\hat{I_r}\\\hat{U}\\\hat{V}
\end{bmatrix}^{(1)} 
\label{eq:General_P_matrix}
\end{equation}

with 
\begin{eqnarray} \label{Eq:Rdef_M}
\mathbf{A} \ = \ \left[
\begin{array}{cccc}
M_{ll}  M^*_{ll}&M_{lr} M^*_{lr} &  \frac{1}{2}(M_{ll}M_{lr}^*+ M_{lr}M_{ll}^*) & \frac{1}{2} i (M_{ll}M_{lr}^*- M_{lr}M_{ll}^*) \\
  M_{rl}  M^*_{rl} & M_{rr} M^*_{rr} & \frac{1}{2}(M_{rl}M_{rr}^*+ M_{rr}M^*_{rl}) & \frac{1}{2} i (M_{rl}M^*_{rr}- M_{rr}M^*_{rl}) \\
M_{ll} M_{rl}^*  +M_{rl} M_{ll}^*  & M_{lr}M^*_{ll}+M_{rr} M^*_{lr} &\mathbf{A}_{33} & \mathbf{A}_{34} \\
- i (M_{ll} M_{rl}^*  - M_{rl} M_{ll}^* ) &- i (M_{lr}M^*_{ll}-M_{rr} M^*_{lr}) & \mathbf{A}_{43} &\mathbf{A}_{44} \\
\end{array}
\right]
\end{eqnarray} 
and 
\begin{align}
    \mathbf{A}_{33}=& \frac{1}{2} (M_{ll}  M^*_{rr} + M_{lr}M^*_{rl}+M_{rl} M^*_{lr}+ M_{rr} M^*_{ll} ) \,,\nonumber\\
    \mathbf{A}_{34}=&\frac{1}{2} i (M_{ll}  M^*_{rr} -M_{lr}M^*_{rl}+M_{rl} M^*_{lr}-M_{rr} M^*_{ll} )\,,\nonumber\\
    \mathbf{A}_{43}=&-\frac{1}{2} i(M_{ll}  M^*_{rr} + M_{lr}M^*_{rl}-M_{rl} M^*_{lr}- M_{rr} M^*_{ll} )\,,\nonumber\\
    \mathbf{A}_{44}=& \frac{1}{2} (M_{ll}  M^*_{rr} - M_{lr}M^*_{rl}-M_{rl} M^*_{lr}+ M_{rr} M^*_{ll} ) \,.
\end{align}

We note that, at low energy, the above expression can be used  to derive both $\mathbf{P}$ and $\mathbf{R}$--matrices, depending on the choice of kinematics. Using the rest frame kinematics leads to the expression of the $\mathbf{R}$--matrix while the fixed  frame kinematics  (see Eq.~(\ref{eq:General_P_matrix})) leads to the $\mathbf{P}$--matrix. Finally we have checked that these expressions are consistent with Chandrasekhar's definitions of the $\mathbf{R}$ and $\mathbf{P}$ matrices in the case of Thomson interactions in Appendix~\ref{sec:comparisonCQThomson}.

\section{Compton interactions \label{sec:compton}}
We are now equipped to determine the change in polarisation of $\gamma$-rays after they scatter off electrons, whatever the energy regime (and in particular  when the initial energy exceeds the electron mass, that is  $\sqrt{s} > m_e$).  Our formalism is general enough to study dark photon scattering off electrons or photon scattering off Beyond Standard Model particles. Unlike Thomson interactions which do not flip the spin of the electron, Compton interactions can affect both the photon polarisation and the electron spin configuration due to the energies at play. As a result, we expect the relation between the outgoing and incoming Stokes parameters to be much more complex than in the case of Thomson interactions, and to depend on both the momentum and energy of the incoming particles.

We note that other works have attempted to describe the change in linear polarisation of high energy gamma-rays after scattering. However there is a number of issues. For example, Ref.~\cite{FERNANDEZ1993579} uses Chandrasekhar's geometrical approach to describe the scattering of X-rays and gamma--rays but the geometrical approach does not capture the complexity of Compton scattering interactions at high energy (in particular the presence of helicity-flip processes).  The Klein—Nishina formula has also been used in \cite{RevModPhys.33.8} to describe the linear polarisation of high energy gamma-rays but it is only valid when the electrons are strictly at rest and thus does only apply in very specific circumstances.  Finally,  other works have used the photon density matrix to described the process of polarisation transfer \cite{Mao_2017,PhysRevSTAB.18.110701,Stock:2015yha} but have not folded in the information about the cross section which is critical at high energy ($\sqrt{s} > m_e$). The  formalism that we have developed thus aims to provide a consistent treatment of polarisation after scattering, whatever the energy regime one is considering.

In the previous section, we have defined the $\mathbf{R}$--matrix in the ($l,r$) basis and showed how to convert it in the $\pm$ basis. Defining the $\mathbf{R}$--matrix in the ($l,r$) basis was straightforward because at low energy, the scattering only induces a change in the photon direction (i.e. $I_l^{(2)} = \cos^2\theta \ I_l^{(1)}$ and $I_r^{(2)} =  I_r^{(1)}$). However at high energies, the effect of the scattering is much more complex and one needs to account for helicity-flip processes. This means that i) one needs to use a QFT approach and ii)  the $\mathbf{R'}$--matrix in the $\pm$ basis gives more information about the physical process than the $\mathbf{R'}$--matrix in the $(l,r)$ basis.

For clarity, we  remind the reader of our notations: 
\begin{itemize}
    \item $\mathbf{A'}$\textbf{--matrix}: This the most generic relation between the incoming and outgoing ($I,Q,U,V$)  parameters. It is valid for any photon energy and scattering off any type of particle and can be computed using both the $\pm$ and the ($l,r$) photon helicity states (see Eq. \eqref{eq:stokes2_stokes1_amp_def} and Eq. \eqref{eq:stokes2_stokes1_amplr_def} respectively). 
    \item $\mathbf{A}$\textbf{--matrix}: Similarly, the $\mathbf{A}$--matrix is the most generic relation between the incoming and scattered $(I_l,I_r,U,V)$ parameters. Due to the definition of the modified Stokes parameters, the $\mathbf{A}$-matrix is only expressed in terms of the $(l,r)$ photon polarisation states.
    \item $\mathbf{R'}$\textbf{--matrix}   \textbf{and} $\mathbf{P}'$\textbf{--matrix}: These are the $\mathbf{A'}$-matrices expressed in the rest-frame (scattering plane)  and fixed frame kinematics (3D plane) respectively. 
    \item $\mathbf{R}$\textbf{--matrix} \textbf{and} $\mathbf{P}$\textbf{--matrix}: These are the $\mathbf{A}$-matrices expressed in the rest-frame (scattering plane)  and fixed frame kinematics (3D plane) respectively. 
    \end{itemize}

\subsection{$\mathbf{A'}$--matrix for Compton interactions \label{ssec:VIA}} 
We can now determine how the Stokes parameters change after $\gamma$-rays scatter off electrons by inserting the Compton scattering matrix elements in Eq.~\eqref{eq:M_def}. 
Using the definitions $M_{i'i} \equiv \mathcal{M}(e_\alpha \gamma_i \to e_\beta \gamma_i')$ (where $\alpha,\beta=\pm$ denote the electron spin configurations) and 
\begin{equation}
M_{i'i} M^\ast_{j'j}\equiv \frac{1}{2}\sum_{\alpha,\beta=\pm} \mathcal{M}(e_\alpha \gamma_i \to e_\beta \gamma_i') \, \mathcal{M}^\ast(e_\alpha \gamma_j \to e_\beta \gamma_j')\,  \label{eq:General_amplitude}
\end{equation}
we find that the $\mathbf{A'}$--matrix simplifies to
\begin{equation}
\mathbf{A'} = 
\begin{bmatrix}
  \mathbf{A'}_{11} &\mathbf{A'}_{12} & 0 & 0 \\
    \mathbf{A'}_{21} &\mathbf{A'}_{22} & 0 & 0  \\
    0 & 0 &\mathbf{A'}_{33} & 0  \\
    0 & 0 &0 &\mathbf{A'}_{44} 
\end{bmatrix} 
\label{eq:A_matrix_Compton}
\end{equation}
with
\begin{align}
&\mathbf{A'}_{11} = 2 \, \left(\frac{\piki}{\piko}+\frac{\piko}{\piki}\right)+ 4 \, m_e^2\left(\frac{1}{\piki}-\frac{1}{\piko}\right)+  2 \, m_e^4 \left(\frac{1}{\piki}-\frac{1}{\piko}\right)^2   \,, \nonumber\\
&\mathbf{A'}_{12}  =\mathbf{A'}_{21} = 4 \, m_e^2\left(\frac{1}{\piki}-\frac{1}{\piko}\right)+  2 \, m_e^4 \left(\frac{1}{\piki}-\frac{1}{\piko}\right)^2\,, \nonumber\\
&\mathbf{A'}_{22} = 2 \, + \, 2 \, \left(1 + m_e^2\left(\frac{1}{\piki}-\frac{1}{\piko}\right)\right)^2\,, \nonumber\\
&\mathbf{A'}_{33} = 4 \, + 4 \, m_e^2\left(\frac{1}{\piki} - \frac{1}{\piko}\right) \,, \nonumber\\
&\mathbf{A'}_{44} = 2 \, \left(\frac{\piki}{\piko} +\frac{\piko}{\piki} \right) + 2 \, m_e^2 \, \left( \frac{\piki}{\piko}+\frac{\piko}{\piki}\right)\left(\frac{1}{\piki}-\frac{1}{\piko}\right)\,,
\label{eq:PinTermsOfPk}
\end{align}
where $p_1$ is the 4--momentum of the incoming electron and $k_1, k_2$ are the 4--momentum of the incoming and outgoing photons respectively.


\subsection{$\mathbf{R'}$ and $\mathbf{P'}$--matrices for Compton interactions }

We can now express the $\mathbf{R'}$ and  $\mathbf{P'}$--matrices for Compton scattering by taking the appropriate kinematics. 
In the rest frame of the electron,  the $\mathbf{R'}$-matrix reads as 
\begin{eqnarray}
 \mathbf{R'}=\frac{2}{m_e}
 \begin{bmatrix}
\Delta E_\gamma(1-\cos\theta)+m_e(1+\cos^2\theta)&-m_e\,\sin^2\theta &0&0\\
 -m_e\,\sin^2\theta&m_e(1+\cos^2\theta)&0&0\\
 0&0&2m_e\,\cos\theta&0\\
 0&0&0&[2m_e+\Delta E_\gamma(1-\cos\theta)] \cos\theta \label{eq:Rprime_matrix}
 \end{bmatrix} 
 \end{eqnarray}
with $\Delta E_\gamma\equiv(E_{\gamma,1}-E_{\gamma,2})$, $\theta$  the angle between the incoming and outgoing photons. Here $E_{\gamma,1}$ and $E_{\gamma,2}$ are the energies of the incoming and outgoing photons, respectively. 
The $\mathbf{P}'$--matrix can  be calculated using the fixed frame kinematics and is also included in the Mathematica notebook submitted with this paper.


\subsection{$\mathbf{R}$ and $\mathbf{P}$ --matrices for Compton interactions}

Similarly we can find the $\mathbf{R}$--matrix using the expression of the $\mathbf{A}$--matrix in the $(l,r)$ basis, that is
\begin{equation}
\mathbf{A} = 
\begin{bmatrix}
  \mathbf{A}_{11} &\mathbf{A}_{12} & 0 & 0 \\
    \mathbf{A}_{21} &\mathbf{A}_{22} & 0 & 0  \\
    0 & 0 &\mathbf{A}_{33} & 0  \\
    0 & 0 &0 &\mathbf{A}_{44} 
\end{bmatrix} 
\label{eq:Alr_matrix_Compton}
\end{equation}
with
\begin{align}
&\mathbf{A}_{11} = 4+\frac{2 \,  m_e^2 \, (\piki-\piko) \left(m_e^2 \  (\piki-\piko) \,  - \,  2 \  \piki \,  \piko\right)}{(\piki)^2 (\piko)^2}, \nonumber\\
&\mathbf{A}_{12}  =\mathbf{A}_{21} = \frac{2 \,  m_e^2 \ (\piki-\piko) \left(m_e^2 \  (\piki-\piko) \,  - \,  2 \,  \piki \ \piko\right)}{(\piki)^2 \, (\piko)^2}, \nonumber\\
&\mathbf{A}_{22} = \frac{2 \,  \left(m_e^4 \,  (\piki-\piko)^2 + \,  2 \,  m_e^2 \  \piki \ \piko \  (\piko-\piki) \, + \, \piki \ \piko \left(\piki^2 + \piko^2\right)\right)}{(\piki)^2 (\piko)^2}, \nonumber\\
&\mathbf{A}_{33} = \frac{2 \left((\piki)^2 + (\piko)^2\right) \left(m_e^2 \  (\piko-\piki) + \piki \  \piko\right)}{(\piki)^2 (\piko)^2}, \nonumber\\
&\mathbf{A}_{44} = 4 \,  + 4 \,  m_e^2 \,  \left(\frac{1}{\piki}-\frac{1}{\piko}\right).
\label{eq:Alr_inTermsOfPk}
\end{align}

The $\mathbf{R}$--matrix then reads as 
\begin{eqnarray}
  \begin{bmatrix}
 \hat{I_l}  \\
 \hat{I_r} \\
 \hat{U} \\ 
  \hat{V} \\ 
\end{bmatrix}
  ^{(2)}
  =
 \begin{bmatrix}
   \mathbf{R}_{11} & \mathbf{R}_{12}  &  0 &  0 \\
   
  \mathbf{R}_{21} & \mathbf{R}_{22}  &  0 &  0\\
   
 0 &0 &  \mathbf{R}_{33} &  0 \\
  
   0 & 0 & 0  &  \mathbf{R}_{44} \\

  \end{bmatrix}
 \begin{bmatrix}
 \hat{I_l}  \\
 \hat{I_r} \\
 \hat{U} \\ 
  \hat{V} \\ 
\end{bmatrix}^{(1)}
\end{eqnarray}
with
\begin{empheq}[box=\fbox]{align}
\mathbf{R}_{11} &= \tfrac{1}{2}  \left(   \mathbf{A}_{11} +   \mathbf{A}_{12} +   \mathbf{A}_{21} +   \mathbf{A}_{22} \right)\!&\!\mathbf{R}_{12} &=\tfrac{1}{2}  \left(   \mathbf{A}_{11} -   \mathbf{A}_{12} +   \mathbf{A}_{21} - \  \mathbf{A}_{22} \right) \nonumber \\ \!
\mathbf{R}_{21} &=\tfrac{1}{2}  \left(   \mathbf{A}_{11} +   \mathbf{A}_{12} -   \mathbf{A}_{21} -  \mathbf{A}_{22} \right) \!&\! \mathbf{R}_{33} &=   \mathbf{A}_{33} \label{Rmatrix} \\ \!
\mathbf{R}_{22} &= \tfrac{1}{2}  \left(   \mathbf{A}_{11} -   \mathbf{A}_{12} -  \mathbf{A}_{21} + \  \mathbf{A}_{22} \right) 
\!&\!\mathbf{R}_{44} &= \mathbf{A}_{44}\nonumber\,  
\end{empheq}
which leads in the rest frame of the electron (see Appendix \ref{Ap:kinematics})  
\begin{eqnarray}
 \mathbf{R}=\frac{2}{m_e}
 \begin{bmatrix}
 \sin^2\frac{\theta}{2}\,\Delta E_\gamma+2m_e\,\cos^2\theta&\sin^2\frac{\theta}{2}\,\Delta E_\gamma&0&0\\
 \sin^2\frac{\theta}{2}\,\Delta E_\gamma&\sin^2\frac{\theta}{2}\,\Delta E_\gamma +2m_e&0&0\\
 0&0&2m_e\,\cos\theta&0\\
 0&0&0&[2m_e+\Delta E_\gamma\,(1-\cos\theta)]\cos\theta
 \end{bmatrix} \,.
 \label{eq:R_matrix_result_compton}
 \end{eqnarray}

The expression of the $\mathbf{P}$-matrix for Compton scattering (3D, fixed frame) is too long to be given in this paper but is provided in the Mathematica notebook accompanying this paper. 

As one can see, the expression of the $\mathbf{R}$-matrix now involves $\Delta E_{\gamma}$ and a less straightforward combination of the scattering angle which is not just purely a geometrical factor.

\subsection{Cross-check in the low energy limit}
Using Eq.~(\ref{eq:R_matrix_result_compton}) (as well as the expression of the $\mathbf{P}$--matrix) and taking the low energy limit ($E_{\gamma,1} \simeq E_{\gamma,2} \ll m_e$), we could verify that we recover the same $\mathbf{R}$ and $\mathbf{P}$ --matrices as in \cite{chandrasekhar1960radiative} (see Eq.~(\ref{eq:R_mat_Chandra})), thus confirming that using a QFT approach and taking different kinematics is indeed an alternative to Chandrasekhar's geometrical approach for deriving the 
$\mathbf{P}$ --matrix. The novelty of this technique though is that it allows us to compute the relationship between the Stokes (or modified Stokes) parameters -- before and after scattering --  whatever the incident photon energy, interaction and type of scattering material (i.e. whether the particles belong to the Standard Model or to some extensions).

\section{Circular polarisation \label{sec:CP}}

Now that we have described how the polarisation of high energy photons could change after scattering, we can focus on either linear or circular polarisation. 
Circular polarisation is given by the $V$ parameters and, like observed by Chandrasekhar for Rayleigh scattering, we note that in the rest frame, there is no transfer from linear to circular polarisation and vice versa even for high energy photons. 
Indeed under these conditions, $\mathbf{A}'_{44}$ (and therefore    $\mathbf{R}'_{44}$) is secluded.  In other words,  a net circular polarisation signal cannot be converted into a linear polarisation signal and vice versa. This means that no circular polarisation signal can be created by the scattering of linearly polarised light (i.e. $V^{(2)} \neq 0$ requires that $V^{(1)} \neq 0$) and  implies that a change in circular polarisation can only occur if there is a change in the number of photons with a given polarisation state. These results have also been checked using a different approach to calculate the scattering amplitude based on its decomposition in terms of different photon polarisations. We refer the reader to Appendix \ref{App:SecondAppr} for further details.  Below is a list of possible mechanisms that may give rise to a circular polarisation signal (i.e. $V^{(1)} \neq 0$).

\begin{itemize}
\item \textit{\underline{Faraday Conversion}}

Circular polarisation can be generated by Faraday conversion of linear polarisation. This mechanism takes place when linearly polarised photons pass through a strong magnetic field. A phase shift is generated in the photon linear polarisation components, and eventually leads to a circular polarisation signal.

\item \textit{\underline{Bi-refringence}}

Another way to produce circular polarisation signals is through birefringence, see \cite{Montero-Camacho:2018vgs} for details. Birefringence occurs when a linear polarisation signal passes through a medium of aligned grains whose alignment twists along the line of sight. When the amount of linear polarisation is high (as this could occur for example in reflection  nebulae), a high degree of circular polarisation  could be produced \cite{1979ApJ229954W}.

\item \textit{\underline{Synchrotron emission}}

Synchrotron emission can emit polarised light ~\cite{Zhang:2013bna, Zhang:2014pza,Beckert:2001az,Ensslin:2002gn,0004-637X-556-1-113,1998Nature_395_457w,0004-637X-521-2-582,1538-4357-523-1-L29,1538-4357-526-2-L85,0004-637X-571-2-843} and was proposed as a source of circular polarisation in e.g. Ref~\cite{Westfold_1959} and then developed by Legg~\cite{Legg_1968}. This mechanism continues to be studied, in particular in light of the recent progress regarding strong magnetic fields as well as homogeneous and inhomogeneous magnetic fields see Ref.~\cite{deBurca:2015kea,osti_7310672,19770055569}. We note in addition that a circular polarisation  signal may be of intrinsic origin~\cite{1538-4357-530-1-L29,Brunthaler:2001dq} (i.e., it could be generated in absence of \FC of linear polarisation in sites where there exists a large-scale magnetic field).

 \item \textit{\underline{Parity-violating interactions and charge asymmetry}}

Circular polarisation could be produced in cosmic accelerators when Standard Model particles of the same charge (for example proton-proton) collide with each other and produce an excess of positive (negative) mesons and muons with respect to their negative (positive) counterparts \cite{Boehm:2019yit}, which eventually decay radiatively through parity-violating interactions.  Electroweak (loop-induced) interactions of photons with the cosmological neutrino background could also create a circular polarisation signal~\cite{Mohammadi:2013dea}, although the signal is expected to be very small.  A circular polarisation signal could also be generated by beyond the Standard Model interactions, see for example \cite{Alexander:2008fp, Shakeri:2017iph, Sadegh:2017rnr, Montero-Camacho:2018vgs,Giovannini:2009ru,Gorbunov:2016zxf,Boehm:2017nrl,Elagin:2017cgu,Huang:2018qui,Kumar:2016cum}.
 
 \item \textit{\underline{21 cm}}

Finally  a circular polarisation signal is expected in conjunction with the 21 cm line~\cite{Hirata:2017dku}. The Hydrogen excitation that generates the circular polarisation signal is produced owing to the interaction of the Hydrogen with the CMB quadrupole moment and could be measured in the future with an array of dipole antennas. Such a signal would indicate  the existence of primordial gravitational waves \cite{Mishra:2017lpz}.
\end{itemize}

\subsection{Scattering with unpolarised electrons} 
We can now study the change of polarisation when high energy $\gamma$--rays hit electrons. Since $\mathbf{A'}_{i4, i \neq 4} = 0$  (i.e. there is no transfer of polarisation from linear to circular polarisation and vice versa), 
a possible change in the magnitude of the circular polarisation signal after scattering, $V^{(2)} \neq V^{(1)}$, has to reflect the number of photons whose polarisation state is changed by the scattering process. If the scattering is as likely to change the photon polarisation as to maintain it,  we would expect no net circular polarisation -- that is $V^{(2)} =0$ -- regardless of the initial net polarisation. Therefore the change in net circular polarisation ($\Delta_{V} $) can be formulated in terms of the scattering matrix amplitudes, as 

\begin{eqnarray} 
\Delta_{V} = \frac{|\mathcal{M}(e\gamma_+ \to e\gamma_+)|^2 + |\mathcal{M}(e\gamma_- \to e\gamma_-)|^2 - |\mathcal{M}(e\gamma_+ \to e\gamma_-)|^2 - |\mathcal{M}(e\gamma_- \to e\gamma_+)|^2}{ |\mathcal{M}(e\gamma_+ \to e\gamma_+)|^2 + |\mathcal{M}(e\gamma_- \to e\gamma_-)|^2 +|\mathcal{M}(e\gamma_+ \to e\gamma_-)|^2 + |\mathcal{M}(e\gamma_- \to e\gamma_+)|^2 }=\frac{\mathbf{A'}_{44}}{\mathbf{A'}_{11}} \,, 
\label{eq:PolPer}
\end{eqnarray}
where
\begin{eqnarray}
|\mathcal{M}(e\gamma_\pm \to e\gamma_\pm)|^2
&=&2 \left(\frac{\piki}{\piko}+\frac{\piko}{\piki}\right)+ m_e^2 \left(2+\frac{\piki}{\piko}+\frac{\piko}{\piki}\right) \left(\frac{1}{\piki}-\frac{1}{\piko}\right)+  m_e^4 \left(\frac{1}{\piki}-\frac{1}{\piko}\right)^2,\nonumber\\ 
|\mathcal{M}(e\gamma_\pm \to e\gamma_\mp)|^2 
&=& m_e^2 \left(2-\frac{\piki}{\piko}-\frac{\piko}{\piki}\right) \left(\frac{1}{\piki}-\frac{1}{\piko}\right) + m_e^4 \left(\frac{1}{\piki}-\frac{1}{\piko}\right)^2.
\label{eq:lorentz_invariant_amp}
\end{eqnarray}
From this expression, one readily sees that the value of $\Delta_{V}$ is limited to the range $[-1,1]$ and different values can be interpreted as follows.
\begin{itemize}
    \item $\Delta_{V}=1$: There is no change in the initial value of the $V$ parameter (i.e., no change in the amount of circular polarisation left after scattering).
    \item $0<\Delta_{V}<1$: The initial circular polarisation is partly washed out by the scattering.  
    \item $\Delta_{V}=0$: Any net circular polarisation will be erased completely after one single scattering.
    \item $-1<\Delta_{V}<0$: The sign of the circular polarisation is changed and most polarisation states have flipped.
    \item $\Delta_{V}=-1$: All polarisation states have flipped ($V^{(2)}=-V^{(1)}$).
\end{itemize}

A similar information can be defined in terms of the total cross section corresponding to each amplitude (see Appendix \ref{App:xsec_frames} for details). 
We can now determine how likely a net circular polarisation signal is expected to change after Compton scattering  as a function of the incoming and outgoing kinematics.
\begin{figure}[ht!]
\centering
\subfloat[Centre of Mass Frame]{\label{fig:a}\includegraphics[width=0.45\linewidth]{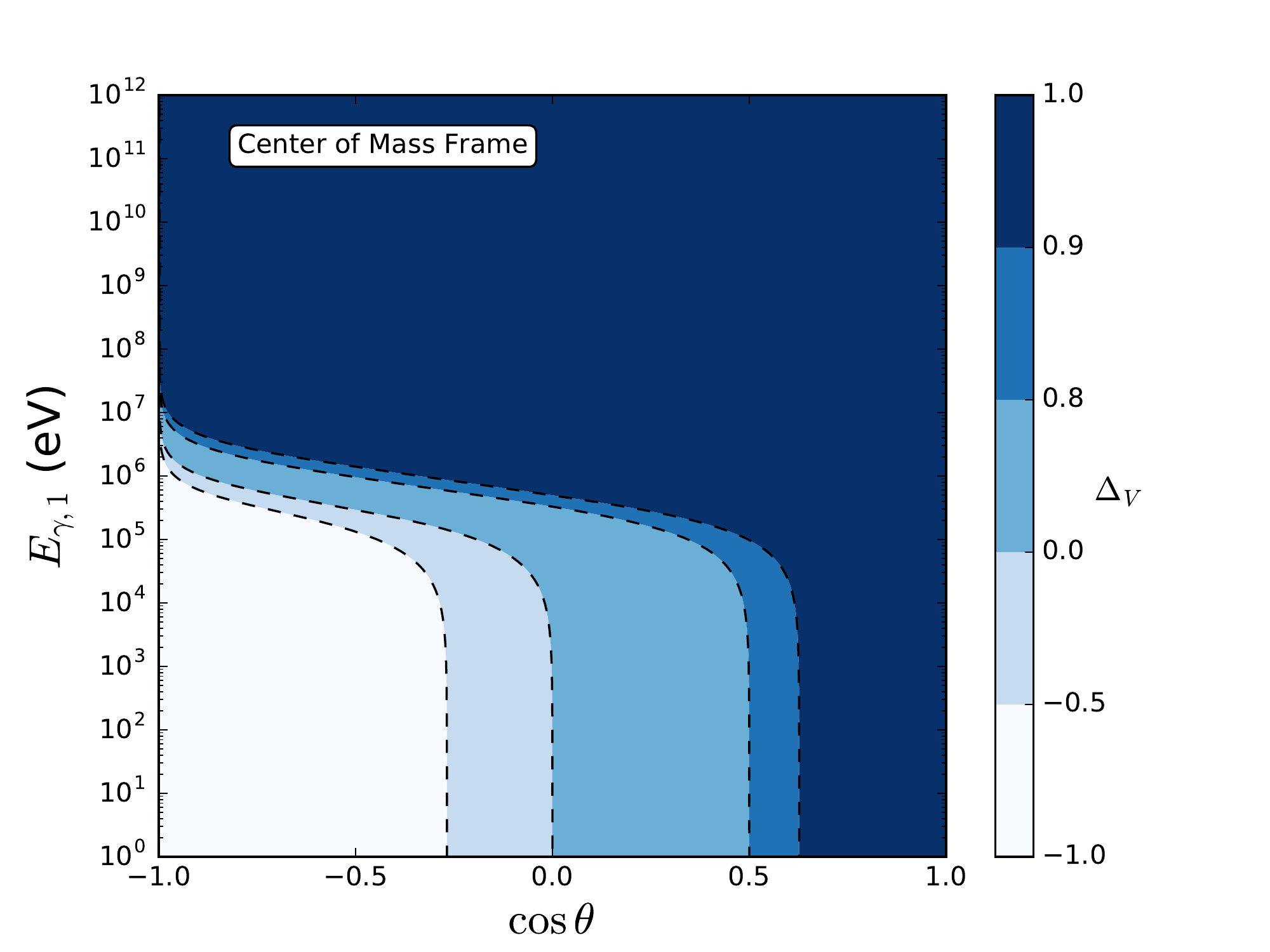}}\qquad
\subfloat[Rest Frame]{\label{fig:b}\includegraphics[width=0.45\linewidth]{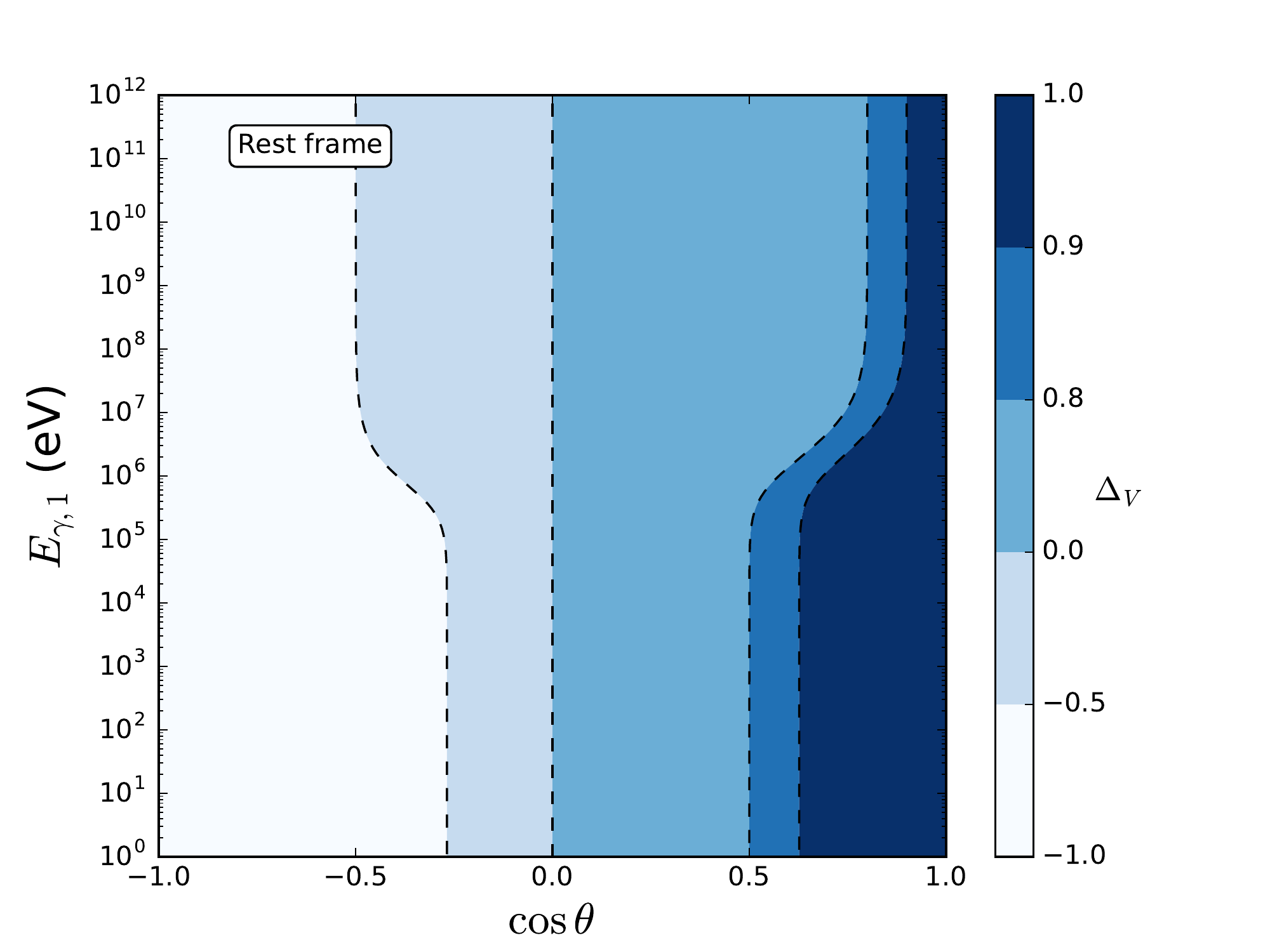}}\\
\subfloat[Spin Frame]{\label{fig:c}\includegraphics[width=0.45\textwidth]{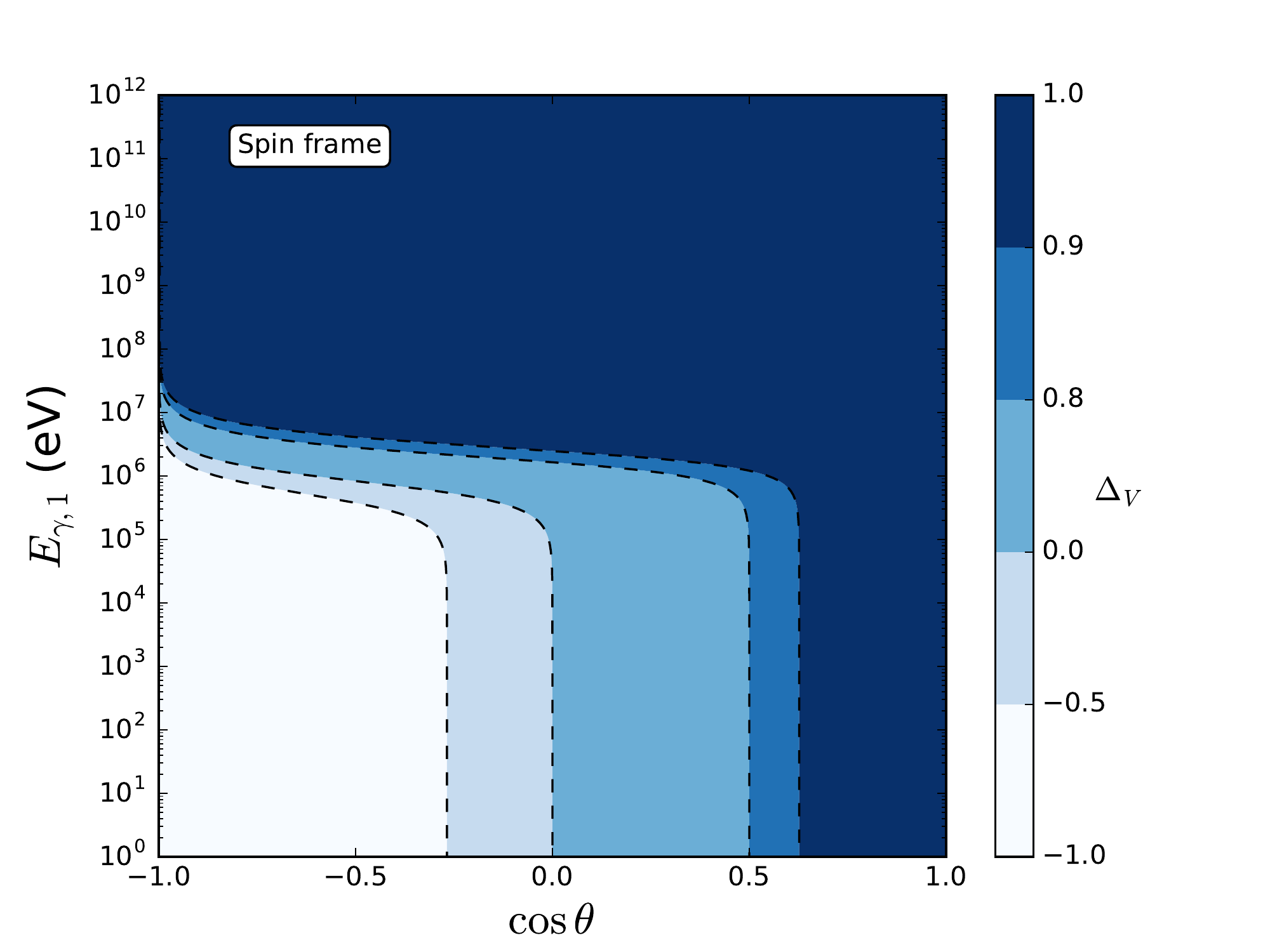}}\qquad%
\subfloat[Fixed Frame]{\label{fig:d}\includegraphics[width=0.45\textwidth]{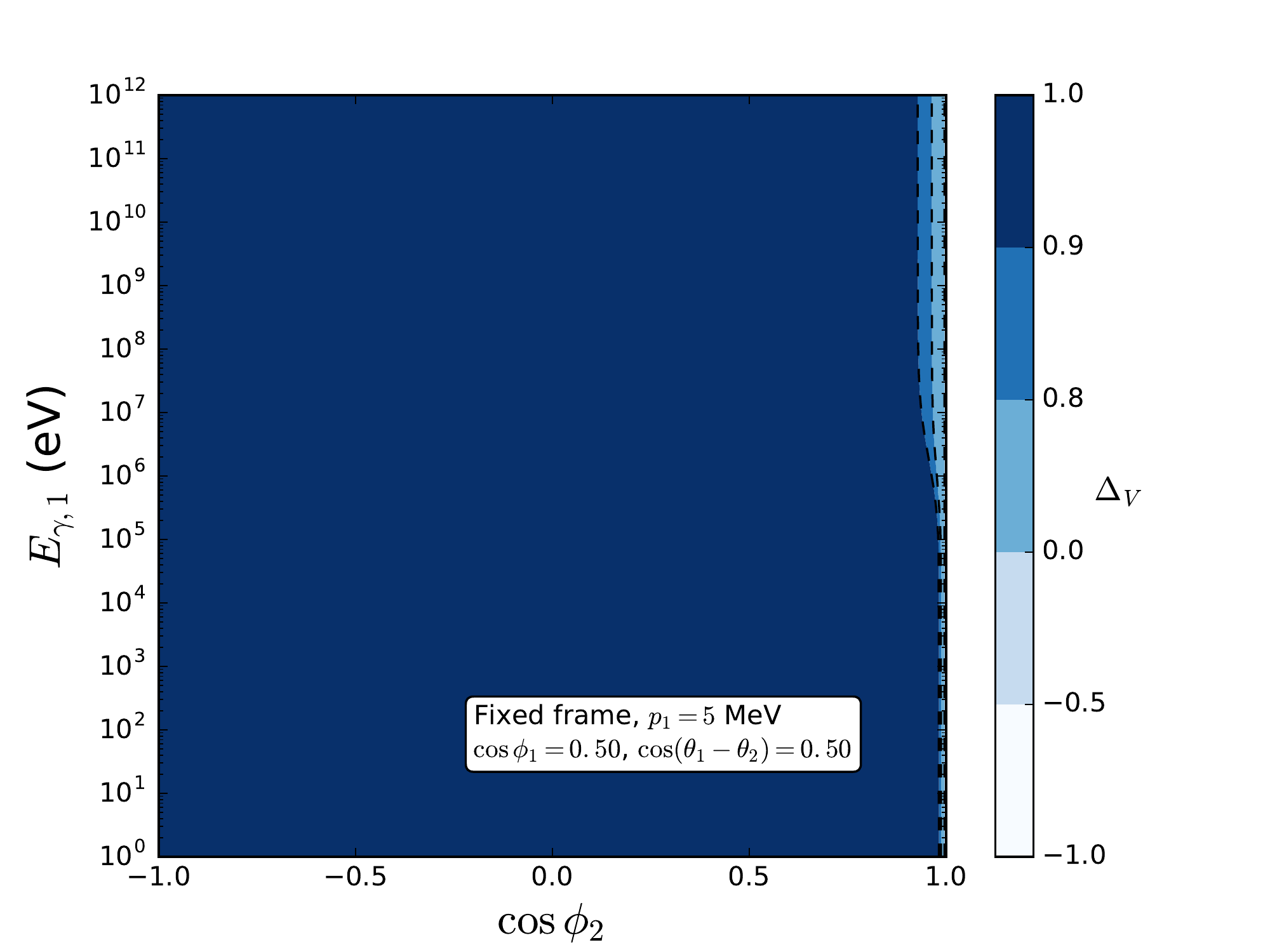}}%
\caption{The change of net circular polarisation after a single $e-\gamma$ scattering in (a) the centre of mass frame, (b) the electron rest frame, (c) the spin frame, and (d) the fixed frame. In the former three frames, we choose the incoming photon energy $E_{\gamma, 1}$ and the angle between outgoing and incoming photons $\theta$ as variables. In the last frame, the incoming photon energy $E_{\gamma, 1}$ and the angle between outgoing photon and incoming electron are chosen as variables. }
\label{fig:PercResults}
\end{figure}
In Fig. \ref{fig:PercResults} we present the asymmetry $\Delta_{V}$ defined in Eq. (\ref{eq:PolPer}) as a function of the  incoming photon energy and the angle between the incoming and outgoing photon ($\theta$). The results are shown in the centre of mass frame (COM), the rest frame, the spin frame and the fixed frame. The rest frame is very useful for energetic photons propagating through a medium and scattering with very low energy (background) electrons. The spin frame is a new frame that we define in this paper to reflect the fact that for incoming and outgoing electrons travelling in the $\mp z$, the spinor definitions that we are using to calculate the matrix amplitude (see Appendix 
\ref{App:C}) correspond to the spin eigenstates of the electrons. The results in this frame thus match 
the result in the COM frame. The fixed frame gives the most general description of electron photon scattering. The only constraint is that the momentum of the incoming electron is fixed along the $+z$ direction. This gives us more freedom about the angular configuration of the scattering. In this frame, it is not very intuitive to show $\Delta_{V}$ as a function of $\theta$ so we will present it as a function of the angle between  the outgoing photon and the incoming electron (i.e., $ \phi_2$),  as in  Fig. \ref{fig:d}. In this figure, we have fixed the incoming electron momentum at $p_1=5$ MeV, as well as the angles $\phi_1=\theta_1-\theta_2=\pi/3$, where $\phi_1$ is the angle between incoming electron and photon and $\theta_1-\theta_2$ is the difference of the angles of the incoming and outgoing photons projected on the plane perpendicular to the $z$ direction. Note that this plot is just shown as an example, more results for the fixed frame can be found in Appendix \ref{App:FixedFramePlots}. The polarisation behaviour is dependent on all these kinetic variables. If one of them changes, $\Delta_{V}$  changes quite a bit too. For more information about how $\Delta_{V}$ varies in the different frames of reference, we refer the reader to Appendix \ref{Ap:kinematics}.

A common feature among the first three frames is that, for a low energy incoming photon, $E_{\gamma,1} \ll m_e$, the asymmetry $\Delta_{V}$ crucially depends on the direction of the outgoing photon after scattering. The fixed frame does not follow this feature simply because we have assumed a relativistic incoming electron by fixing $p_1$ at 5 MeV. If the incoming electron is non-relativistic (as can be seen in the appendix \ref{App:FixedFramePlots}) we get similar results as in the other frames.

In the high energy regime of the incoming photon ($E_{\gamma,1} \gg m_e$), Compton scattering preserves the polarisation states of most of the outgoing photons regardless of the direction of the incoming photon. However, this is not true in the rest frame (Fig. \ref{fig:b}), where $\Delta_{V}$ strongly depends on the scattering direction of the outgoing photon with respect to the incoming photon and is independent on the incoming photon energy. When the initial electron is at rest, the only way that a ``$+$/$-$'' polarised photon can conserve angular momentum requires the photon to be scattered in the forward direction. Otherwise, it would have to flip its polarisation. On the contrary, when the electron has some initial energy, (like in the spin or COM frames for $E_{\gamma,1} \gg m_e$), the photon can scatter in any direction while keeping its initial polarisation without violating angular momentum conservation. 
One common feature of all the frames in this regime is that the change of the net polarisation no longer depends on the incoming photon energy. However, when the energy of the incoming photon is around or slightly above the electron mass, the value of $\Delta_{V}$ then becomes strongly frame dependent.

\subsection{Scattering off polarised electrons} 

So far we have averaged over the electron spin configuration. However it may be that the light propagates in a medium where the electrons have one particular spin configuration. To study whether the net circular polarisation will be affected in such a medium, we define the change of the net circular polarisation asymmetry by: 
\begin{equation}
\Delta_{V,e_+}=\frac{|\mathcal{M}(e_+ \gamma_+ \to e_+ \gamma_+)|^2 + |\mathcal{M}(e_+ \gamma_+ \to e_- \gamma_+)|^2-|\mathcal{M}(e_+ \gamma_+ \to e_+ \gamma_-)|^2 - |\mathcal{M}(e_+ \gamma_+ \to e_- \gamma_-)|^2}{|\mathcal{M}(e_+ \gamma_+ \to e_+ \gamma_+)|^2 + |\mathcal{M}(e_+ \gamma_+ \to e_- \gamma_+)|^2+|\mathcal{M}(e_+ \gamma_+ \to e_+ \gamma_-)|^2 + |\mathcal{M}(e_+ \gamma_+ \to e_- \gamma_-)|^2}\label{eq:spin_CCP}
\end{equation}
for the cases of the electron spin $J_z(e_1)=+ \frac{1}{2}$ in the rest frame.  Similarly, one can define $\Delta_{V,e_-}$ for the case of $J_z(e_1)=- \frac{1}{2}$ by exchanging $e_+$ and $e_-$. Here we only show the results in the rest frame as a case of study, but it is worth noting that this definition is valid in all four frames of reference as long as the initial electron momentum is zero or along the $z$--axis. When the electron travels at an angle with respect to the $z$--axis, the spinor defined in Appendix~\ref{App:C} is no longer an eigenstate of the spin operator and consequently, does not carry any physical meaning. 

\begin{figure}[ht!]%
    \centering
\subfloat[$J_z(e_1)=+\frac{1}{2}$.]{\label{fig:1}{\includegraphics[width=0.45\textwidth]{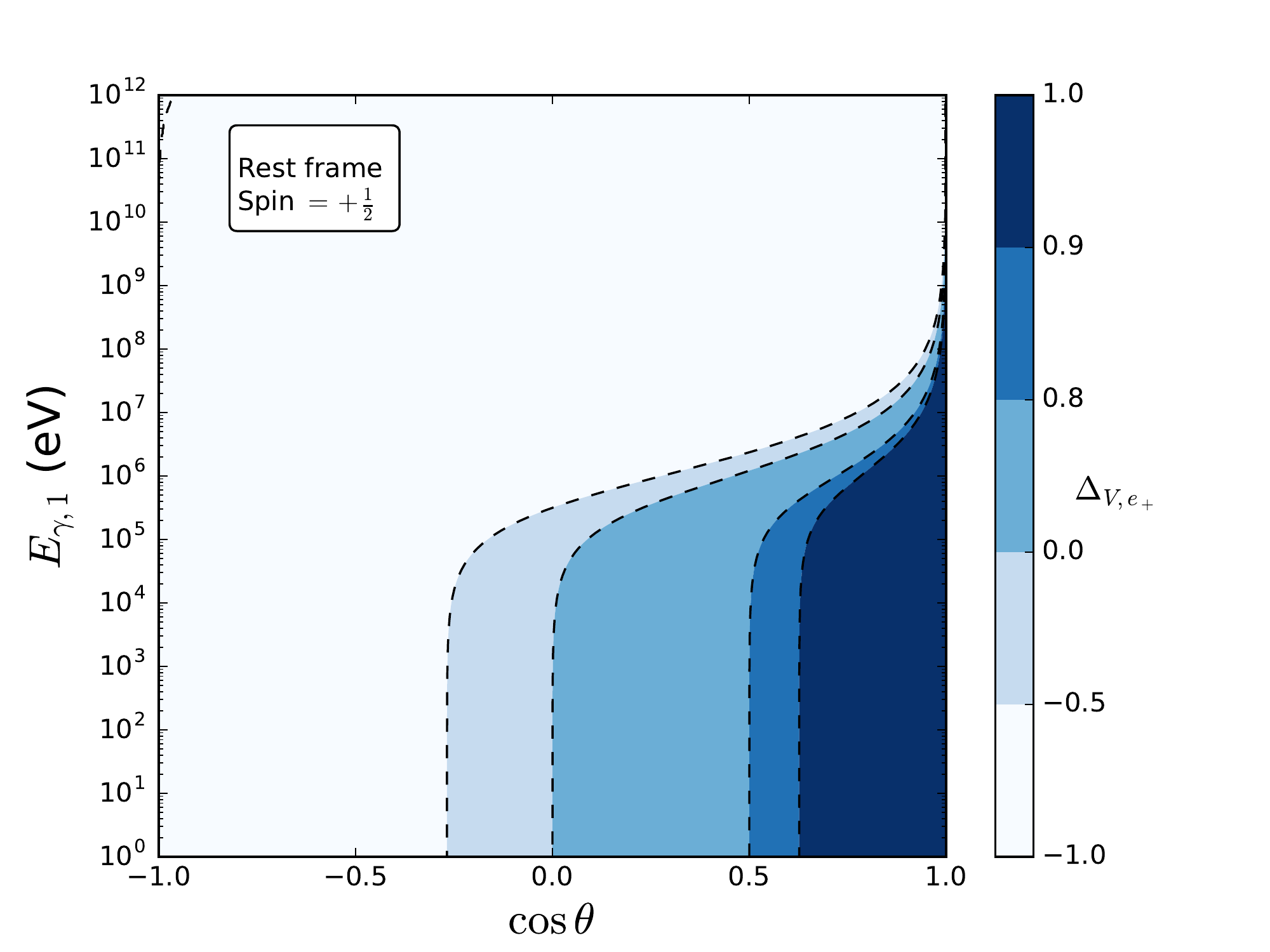} }}%
    \qquad
 \subfloat[$J_z(e_1)=-\frac{1}{2}$.]{\label{fig:2}{\includegraphics[width=0.45\textwidth]{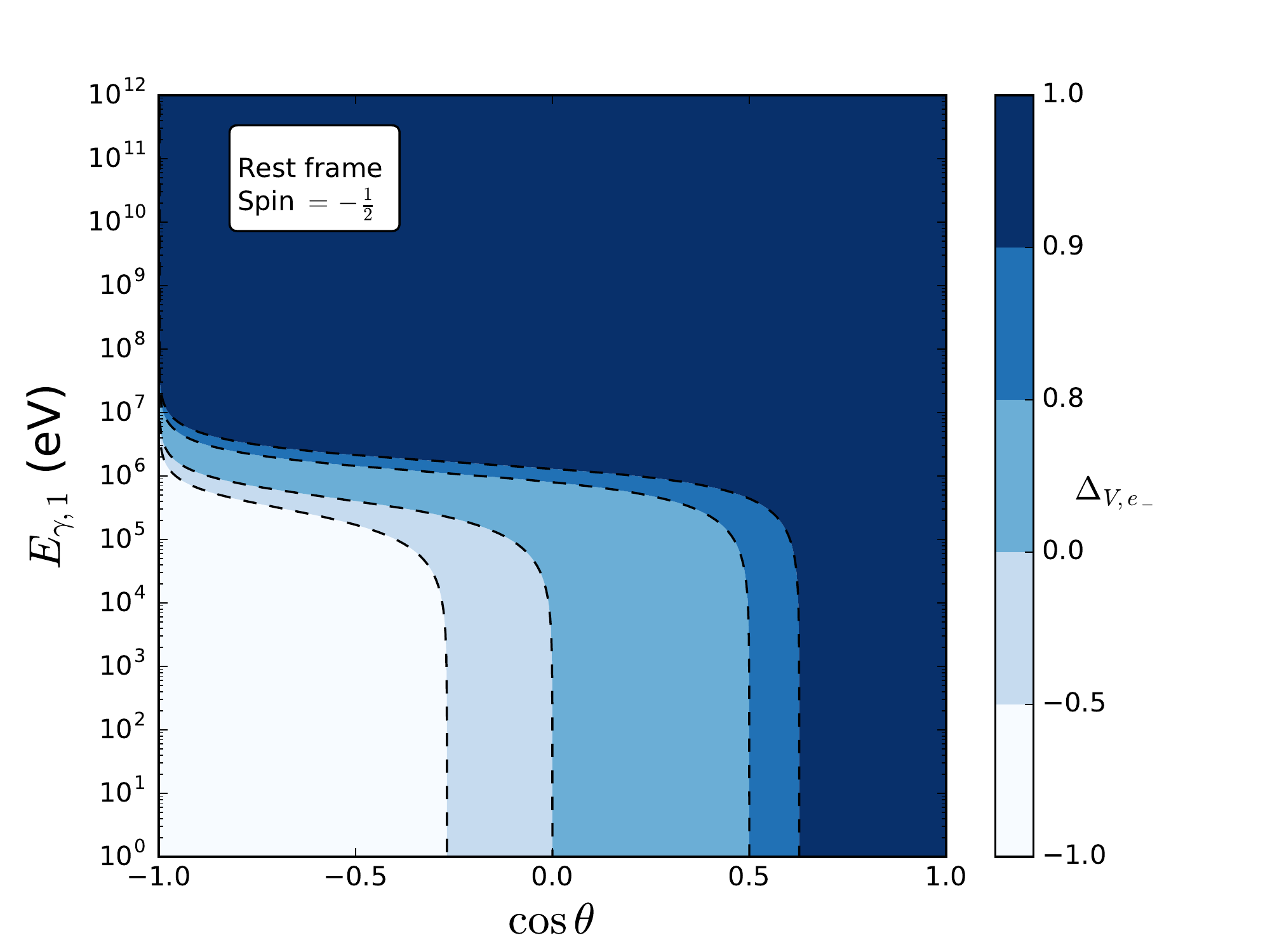} }}%
    \caption{Percentage of photon circular polarisation conservation after a single scattering with incoming polarised electron with spin $+\frac{1}{2}$ (a) or spin $-\frac{1}{2}$ (b) in the rest frame. The axis are the same as in Fig. \ref{fig:c}.}%
    \label{fig:spin_RestF}%
\end{figure}

The behaviour of $\Delta_{V,e_\pm}$ is numerically shown in Fig.~\ref{fig:spin_RestF}. In the low energy regime ($E_{\gamma,1}\ll m_e$), $\Delta_{V,e_+}$ and $\Delta_{V,e_-}$ follow the same distribution as that in the scattering with an unpolarised electron. Consequently, the change in polarisation of low energy photons scattering with electrons at rest is the same regardless the electron initial spin state. On the other hand, the results for high energy incoming photons present different features. For a photon with energy $E_{\gamma,1}> m_e$ and positive state, its polarisation is flipped if the spin of the incoming electron is $+ \frac{1}{2}$, see Fig. \ref{fig:1}, and it is conserved after scattering if the spin of the incoming electron is $- \frac{1}{2}$, see Fig. \ref{fig:2} independently of the direction of scattering of the outgoing photon. We note that, unlike the results for photon scattering with an unpolarised electron in the rest frame, the mass of the electron is a relevant energy scale when fixing the electron spin since in this case, the change of net circular polarisation behaves very differently when the incoming photon energy is above or below the electron mass. For instance,  given an incoming spin $+\frac{1}{2}$ electron with $\cos\theta$ around 1, i.e., backward scattering, the photon needs more energy to flip its polarisation.

\section{Boltzmann Equation \label{sec:boltzman}}

Now that we know how the Stokes parameters are modified after scattering, see Eq.~\eqref{eq:R_matrix_result_compton}, we can study their evolution as  light propagates through space, using the Boltzmann equation. The latter is given by 
\begin{equation}
\frac{dn}{dt}=C[n]\,,\label{Def:Boltzmann}
\end{equation}
where $n$ is the phase space photon distribution function and $C$ is the collisional term, i.e., a functional of the photon distribution function $n$ describing the scattering of the photon with any other particles $\psi$ in the medium. The latter reads as 
\begin{align}
C[n]=\int d\mathbf{p}_{1} d\mathbf{p}_{2} d\mathbf{k}_2 \ \ &\vert\mathcal{M}(\psi \gamma \to \psi \gamma)\vert^2 \ \ (2\pi)^4 \ \delta^{4}(p_1+k_1-p_2-k_2)
\ [n_{\psi}(\mathbf{p}_2)n_\gamma(\mathbf{k}_2)-n_{\psi}(\mathbf{p}_1)n_\gamma(\mathbf{k}_1)] \,,
\end{align}
where $p_{1(2)} \equiv (E_{\psi,1(2)},~\mathbf{p}_{1(2)})$ is the four-momentum of the incoming (outgoing) particle $\psi$ in the medium, $k_{1(2)} \equiv (E_{\gamma,1(2)},~\mathbf{k}_{1(2)})$ is the four-momentum of the incoming (outgoing) photon,  $\displaystyle d\mathbf{k}_2\equiv \frac{d^3\mathbf{k}_2}{(2\pi)^32E_{\gamma,2}}$, $\displaystyle d\mathbf{p}_{1(2)}\equiv \frac{d^3\mathbf{p}_{1(2)}}{(2\pi)^3 } \frac{m_{\psi}}{E_{\psi,1(2)}}$, $n_{\psi, \gamma}$ is the distribution function of the particles $\psi$ and $\gamma$ respectively, and $\vert\mathcal{M}(\psi\gamma \to \psi \gamma)\vert^2$ is the squared scattering matrix amplitude. In the current form, this equation  \textcolor{red}{ is for} any particle physics process involving photon scattering.

\subsection{Boltzmann formalism for generic interactions 
\label{sec:timeevolution}}

In order to study the evolution of the polarisation, we need to relate the photon energy distribution to the Stokes parameters.   In Eq.~(\ref{eq:Stoke_definition}), we saw that the Stokes parameters  could be expressed in terms of the different photons states. 
The next step is to relate them to 
 the density matrix. Combining Eq.~(\ref{eq:Stoke_definition}) with the definition of  density matrix 
 \begin{equation}
     \rho_{ij} = \frac{| \epsilon_i \rangle \langle \epsilon_j|}{\text{Tr}(\rho)},
 \end{equation} 
 where $i,j=\{l,r\}$ or $\{+,-\}$,  and making use of 
\begin{eqnarray}
  \langle S \rangle = \text{Tr}( \rho \hat{S} ) \,, 
\end{eqnarray}
with $S = I, Q, U, V$, the Stokes parameters in the $\pm$ basis can be expressed as 
\begin{align} \label{eq:PolBasis}
\langle I \rangle =&\ \rho_{++} + \rho_{--}  \,, \nonumber\\
\langle Q \rangle=&\ -\left(\rho_{+-} + \rho_{-+}\right) \,, \nonumber \\
\langle U \rangle=&\ i \left( \rho_{+-} - \rho_{-+}\right) \,, \nonumber \\
\langle V \rangle =& \ \rho_{++} - \rho_{--} \, 
\end{align}
To continue further, we need to define the time evolution of the different matrix density elements.  This can be done by expressing the photon number operator $\mathcal{D}_{ij}(\mathbf{k})\equiv a^\dagger_i(\mathbf{k})a_j(\mathbf{k})$  in terms of the density matrix associated with the different photon polarisation states~\cite{Kosowsky:1994cy}.  Making use of the relation
\begin{equation}
\langle \mathcal{D}_{ij} \rangle = (2\pi)^3\delta(0)2k^0 \rho_{ij}(\mathbf{k})
\end{equation}
and assuming that the collision time scale is smaller than the time scale for the variation of the density matrix (which is true for weak scale processes), we can then express the density matrix in terms of the photon number operator~\cite{Sigl:1992fn}
\begin{equation}
(2\pi)^3\delta(0)2E_k\frac{d}{dt}\rho_{ij}(\mathbf{k})=-\frac{1}{2}\int_{-\infty}^\infty dt \langle [H_I^0(t),[H_I^0(0),\mathcal{D}^0_{ij}(\mathbf{k})]]\rangle\label{generalized_blmnn_eq} \,,
\end{equation}
where $H^0_I$ is the interaction Hamiltonian to first order.

Assuming that most of the particles in space are not polarised since there is no left/right asymmetry,   Eq.~\eqref{generalized_blmnn_eq} can be rewritten as 
\begin{align}
2E_{\gamma,1} \frac{d}{dt}\rho_{ij}(\mathbf{k}_1)&=-\frac{1}{4}\int d\mathbf{p}_1 \ d \mathbf{p}_2 \  d \mathbf{k}_2 \ (2 \pi)^4 \ \delta^4(p_2+k_2-p_1-k_1) \ \ {M}_{\alpha\alpha'}{M}^\ast_{\beta' \beta}\nonumber\\
&\times[n_{\psi}(\mathbf{p}_1)  \delta_{\beta \alpha'}(\delta_{i\alpha}\rho_{\beta'j}(\mathbf{k}_1)+\delta_{j\beta'}\rho_{i\alpha}(\mathbf{k}_1))-2n_{\psi}(\mathbf{p}_2)\delta_{i\alpha}\delta_{j\beta'}\rho_{\alpha'\beta}(\mathbf{k}_2)] \,,
\end{align}
where ${M}_{\alpha \alpha'}\equiv \mathcal{M}(\psi \gamma_{\alpha} \to \psi \gamma_{\alpha'})$ with $\alpha, \alpha'=\pm$ being different polarisation states of the photon.  This eventually leads to 
\begin{align}
\frac{d}{dt}\rho_{++}(\mathbf{k}_1)&=-\frac{1}{8E_{\gamma,1}}\int d\mathbf{p}_1d\mathbf{p}_2d\mathbf{k}_2(2 \pi)^4\delta^4(p_2+k_2-p_1-k_1)\nonumber\\
&\times\Big(2|{M}_{++}|^2\left[n_{\psi}(\mathbf{p}_1)\rho_{++}(\mathbf{k}_1)-n_{\psi}(\mathbf{p}_2)\rho_{++}(\mathbf{k}_2)\right] + 2|{M}_{+-}|^2\left[n_{\psi}(\mathbf{p}_1)\rho_{++}(\mathbf{k}_1)-n_{\psi}(\mathbf{p}_2)\rho_{--}(\mathbf{k}_2)\right]\nonumber\\ 
&\left.+ n_{\psi}(\mathbf{p}_1)\left(\rho_{-+}(\mathbf{k}_1)\left(M_{++}M^*_{-+}+M_{+-}M^*_{--} \right)+\rho_{+-}(\mathbf{k}_1)\left(M_{-+}M^*_{++}+M_{--}M^*_{+-} \right)\right) \right. \nonumber \\ 
&-2n_{\psi}(\mathbf{p}_2)[{M}_{++}{M}^*_{+-}\rho_{+-}(\mathbf{k}_2)+{M}_{+-}{M}^*_{++}\rho_{-+}(\mathbf{k}_2)]\Big) \,, \nonumber \\[1mm] 
\frac{d}{dt}\rho_{--}(\mathbf{k}_1)&=-\frac{1}{8E_{\gamma,1}}\int d\mathbf{p}_1d\mathbf{p}_2d\mathbf{k}_2(2 \pi)^4\delta^4(p_2+k_2-p_1-k_1)\nonumber\\ 
&\times\Big(2|{M}_{-+}|^2\left[n_{\psi}(\mathbf{p}_1)\rho_{--}(\mathbf{k}_1)-n_{\psi}(\mathbf{p}_2)\rho_{++}(\mathbf{k}_2)\right] + 2|{M}_{--}|^2\left[n_{\psi}(\mathbf{p}_1)\rho_{--}(\mathbf{k}_1)-n_{\psi}(\mathbf{p}_2)\rho_{--}(\mathbf{k}_2)\right] \nonumber \\
&\left.+ n_{\psi}(\mathbf{p}_1)\left(\rho_{+-}(\mathbf{k}_1)\left(M_{-+}M^*_{++}+M_{--}M^*_{+-} \right)+\rho_{-+}(\mathbf{k}_1)\left(M_{++}M^*_{-+}+M_{+-}M^*_{--} \right)\right) \right. \nonumber \\ 
&-2n_{\psi}(\mathbf{p}_2)[{M}_{-+}{M}_{--}^*\rho_{+-}(\mathbf{k}_2)+{M}_{--}{M}_{-+}^*\rho_{-+}(\mathbf{k}_2)]\Big) \,, \nonumber \\[1mm]
\frac{d}{dt}\rho_{+-}(\mathbf{k}_1)&=-\frac{1}{8E_{\gamma,1}m_{\psi}^2}\int d\mathbf{p}_1d\mathbf{p}_2d\mathbf{k}_2(2 \pi)^4\delta^4(p_2+k_2-p_1-k_1)\times\,,\nonumber\\
&\times\Big(n_{\psi}(\mathbf{p}_1)\left[\right.\rho_{+-}(\mathbf{k}_1)(|M_{++}|^2+|M_{-+}|^2+|M_{+-}|^2+|M_{--}|^2)\nonumber\\
&  +(\rho_{++}(\mathbf{k}_1)+\rho_{--}(\mathbf{k}_1))(M_{++}M^\ast_{-+}+M_{+-}M^\ast_{--})\left.\right]\nonumber\\
&-2n_{\psi}(\mathbf{p}_2)[M_{++}M^\ast_{-+}\rho_{++}(\mathbf{k}_2)+M_{+-}M^\ast_{--}\rho_{--}(\mathbf{k}_2)+M_{+-}M^\ast_{-+}\rho_{-+}(\mathbf{k}_2)+M_{++}M^\ast_{--}\rho_{+-}(\mathbf{k}_2)]\Big) \,,\nonumber\\[1mm]
\frac{d}{dt}\rho_{-+}(\mathbf{k}_1)&=-\frac{1}{8E_{\gamma,1}m_{\psi}^2}\int d\mathbf{p}_1d\mathbf{p}_2d\mathbf{k}_2(2 \pi)^4\delta^4(p_2+k_2-p_1-k_1)\nonumber\\
&\times\Big(n_{\psi}(\mathbf{p}_1)\left[\rho_{-+}(\mathbf{k}_1)(|M_{++}|^2+|M_{-+}|^2+|M_{+-}|^2+|M_{--}|^2)\right.\nonumber\\
&+(\rho_{++}(\mathbf{k}_1)+\rho_{--}(\mathbf{k}_1))(M_{-+}M^\ast_{++}+M_{--}M^\ast_{+-})\left.\right]\nonumber\\
&-2n_{\psi}(\mathbf{p}_2)[M_{-+}M^\ast_{++}\rho_{++}(\mathbf{k}_2)+M_{--}M^\ast_{+-}\rho_{--}(\mathbf{k}_2)+M_{-+}M^\ast_{+-}\rho_{+-}(\mathbf{k}_2)+M_{--}M^\ast_{++}\rho_{-+}(\mathbf{k}_2)]\Big) \,.
\label{eq:Final}
\end{align}

Expressing the scattering matrix elements in terms of the $\mathbf{A'}$--matrix elements, we then get 

\begin{empheq}[box=\fbox]{align}
|{M}_{++}|^2 &=\tfrac{1}{2} \left(\mathbf{A'}_{11}+\mathbf{A'}_{14}+\mathbf{A'}_{41}+\mathbf{A'}_{44} \right)&{M}_{++}{M}^\ast_{-+}&=\tfrac{1}{2} \left(-\mathbf{A'}_{21}- \mathbf{A'}_{24}+i\mathbf{A'}_{31}+i\mathbf{A'}_{34} \right) \, \nonumber \\
|{M}_{+-}|^2 &=\tfrac{1}{2} \left(\mathbf{A'}_{11}-\mathbf{A'}_{14}+\mathbf{A'}_{41}-\mathbf{A'}_{44} \right) &{M}_{--}{M}^\ast_{+-}&=\tfrac{1}{2} \left(-\mathbf{A'}_{21}+ \mathbf{A'}_{24}-i\mathbf{A'}_{31}+i\mathbf{A'}_{34} \right) \nonumber \\
|{M}_{-+}|^2 &=\tfrac{1}{2} \left(\mathbf{A'}_{11}+\mathbf{A'}_{14}-\mathbf{A'}_{41}-\mathbf{A'}_{44} \right) \, & {M}_{++}{M}^*_{+-}&={\tfrac{1}{2}\left(-\mathbf{A'}_{12}-i\mathbf{A'}_{13}-\mathbf{A'}_{42}-i\mathbf{A'}_{43}\right)} \\
|{M}_{--}|^2 &=\tfrac{1}{2} \left(\mathbf{A'}_{11}-\mathbf{A'}_{14}-\mathbf{A'}_{41}+\mathbf{A'}_{44} \right) &{M}_{-+}{M}^\ast_{+-}&= \tfrac{1}{2}\left(\mathbf{A'}_{22}+i\mathbf{A'}_{23}+i\mathbf{A'}_{32}-\mathbf{A'}_{33}\right)\, \nonumber \\
{M}_{++}{M}^\ast_{--}&=\tfrac{1}{2}\left(\mathbf{A'}_{22}+i\mathbf{A'}_{23}+\mathbf{A'}_{33}-i\mathbf{A'}_{32}\right)&
{M}_{-+}{M}^\ast_{--}&=\tfrac{1}{2}\left(-\mathbf{A'}_{12}+\mathbf{A'}_{42}-i\mathbf{A'}_{13}+i\mathbf{A'}_{43}\right)\nonumber
\end{empheq}
Therefore, using Eqs.~(\eqref{eq:PolBasis}) and (\eqref{eq:Final}), we  obtain the time evolution of the Stokes parameters, namely 
\begin{align} 
\frac{d}{dt} I(\mathbf{k}_1)  &=-\frac{m_{\psi}^2}{8\pi E_{\psi,1}E_{\gamma,1}E_{\psi,2}} \int^\infty_0 dE_{\gamma,2} E_{\gamma,2} \int \frac{d\Omega}{4\pi}\delta \left( E_{\psi,2} + E_{\gamma,2} - E_{\psi,1}-E_{\psi,1}\right) \times \Bigg{[} \Big(n_{\psi,1}I(\mathbf{k}_1)- n_{\psi,2}I(\mathbf{k}_2)\Big)\mathbf{A'}_{11}\nonumber \\
&  +n_{\psi,1}\Big(V(\mathbf{k}_1)\mathbf{A'}_{41} +Q(\mathbf{k}_1)\mathbf{A'}_{21}-U(\mathbf{k}_1)\mathbf{A'}_{31}\Big)-n_{\psi,2}\Big(V(\mathbf{k}_2)\mathbf{A'}_{14}+Q(\mathbf{k}_2)\mathbf{A'}_{12} - U(\mathbf{k}_2)\mathbf{A'}_{13}
\Big)\Bigg{]} \, \nonumber\\[1mm]
\frac{d}{dt} Q(\mathbf{k}_1)  &=-\frac{m_{\psi}^2}{8\pi E_{\psi,1}E_{\gamma,1}E_{\psi,2}} \int^\infty_0 dE_{\gamma,2} E_{\gamma,2} \int \frac{d\Omega}{4\pi}\delta \left( E_{\psi,2} + E_{\gamma,2} - E_{\psi,1}-E_{\psi,1}\right)\nonumber \\
&\times \Bigg{[} \Big(n_{\psi,1}I(\mathbf{k}_1)- n_{\psi,2}I(\mathbf{k}_2)\Big)\mathbf{A'}_{21}  +n_{\psi,1}Q(\mathbf{k}_1)\mathbf{A'}_{11}-n_{\psi,2}\Big(V(\mathbf{k}_2)\mathbf{A'}_{24}+Q(\mathbf{k}_2)\mathbf{A'}_{22} - U(\mathbf{k}_2)\mathbf{A'}_{23}
\Big)\Bigg{]} \, \nonumber\\[1mm] 
\frac{d}{dt} U(\mathbf{k}_1)  &=-\frac{m_{\psi}^2}{8\pi E_{\psi,1}E_{\gamma,1}E_{\psi,2}} \int^\infty_0 dE_{\gamma,2} E_{\gamma,2} \int \frac{d\Omega}{4\pi}\delta \left( E_{\psi,2} + E_{\gamma,2} - E_{\psi,1}-E_{\psi,1}\right) \nonumber \\
& \times \Bigg{[} -\Big(n_{\psi,1}I(\mathbf{k}_1)- n_{\psi,2}I(\mathbf{k}_2)\Big)\mathbf{A'}_{31} +n_{\psi,1}U(\mathbf{k}_1)\mathbf{A'}_{11}+n_{\psi,2}\Big(V(\mathbf{k}_2)\mathbf{A'}_{34}+Q(\mathbf{k}_2)\mathbf{A'}_{32} - U(\mathbf{k}_2)\mathbf{A'}_{33}
\Big)\Bigg{]} \, \nonumber\\[1mm]
\frac{d}{dt} V(\mathbf{k}_1)  &=-\frac{m_{\psi}^2}{8\pi E_{\psi,1}E_{\gamma,1}E_{\psi,2}} \int^\infty_0 dE_{\gamma,2} E_{\gamma,2} \int \frac{d\Omega}{4\pi}\delta \left( E_{\psi,2} + E_{\gamma,2} - E_{\psi,1}-E_{\psi,1}\right)\nonumber \\
& \times \Bigg{[} \Big(n_{\psi,1}I(\mathbf{k}_1)- n_{\psi,2}I(\mathbf{k}_2)\Big)\mathbf{A'}_{41} +n_{\psi,1}V(\mathbf{k}_1)\mathbf{A'}_{11}-n_{\psi,2}\Big(V(\mathbf{k}_2)\mathbf{A'}_{44}+Q(\mathbf{k}_2)\mathbf{A'}_{42} - U(\mathbf{k}_2)\mathbf{A'}_{43}
\Big)\Bigg{]}\,   \label{eq:General_Boltz} 
\end{align}
where $\mathbf{p}_2=\mathbf{p}_1+\mathbf{k}_1-\mathbf{k}_2$ and where it is assumed that the particles in the medium follow a thermal Maxwell-Boltzmann distribution so that $\displaystyle n_{\psi,1 (2)}\equiv n_{\psi,1 (2)}(\mathbf{x})=\int \frac{d^3 \mathbf{p}_{1(2)}}{(2\pi)^3}f_{\psi}(\mathbf{x},\mathbf{p}_{1(2)})$.
We are now ready to compute the time evolution of the circular polarisation component by inputting the appropriate electron densities and $\mathbf{A'}$--matrix elements. This has been done for low energy photons in \cite{Bartolo:2019eac}.

\subsection{ Boltzmann formalism for photon-electron scattering\label{subsec:Boltz}}
For the study of more than one process we now apply the Boltzmann formalism to the photon-electron scattering. 
 Using the general results in Eq.~\eqref{eq:General_Boltz} and Eq.~\eqref{eq:A_matrix_Compton} the  Boltzmann equation for photon-electron scattering simplifies and can then be expressed in terms of the $\mathbf{A'}$--matrix elements. Consequently, for the specific case of Compton scattering, we get 
\begin{align} 
\frac{d}{dt} I(\mathbf{k}_1)  =-\frac{m_e^2}{8\pi E_{e,1}E_{\gamma,1}E_{e,2}}& \int^\infty_0 dE_{\gamma,2} E_{\gamma,2} \int \frac{d\Omega}{4\pi}\delta \left( E_{e,2} + E_{\gamma,2} - E_{e,1}-E_{e,1}\right) \nonumber\\&\times\Bigg{[} \Big(n_{e,1}I(\mathbf{k}_1)- n_{e,2}I(\mathbf{k}_2)\Big)\mathbf{A'}_{11} +\Big(n_{e,1}Q(\mathbf{k}_1)-n_{e,2}Q(\mathbf{k}_2)\Big)\mathbf{A'}_{12}\Bigg{]} \,, \nonumber\\
\frac{d}{dt} Q(\mathbf{k}_1)=-\frac{m_e^2}{8\pi E_{e,1}E_{\gamma,1}E_{e,2}}& \int^\infty_0 dE_{\gamma,2} E_{\gamma,2} \int \frac{d\Omega}{4\pi}\delta \left( E_{e,2} + E_{\gamma,2} - E_{e,1}-E_{e,1}\right)\nonumber\\
&\times \Bigg{[} \Big(n_{e,1}I(\mathbf{k}_1)- n_{e,2}I(\mathbf{k}_2)\Big)\mathbf{A'}_{12}+n_{e,1}Q(\mathbf{k}_1)\mathbf{A'}_{11} -n_{e,2}Q(\mathbf{k}_2)\mathbf{A'}_{22}\Bigg{]} \,, \nonumber\\
\frac{d}{dt} U(\mathbf{k}_1)=-\frac{m_e^2}{8\pi E_{e,1}E_{\gamma,1}E_{e,2}} &\int^\infty_0 dE_{\gamma,2} E_{\gamma,2} \int \frac{d\Omega}{4\pi}\delta \left( E_{e,2} + E_{\gamma,2} - E_{e,1}-E_{e,1}\right)
 \Big{[}n_{e,1}U(\mathbf{k}_1)\mathbf{A'}_{11}-n_{e,2} U(\mathbf{k}_2)\mathbf{A'}_{33}\Big{]} \,, \nonumber\\  \nonumber\\
\frac{d}{dt} V(\mathbf{k}_1)  =-\frac{m_e^2}{8\pi E_{e,1}E_{\gamma,1}E_{e,2}}& \int^\infty_0 dE_{\gamma,2} E_{\gamma,2} \int \frac{d\Omega}{4\pi}\delta \left( E_{e,2} + E_{\gamma,2} - E_{e,1}-E_{e,1}\right)\Big{[}n_{e,1}V(\mathbf{k}_1)\mathbf{A'}_{11}-n_{e,2}V(\mathbf{k}_2)\mathbf{A'}_{44}
\Big{]} \,,  \label{eq:Comptom_Boltz}
 \end{align}
where $c_2\equiv\cos2\phi_2$ and $c_1\equiv\cos2\phi_1$ and
the explicit form of the $\mathbf{A'}$--matrix elements are given by Eq.~(\ref{eq:PinTermsOfPk}).

We note that the evolution of the intensity $I(\mathbf{k}_1)$ and linear polarisation parameter $Q(\mathbf{k}_1)$  are independent of the evolution of the $U(\mathbf{k}_1)$ and $V(\mathbf{k}_1)$ parameters. Consequently, for Compton scattering, there is no conversion between circular and linear polarisation over time as expected.

\section{Conclusion}\label{sec:conclusion}

Polarisation is a critical measurement in Astrophysics. It strongly relies on theoretical estimates of fundamental processes such as Synchrotron radiation and Faraday conversion. So far the literature has focused on the polarisation of visible light and mm radiation, and there are some efforts to describe the polarisation of X-ray radiation \cite{De:2014qza}. However $\gamma$--ray signals may also be polarised and the question we address in this paper is how to describe the possible change of polarisation as photons of any energy propagate through space or the atmosphere. Here we derive the formalism for such studies assuming generic interactions and eventually focus on Compton interactions. 

The classical radiative transfer introduced by Chandrasekhar to describe Rayleigh scattering of visible light cannot be used to describe the change in polarisation of these signals because it only provides a description of the geometry of the scattering. Yet the \P--matrix introduced by Chandrasekhar to relate the Stokes parameters in the $(I_l,I_r,U,V)$ basis before and after scattering can still be defined in the case of Compton interactions. Our formalism generalises Chandrasekhar's results to any type of photon interaction at any energy. This is done by relating the incoming and outgoing Stokes parameters in the ($I,Q,U,V$) basis using the $\mathbf{A'}$--matrix, which is given in Eq.~\eqref{eq:A_matrix_general}. 

As it is expected, the $\mathbf{A'}$--matrix elements at high energy (Eq.~\eqref{eq:A_matrix_Compton}) are significantly different from the ones at low energy. Some of the elements which were vanishing in the low energy limit do not vanish at high energy. Furthermore, unlike in the low energy case, the change of the Stokes parameters after scattering also depends on the photon energies in the initial and final states. The relationship between the Stokes parameters before and after scattering is therefore more complex at high energy than at low energy. Nevertheless, just like for the interactions at now energies, circular polarisation is secluded. Consequently, if the $V$-parameter changes after scattering, this means that a number of photons with a given helicity state were converted into photons with the opposite helicity. Therefore, if one kind of circular polarisation dominates over the other one, we will be able to observe a net circular polarisation signal.

To study  the change in the  net circular polarisation after photon--electron scattering we defined $\Delta_V$, which is given in terms of the scattering matrix amplitude. According to Eq.~\eqref{eq:PolPer}, $\Delta_V=0$ implies no circular polarisation after scattering, meaning that the number of photon states with opposite polarisations is equal. Therefore when $\Delta_V\neq0$, we will be able to observe a circularly polarised signal, regardless of whether the polarisation states are all preserved ($\Delta_V=1$) or all flipped ($\Delta_V=-1$),

We also determine the conditions for which a net circular polarisation signal would be preserved after scattering at low/high energies. This was done in four different frames: centre of mass frame, rest frame, spin frame and fixed frame.  We observed that, for the first three frames mentioned before, a common characteristic is that for low energy incoming photon, the change on the net circular polarisation depends on the scattering direction. The fixed frame does not have this characteristic because the incoming electron is relativistic. On the other hand, for high energies of the incoming photon in the centre of mass and spin frame, the circular polarisation is conserved independently of the scattering direction. This is not true for the rest frame, where the change in the net circular polarisation depends on the scattering direction of the outgoing photon. The only way the polarisation changes in this frame is when the photon scatters in the forward direction. 
To complement this work, we also studied the effect of the electron spin on the conservation of circular polarisation. In this case, we define the change of net circular polarisation as $\Delta_{V,e_\pm}$ for an initial electron spin $J_z(e_1)=\pm\tfrac{1}{2}$ in the rest frame. We observed that by fixing the electron spin, the mass of the electron becomes relevant since the  change of circular polarisation conservation behaves very different when the incoming photon energy is above or below it. Indeed, in the low energy limit, the conservation of net circular polarisation depends on the scattering direction, as in the case of unpolarised electrons, and it is independent of the initial electron spin. However, this is not the case for high energy photons, where $\Delta_{V,e_\pm}$ depends on the initial electron spin and it does not depend on the scattering direction. For instance, for an incoming high energy photon with positive helicity, its polarisation changes when the spin of the incoming electron is positive, otherwise the photon polarisation is conserved independently of the scattering direction.

We also developed a general formalism to study the time evolution of the Stokes parameters in the ($I,Q,U,V$) basis in terms of the scattering matrix elements. This formalism is valid for any type of scattering at any energy and can be simply computed using the entries of the $\mathbf{A'}$--matrix given in Eq.~\eqref{eq:A_matrix_general}. For the particular case of Compton scattering, we found that the time evolution of the $V$--parameter is independent of the other Stokes parameters. This means that even after multiple scatterings, while the amount of circular polarisation might change (i.e., the difference between left or right helicity states), circularly polarised light will never become linearly polarised or vice versa.

The formalism developed in this paper provides with a powerful tool to study the changes in circular polarisation as light propagates through any type of medium. This implies that observations of circularly polarised light can be used to deepen our understanding of the nature of dark matter or other theories beyond the Standard Model.

\section*{Acknowledgments}
AO and MRQ thank the University of Sydney for kind hospitality during the completion of this work. MRQ would like to thank Alexis Plascencia and Oscar Ochoa for useful discussion. AO would like to thank Kristian Moffat for very useful discussion about the spin diagonalization. AO is  supported  by  the  European  Research  Council under ERC Grant “NuMass” (FP7-IDEAS-ERC ERC-CG 617143). MRQ is supported by Consejo Nacional de Ciencia y Tecnologia, Mexico (CONACyT) under grant 440771. YLZ acknowledges the STFC Consolidated Grant ST/L000296/1 and the European Union's Horizon 2020 Research and Innovation programme under Marie Sk\l{}odowska-Curie grant agreements Elusives ITN No.\ 674896 and InvisiblesPlus RISE No.\ 690575.
\vspace{1cm}
\newpage
\appendix


\section{Formalism for the Thomson interactions}\label{App:Low_energy}

In this appendix, we briefly review the change of photon polarisation during Thomson scattering. Most of the result are well-known and are convenient to be compared with our results of scattering in the high energy limit.

\subsection{Deriving the Thomson \P--matrix directly from the Stokes parameters\label{Ap:Pmatrix_direct}}

For the Thomson scattering, the \P--matrix can be directly obtained by following the original definition of Stokes parameters. Applying the formula in Eq.~\eqref{eq:stokes_linear_parameters} to Stokes parameters for both incoming photon and outgoing photon, we obtain
\begin{eqnarray}
I_l^{(2)} &=& (\epsv_{l}^{\ (2)}\cdot\ \epsv_{r}^{\ (1)})^2 \ a_r^2 \ + 
\ (\epsv_{l}^{\ (2)}\cdot\ \epsv_{l}^{\ (1)})^2 \ a_l^2\ 
+ \ 2\, a_r\,a_l \, \cos(\delta_l-\delta_r) \ (\epsv_{l}^{\ (2)}\cdot\ 
\epsv_{r}^{\ (1)}) (\epsv_{l}^{\ (2)}\cdot\ \epsv_{l}^{\ (1)}), \nonumber \\ 
&=& (\epsv_{l}^{\ (2)}\cdot\ \epsv_{r}^{\ (1)})^2 \ I_r^{(1)} \ + 
\ (\epsv_{l}^{\ (2)}\cdot\ \epsv_{l}^{\ (1)})^2 \ I_l^{(1)} \ 
\ +\  (\epsv_{l}^{\ (2)}\cdot\ \epsv_{r}^{\ (1)})
(\epsv_{l}^{\ (2)}\cdot\ \epsv_{l}^{\ (1)}) \ \ U^{(1)}\,, \nonumber\\
\nonumber  \\ 
I_r^{(2)}&=& (\epsv_{r}^{\ (2)}\cdot\ \epsv_{r}^{\ (1)})^2 \ a_r^2 \ + 
\ (\epsv_{r}^{\ (2)}\cdot\ \epsv_{l}^{\ (1)})^2 \ a_l^2 \ + 
\ 2\,  a_r\,a_l \, \cos(\delta_l-\delta_r)\,(\epsv_l^{\ (2)}\cdot\ 
\epsv_r^{\ (1)}) (\epsv_{r}^{\ (2)}\cdot\ \epsv_{l}^{\ (1)})\,, \nonumber\\
&=& (\epsv_{r}^{\ (2)}\cdot\ \epsv_{r}^{\ (1)})^2 \ \ I_r^{(1)} \ + 
\ (\epsv_{r}^{\ (2)}\cdot\ \epsv_{l}^{\ (1)})^2 \ \ I_l^{(1)} \ + 
\   (\epsv_l^{\ (2)}\cdot\ \epsv_r^{\ (1)}) (\epsv_{r}^{\ (2)}\cdot\ 
\epsv_{l}^{\ (1)})  \ U^{(1)}\,, \nonumber\\ 
\nonumber \\ 
U^{(2)} &=&  \,2(\epsv_{r}^{\ (2)}\cdot\  \epsv_{r}^{\ (1)}) \ 
(\epsv_{l}^{\ (2)}\cdot\ \epsv_{r}^{\ (1)}) \ a_r^2 \ + \ 2 \, 
(\epsv_{r}^{\ (2)}\cdot\ \epsv_{l}^{\ (1)}) \ (\epsv_{l}^{\ (2)}\cdot\ \epsv_{l}^{\ (1)}) 
\ a_l^2 \ + \ 2 \, a_r \, a_l \,\cos(\delta_l-\delta_r)\times  \nonumber \\
 && \ \times \left[(\epsv_{r}^{\ (2)}\cdot\ \epsv_{r}^{\ (1)})(\epsv_{l}^{\ (2)}
 \cdot\ \epsv_{l}^{\ (1)}) - (\epsv_{r}^{\ (2)}\cdot\  \epsv_{l}^{\ (1)})
 (\epsv_{l}^{\ (2)}\cdot\ \epsv_{r}^{\ (1)}) \right]  \nonumber \\
 &=&  \ 2\, (\epsv_{r}^{\ (2)}\cdot\  \epsv_{r}^{\ (1)}) \ 
 (\epsv_{l}^{\ (2)}\cdot\ \epsv_{r}^{\ (1)}) \ I_r^{(1)} \ 
 + \ 2 \, (\epsv_{r}^{\ (2)}\cdot\ \epsv_{l}^{\ (1)}) \ 
 (\epsv_{l}^{\ (2)}\cdot\ \epsv_{l}^{\ (1)}) \ I_l^{(1)} \ \  \nonumber \\ && + \ 
 \left[(\epsv_{r}^{\ (2)}\cdot\ \epsv_{r}^{\ (1)})
 (\epsv_{l}^{\ (2)}\cdot\ \epsv_{l}^{\ (1)})\ +\ (\epsv_{r}^{\ (2)}\cdot\ \epsv_{l}^{\ (1)})
 (\epsv_{l}^{\ (2)}\cdot\ \epsv_{r}^{\ (1)}) \right]  U^{(1)} \,, \nonumber  \\ 
 \nonumber \\
V^{(2)} &=& 2\, a_r\, a_l \, \sin(\delta_l-\delta_r) \left[(\epsv_{r}^{\ (2)}\cdot\ 
\epsv_{r}^{\ (1)}) \ (\epsv_{l}^{\ (2)}\cdot\ \epsv_{l}^{\ (1)})\ -\ 
(\epsv_{r}^{\ (2)}\cdot\ \epsv_{l}^{\ (1)}) \ (\epsv_{l}^{\ (2)}\cdot\ \epsv_{r}^{\ (1)}) \right] \nonumber \\ 
&=&   \left[(\epsv_{r}^{\ (2)}\cdot\ \epsv_{r}^{\ (1)}) \ (\epsv_{l}^{\ (2)}
\cdot\ \epsv_{l}^{\ (1)})\ -\ (\epsv_{r}^{\ (2)}\cdot\ \epsv_{l}^{\ (1)}) \ 
(\epsv_{l}^{\ (2)}\cdot\ \epsv_{r}^{\ (1)}) \right] V^{(1)}\,. 
\end{eqnarray}

From these relations, one obtains the following expression for the \Pmat in terms of the photon polarisation vectors
 
\begin{equation}
\mathbf{P}
= \begin{bmatrix}
 (\epsv_{l}^{\ (2)}\cdot\ \epsv_{l}^{\ (1)})^2    & \  
  (\epsv_{l}^{\ (2)}\cdot\ \epsv_{r}^{\ (1)})^2  & \ (\epsv_{l}^{\ (2)}\cdot\ \epsv_{r}^{\ (1)}) (\epsv_{l}^{\ (2)}\cdot\ \epsv_{l}^{\ (1)})  & 0 \\

(\epsv_{r}^{\ (2)}\cdot\ \epsv_{l}^{\ (1)})^2   & \ 
   (\epsv_{r}^{\ (2)}\cdot\ \epsv_{r}^{\ (1)})^2  & \  (\epsv_{r}^{\ (2)}\cdot\ \epsv_{r}^{\ (1)}) (\epsv_{r}^{\ (2)}\cdot\ \epsv_{l}^{\ (1)})  &  0 \\
   
  (\epsv_{r}^{\ (2)} \cdot\ \epsv_{l}^{\ (1)}) \ (\epsv_{l}^{\ (2)}\cdot\  \epsv_{l}^{\ (1)}) &  (\epsv_{r}^{\ (2)} \cdot\ \epsv_{r}^{\ (1)}) \ (\epsv_{l}^{\ (2)}\cdot\ \epsv_{r}^{\ (1)})   &   \mathbf{P}_{33}   &  0\\
  
 0 & 0  &  0  & \mathbf{P}_{44}\\
\end{bmatrix}
\label{eq:Stokes_M}
\end{equation}
with 
\begin{align}
   \mathbf{P}_{33} =& \, \left[(\epsv_{r}^{\ (2)} \cdot\ \epsv_{r}^{\ (1)})(\epsv_{l}^{\ (2)}\cdot\  \epsv_{l}^{\ (1)})\ +\  (\epsv_{r}^{\ (2)}\cdot\ \epsv_{l}^{\ (1)})(\epsv_{l}^{\ (2)}\cdot\ \epsv_{r}^{\ (1)}) \right] \,,\nonumber \\
   \mathbf{P}_{44} =&  \,  \left[(\epsv_{r}^{\ (2)} \cdot\ \epsv_{r}^{\ (1)})(\epsv_{l}^{\ (2)}\cdot\ \epsv_{l}^{\ (1)}) \ - \ (\epsv_{r}^{\ (2)}\cdot\ \epsv_{l}^{\ (1)})(\epsv_{l}^{\ (2)}\cdot\ \epsv_{r}^{\ (1)}) \right] \,. 
\end{align}

One can recover Chandrasekhar's expression for the $\mathbf{P}$--matrix in Eq. (\ref{eq:Chandrasekhar_Pmatrix}) by substituting the 3D kinematics in a general (fixed) frame, see Appendix \ref{Ap:kinematics} for the dot products between the incoming (1) and outgoing (2) polarisation vectors, namely
\begin{eqnarray}
\epsv_{l}^{\ (2)}\cdot\ \epsv_{l}^{\ (1)} &=& \sin\phi_1\sin\phi_2 + \cos\phi_1\cos\phi_2  \cos{\left( \theta_2 - \theta_1 \right)}  \nonumber \, ,  \\
\epsv_{r}^{\ (2)}\cdot\ \epsv_{r}^{\ (1)} &=& \cos{\left( \theta_2 - \theta_1 \right)} \, , \nonumber \\
\epsv_{l}^{\ (2)}\cdot\ \epsv_{r}^{\ (1)} &=& \cos\phi_2 \sin{\left( \theta_2 - \theta_1 \right)}\, , \nonumber \\
\epsv_{r}^{\ (2)}\cdot\ \epsv_{l}^{\ (1)} &=& - \cos\phi_1 \sin{\left( \theta_2 - \theta_1 \right)} \, \cdot\ \label{eq:kinematics_R}
\end{eqnarray}

\subsection{$\mathbf{P}$--matrix  for Thomson interactions \label{Ap:Pmatrix_chandra}}
Using the expression of the $\mathbf{R}$--matrix as shown in Eq.~(\ref{eq:R_mat_Chandra}), we  obtain the following \P--matrix elements
\begin{eqnarray}
\mathbf{P}_{\rm Chandrasekhar}=\begin{bmatrix}
 (\cos\theta \cos\Phi_1\cos\Phi_2-\sin\Phi_1\sin\Phi_2)^2&(\cos\Phi_1\sin\Phi_2+\cos\theta\sin\Phi_1\cos\Phi_2)^2&\mathbf{P}_{13}&0\\
 (\cos\theta \cos\Phi_1 \sin\Phi_2+\sin\Phi_1\cos\Phi_2)^2&(\cos\Phi_1 \cos\Phi_2-\cos\theta \sin\Phi_1\sin\Phi_2)^2&\mathbf{P}_{23}&0\\
 \mathbf{P}_{31}&\mathbf{P}_{32}&\mathbf{P}_{33}&0\\
0&0&0&\cos\theta
\end{bmatrix}
\end{eqnarray}
with 
\begin{align}
    \mathbf{P}_{13}&=\tfrac{1}{2} \left(-\cos^2\theta\sin2\Phi_1\cos^2\Phi_2-\cos\theta\cos2\Phi_1 \sin2\Phi_2+\sin2\Phi_1\sin^2\Phi_2\right)\nonumber\\
    \mathbf{P}_{23}&=\tfrac{1}{2} \left(\sin2\Phi_1\left(\cos^2\Phi_2-\cos^2\theta\sin^2\Phi_2\right)+\cos\theta\cos2\Phi_1 \sin2\Phi_2\right)\nonumber\\
    \mathbf{P}_{31}&=\sin2\Phi_2\left(\cos^2\theta\cos^2\Phi_1-\sin^2\Phi_1\right)+\cos\theta\sin2\Phi_1\cos2\Phi_2\nonumber\\ 
    \mathbf{P}_{32}&=\sin2\Phi_2\left(\cos^2\theta\sin^2\Phi_1-\cos^2\Phi_1\right)-\cos\theta\sin2\Phi_1\cos2\Phi_2\nonumber\\
    \mathbf{P}_{33}&=\cos\theta\cos2\Phi_1 \cos2\Phi_2-2 \left(1+\cos^2\theta\right)\sin\Phi_1 \cos\Phi_1\sin\Phi_2\cos\Phi_2
\end{align}

\subsection{$\mathbf{P}'$--matrix for Thomson interactions \label{Ap:Pprimematrix_chandra}}

The $\mathbf{P}'$--matrix denotes the $\mathbf{P}$--matrix in the $(I, Q, U, V)$ basis. Considering a rotation of the plane defined by the incoming polarisation vectors by an angle $\Phi_1$ and a rotation for the final photon with another angle $\Phi_2$, the $\mathbf{P}'$ follows the transformation
\begin{equation}
\mathbf{P}'  =\mathbf{L'}(\pi-\Phi_2) \ \mathbf{R}'  \ \mathbf{L'}(-\Phi_1) \,, \label{eq:rotation_IQ}
\end{equation}
where $\mathbf{L}'(\Phi)$ takes a different form from $\mathbf{L}(\Phi)$ in Eq.~\eqref{eq:rotation_chandra},
\begin{eqnarray} 
\mathbf{L}'(\Phi)= \left(
  \begin{array}{cccc}
   1 & 0  & 0 & 0 \\
   
   0 & \cos 2\Phi  & \sin 2\Phi & 0 \\
   
  0 & -\sin 2\Phi & \cos 2\Phi &  0 \\
  
  0 & 0 &  0 &  1 \\
  \end{array}
  \right)\,. \, \label{eq:rotation2}
\end{eqnarray} 
This eventually leads to the following \P$'$--matrix elements
\begin{empheq}[box=\fbox]{align}
\mathbf{P}'_{11}\! &=\!\mathbf{R}'_{11} 
&\mathbf{P}'_{31} \! &=\! \mathbf{R}'_{31}  \cpt \!+\! \mathbf{R}'_{21}  \spt  \nonumber\\
\mathbf{P}'_{12} \! &=\!\mathbf{R}'_{12}  \cpo\! + \!\mathbf{R}'_{13}  \spo  
&\mathbf{P}'_{32} \! &=\! \cpo\!\! \left(\mathbf{R}'_{32}  \cpt\! \!+\!\mathbf{R}'_{22} \spt \right) \!+\! \spo\! \!\left(\mathbf{R}'_{33} \cpt \!+\! \mathbf{R}'_{23} \spt\! \right)  \nonumber\\
\mathbf{P}'_{13} \! &=\!\mathbf{R}'_{13}  \cpo \!-\! \mathbf{R}'_{12} \spo
&\mathbf{P}'_{33}\! &=\! \cpo\!\! \left(\mathbf{R}'_{33} \cpt \! +\! \mathbf{R}'_{23}  \spt \!\! \right) \!-\!\spo \left(\mathbf{R}'_{32} \cpt \!+\!\mathbf{R}'_{22} \spt \!\right)
 \nonumber\\
\mathbf{P}'_{14} \! &=\! \mathbf{R}'_{14} 
&\mathbf{P}'_{34} \! &=\! \mathbf{R}'_{34} \cpt \!+\! \mathbf{R}'_{24} \spt 
\label{eq:amplitude_matrix_P}\\
\mathbf{P}'_{21} \! &=\! \mathbf{R}'_{21} \cpt \!- \!\mathbf{R}'_{31} \spt 
&\mathbf{P}'_{41} \! &=\! \mathbf{R}'_{41} \nonumber\\
\mathbf{P}'_{22}\! &=\! \cpo \left(\mathbf{R}'_{22} \cpt \!-\!\mathbf{R}'_{32} \spt \right) \!+ \!\spo \left(\mathbf{R}'_{23} \cpt \! -\! \mathbf{R}'_{33} \spt \right)  
&\mathbf{P}'_{42}\! &=\!\mathbf{R}'_{42} \cpo \! + \!\mathbf{R}'_{43} \spo \,, \nonumber\\
\mathbf{P}'_{23}\! &=\! \spo \left( \mathbf{R}'_{32} \spt\! -\!\mathbf{R}'_{22} \cpt \right) \!+\! \cpo \left(\mathbf{R}'_{23} \cpt \! -\! \mathbf{R}'_{33} \spt \right) 
&\mathbf{P}'_{43} \! &=\! \mathbf{R}'_{43}  \cpo\! - \!  \mathbf{R}'_{42} \spo  \,, \nonumber\\
\mathbf{P}'_{24}\! &=\!\mathbf{R}'_{24} \cpt \! -\! \mathbf{R}'_{34} \spt \,, 
&\mathbf{P}'_{44}\! &=\!\mathbf{R}'_{44} \nonumber
\end{empheq}
with $\cpo \equiv  \cos (2 \Phi_1) $, $\cpt \equiv  \cos (2 \Phi_2) $, $\spo \equiv  \sin (2 \Phi_1) $, $\spt \equiv  \sin (2 \Phi_2) $.

\subsection{Deriving the $\mathbf{P}$--matrix using  the Quantum formalism \label{sec:comparisonCQThomson}} 

We can now compare both the geometrical formalism derived by Chandrasekhar and our quantum formalism by studying the special case of  Thomson interactions. 
We first need to replace each of the matrix amplitude elements, as defined in Eq.(\ref{Eq:Rdef_M}), in the quantum formalism by their QFT definition. 
Without loss of generality, the matrix elements for  Thomson scattering can be parameterized as
\begin{equation}
    M_{i'i} \, = \, M_{\mu\nu} \ \epsilon_i^{(1)\mu } \, \epsilon_{i'}^{*(2)\nu } \,,
\end{equation}
where $\mu,\nu$ are Lorentz indices, in the Lorentz gauge $\epsilon_{i} = (0,\epsv_{i})$ and $M_{\mu\nu}$ the  polarisation vector-independent amplitude associated with each Feynman diagram. The  amplitude squared for Thomson scattering is given by 
\begin{equation}
     M_{i'i}  \ M_{j'j}^{*} = \sum_{\lambda=s,t,st} \left( {M_{\mu \nu} \, M_{\mu' \nu'}^*}\right)_{\lambda} \ \epsilon_i^{(1)\mu } \epsilon_{i'}^{*(2)  \nu} \epsilon_{j}^{*(1)  \mu'} \epsilon_{j'}^{(2)\nu' } \,,
\end{equation}
where the amplitudes for the different $s,t, st$ channels $ \left(M_{\mu \nu} M_{\mu' \nu'}^*\right)_{{s,t,st}}$ are equal  to 
\begin{align}
    \left({M_{\mu \nu} \, M_{ \mu' \nu'}^*} \right)_{s}  = & \ g_{\nu \nu'} g_{\mu \mu'} \,, \nonumber\\
    \left({M_{\mu \nu} \, M_{ \mu' \nu'}^*}\right)_{t}  = & \ g_{\nu \nu'} g_{\mu \mu'} \,, \nonumber\\
    \left({M_{\mu \nu} \, M_{ \mu' \nu'}^*}\right)_{st}  = & \ 2 \  ( 2 g_{\mu \nu} g_{\mu' \nu'} - g_{\nu \nu'} g_{\mu \mu'}) \,.
\end{align}

Using these definitions together with ${\mathbf{A}}_{i'ij'j} \equiv {\cal{ \mathbf{W}}} \,  M_{i'i} M^\ast_{j'j}   \, {\cal{\mathbf{W}}}^{-1}$ and 
$\mathbb{\mathbf{P}} = \mathbf{C} \, \mathbf{A}_{i'ij'j} \, \mathbf{C}^{-1}$, one finds that  the Stokes parameters after Thomson scattering are given by  

\begin{eqnarray}
  \begin{bmatrix}
 \hat{I_l}  \\
 \hat{I_r} \\
 \hat{U} \\ 
  \hat{V} \\ 
\end{bmatrix}
  ^{(2)}
  =
 \begin{bmatrix}
 |\epsilon_{l}^{(2)}\cdot\ \epsilon_{l}^{(1)}|^2    & \ 
  |\epsilon_{l}^{(2)}\cdot\ \epsilon_{r}^{(1)}|^2  & \     (\epsilon_{l}^{(2)}\cdot\ \epsilon_{r}^{(1)}) (\epsilon_{l}^{(2)}\cdot\ \epsilon_{l}^{(1)})  & 0 \\
   
   |\epsilon_{r}^{(2)}\cdot\ \epsilon_{l}^{(1)}|^2   & \ 
     |\epsilon_{r}^{(2)}\cdot\ \epsilon_{r}^{(1)}|^2  & \  (\epsilon_{r}^{(2)}\cdot\ \epsilon_{r}^{(1)}) (\epsilon_{r}^{(2)}\cdot\ \epsilon_{l}^{(1)})  &  0 \\
   
  (\epsilon_{r}^{(2)}\cdot\ \epsilon_{l}^{(1)}) \ (\epsilon_{l}^{(2)}\cdot\ \epsilon_{l}^{(1)}) &  (\epsilon_{r}^{(2)} \cdot\ \epsilon_{r}^{(1)}) \ (\epsilon_{l}^{(2)}\cdot\  \epsilon_{r}^{(1)})   &   \mathbf{P}_{33}   &  0\\
  
 0 & 0  &  0  & \mathbf{P}_{44}\\
\end{bmatrix}
 \begin{bmatrix}
 \hat{I_l}  \\
 \hat{I_r} \\
 \hat{U} \\ 
  \hat{V} \\ 
\end{bmatrix}^{(1)}
\end{eqnarray}
with
\begin{align}
   \mathbf{P}_{33}=&\left[(\epsilon_{r}^{(2)}\cdot\ \epsilon_{r}^{(1)})(\epsilon_{l}^{(2)} \cdot\ \epsilon_{l}^{(1)}) + (\epsilon_{r}^{(2)} \cdot\ \epsilon_{l}^{(1)})(\epsilon_{l}^{(2)} \cdot\ \epsilon_{r}^{(1)}) \right] \,,\nonumber\\
   \mathbf{P}_{44} =  & \left[(\epsilon_{r}^{(2)} \cdot\ \epsilon_{r}^{(1)})(\epsilon_{l}^{(2)} \cdot\ \epsilon_{l}^{(1)}) - (\epsilon_{r}^{(2)} \cdot\ \epsilon_{l}^{(1)})(\epsilon_{l}^{(2)} \cdot\ \epsilon_{r}^{(1)}) \right] \,,
\end{align}
which indeed agrees with Chandrasekhar's results as displayed in Eq.~(\ref{eq:Stokes_M}). Using the Thomson kinematics in the fix frame, we obtain Eq.~(\ref{eq:Chandrasekhar_Pmatrix}), as expected. Therefore, In the low energy limit of the incoming photon, the two formalisms are equivalent. 

\section{Kinematics and results in the different frames of reference\label{Ap:kinematics}}
\subsection{Reference frames}
To compare the polarisation of light in processes with relativistic electrons from those with non-relativistic electrons we consider some  frames of reference. Such frames are the centre of mass (COM) frame, the rest frame, the fixed frame and the spin frame; each of them is good for describing different physics aspects of the process.

The COM frame is a good approximation for describing thermal photon scattering with thermal electron. The rest frame can be treated as a limit of an energetic photon scattering with electron with a small momentum. The spin frame  specifies the electron spin $J_z = \pm 1/2$ in the $z$ direction. The fixed frame can be applied to an energetic electron scattering with a soft photon. 
\begin{figure}[ht!]
\includegraphics[width=.8\textwidth, angle = 0]{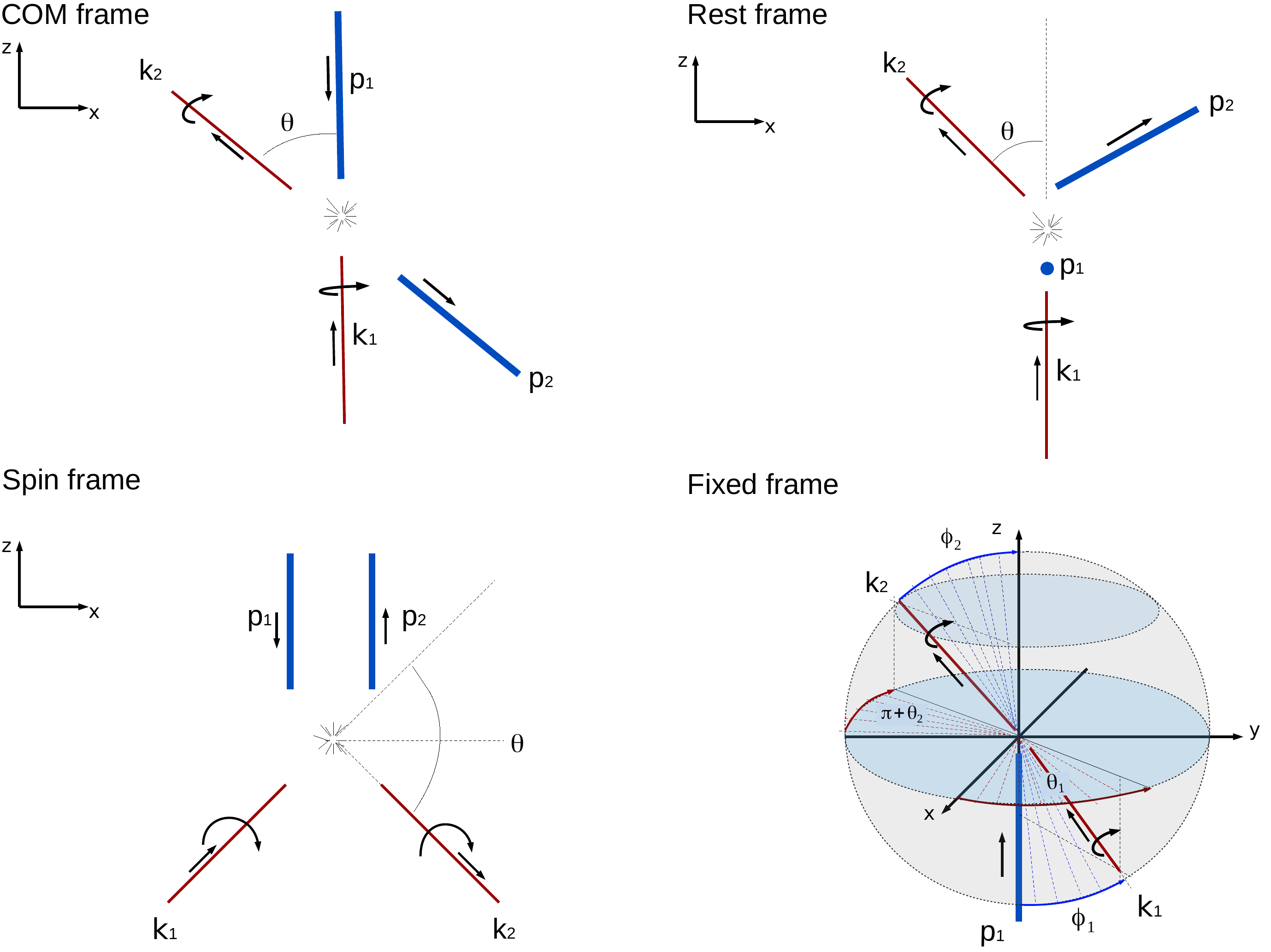}
\caption{Frame of references for the photon-electron scattering. For all four frames, the blue bold line refers to the incoming (outgoing) electron, the red line refers to the incoming (outgoing) photon. The straight arrows indicate the momentum direction for each particle while the deflect arrows indicate the helicity of the photons.  In the upper-left corner we have the centre of mass (COM) frame, upper-right corner shows the rest frame, and lower-left corner shows the spin frame, where both the incoming and outgoing electrons are on the $z$ directions. In each frame, the $\theta$ angle always corresponds to the angle between the outgoing and incoming photon. Finally, the lower-right corner shows the general fixed frame which contains four parameters, $\theta_1$, $\theta_2$, $\phi_1$ and $\phi_2$.}
\label{app:Frame_figs} 
\end{figure}

For every frame of reference shown in Fig. \ref{app:Frame_figs} the four-momenta $p_1$ and $p_2$ correspond to the incoming and outgoing electrons as well as the four-momenta $k_1$ and $k_2$ correspond to the incoming and outgoing photons respectively. In the following we give the explicit kinematics for each of the frames.
\subsubsection{Centre of mass frame}
In this frame the incoming photon and electron move in opposite directions with the same momentum $|\mathbf{p_1}|=E_{\gamma}$, which is the photon energy. The outgoing photon and electron do not change energies but only directions.  We assume the angle between incoming photon and outgoing photon is $\theta$. Finally, in this frame the energy for the incoming electron is $E_e^2 =E_\gamma^2+m_e^2$ and the four momenta reads as follows,
\begin{eqnarray}
\label{electronkine}
p_{1} &=& (E_e, 0, 0, -E_\gamma) \,, \nonumber \\
k_1 &=& (E_\gamma, 0, 0, E_\gamma) \,, \nonumber \\
k_2 &=& (E_\gamma, E_\gamma\sin\theta,0, E_\gamma\cos \theta) \,, \nonumber\\
p_{2}&=&p_{1}+k_1-k_2\,.
\end{eqnarray}

\subsubsection{Rest frame}
In this frame the electron is initially at rest and the photon has energy $E_{\gamma,1}$ and momentum $\mathbf{k}_1=E_{\gamma,1}$ on the $z$ direction. After the scattering the photon acquires an energy $E_{\gamma,2}$ and goes to a direction $\theta$ respect to the $z$--axis with momentum $\mathbf{p}_2$.  The four-momenta are defined as,
\begin{eqnarray}
\label{electronkine_rest}
p_{1} &=& (m_e, 0, 0, 0) \,, \nonumber \\
k_1 &=& (E_{\gamma,1}, 0, 0, E_{\gamma,1}) \,, \nonumber \\
k_2 &=& (E_{\gamma,2}, E_{\gamma,2}\sin\theta,0, E_{\gamma,2}\cos \theta) \,, \nonumber\\
p_{2}&=&p_{1}+k_1-k_2 \,.
\end{eqnarray}
We present the squared matrix element for different polarisation transitions in terms of the initial and final photon energies since it simplifies our results drastically. For this we use
\begin{eqnarray}
\cos \theta=1 - \frac{m_e(E_{\gamma,1}-E_{\gamma,2})}{E_{\gamma,1}E_{\gamma,2}} \,, ~
\sin \theta =\frac{\sqrt{m_e(E_{\gamma,1} - E_{\gamma,2})(2 E_{\gamma,1} E_{\gamma,2} - m_e E_{\gamma,1} + m_eE_{\gamma,2})}}{E_{\gamma,1}E_{\gamma,2}} \,.\label{cossin}
\end{eqnarray} 

\subsubsection{Spin frame}
In this frame of reference the incoming and outgoing electrons are moving along the $\pm z$ direction respectively, and the incoming and outgoing photons are moving in the $x-z$ plane. The four-momentum of the particles in this frame are given by,
\begin{eqnarray}
\label{spin_ref}
p_1 &=& (E_{e}, 0, 0, -k_z) \,,\nonumber \\
k_1 &=& (E_{\gamma}, k_x, 0, k_z) \,,\nonumber \\
k_{2}&=&(E_\gamma,k_x,0,-k_z) \,,\nonumber\\
p_2 &=& (E_{e},0,0,k_z)\,.
\end{eqnarray}
where $E_e=\sqrt{k_z^2+m_e^2}$ and $E_\gamma=\sqrt{k_x^2+k_z^2}$. 
Without of generality, $k_x, k_z>0$ are assumed. Note that $k_z=E_\gamma \sin{\frac{\theta}{2}}$, and $k_x=E_\gamma \cos{\frac{\theta}{2}}$, with $\theta$ being the angle between the incoming and outgoing photon.
\subsubsection{Fixed frame}
In the fixed frame, we chose the incoming electron of energy $E_{e,1}$ and momentum $\mathbf{p}_1$ moving purely along the $z$ direction. The photon in the initial state is coming from any  direction with an energy $E_{\gamma,1}$ and momentum $\mathbf{k}_1$. The outgoing photon gets energy $E_{\gamma,2}$ and momentum $\mathbf{k}_2$ changing its directions with respect to the initial states. We define the four-momentum particles as, 
\begin{eqnarray}
\label{electronkine_fx}
p_1 &=& (E_{e,1}, 0, 0, p_1) \,,\nonumber \\
k_1 &=& (E_{\gamma,1}, E_{\gamma,1} \cos \theta_1 \sin \phi_1, E_{\gamma,1} \sin \theta_1 \sin \phi_1, E_{\gamma,1} \cos \phi_1) \,,\nonumber \\
k_2 &=& (E_{\gamma,2}, E_{\gamma,2} \cos \theta_2 \sin \phi_2, E_{\gamma,2} \sin \theta_2 \sin \phi_2, E_{\gamma,2} \cos \phi_2) \,,\nonumber\\
p_{2}&=&p_{1}+k_1-k_2\,, 
\end{eqnarray}
where $\phi_1$ is the angle between the incident photon and the incoming electron, $\phi_2$ is the angle between the incident electron and the outgoing photon. Additional  $\theta_1$ and $\theta_2$ are the angles between the incoming electron and the incoming photon, and the angle between the direction of the incoming electron and the outgoing photon in the $x-y$ plane respectively. In the fixed frame we require that $p_2^2 = p_1^2$ which leads to 
\begin{eqnarray}
E_{\gamma,2} &=& E_{\gamma,1} \frac{E_{e,1}-p_1 \cos \phi_1}{E_{e,1}-p_1 \cos \phi_2 -E_{\gamma,1} ( \sin\phi_1 \sin\phi_2 \cos (\theta_2-\theta_1) + \cos\phi_1 \cos\phi_2 - 1)} \nonumber\\
&=&E_{\gamma,1}\frac{E_{e,1}-p_1\mu_1}{E_{e,1}-p_1 \mu_2 -E_{\gamma,1} ( \sqrt{1-\mu_1^2} \sqrt{1-\mu_2^2} \cos (\theta_2 - \theta_1) + \mu_1 \mu_2 - 1)} \,,
\end{eqnarray}
where we have used $\cos \phi_1 = \mu_1$ and $\cos \phi_2 = \mu_2$. In the above, one may use the angle between incoming photon and outgoing photon $\theta$, which is given by
\begin{eqnarray}
\cos\theta =  \sin\phi_1 \sin\phi_2 \cos(\theta_2-\theta_1) + \cos\phi_1 \cos\phi_2 \,,
\end{eqnarray}
to simplify the formula.

\subsection{Polarised squared amplitudes for the different frames }
Using the general result of the squared amplitude in Eq.~\eqref{eq:lorentz_invariant_amp}, we have calculated the amplitudes of photon-electron scattering with specified polarisations $e\gamma_{\pm} \to e\gamma_{\pm}$ and $e\gamma_{\pm} \to e\gamma_{\mp}$ in our four different frames. In the non-relativistic limit ($m_e \gg E_\gamma, \sqrt{E_e^2-m_e^2}$), all four frames approximate to Thomson scattering. However, in the relativistic limit, different frames can be applied to different physical contexts.

\subsubsection{Centre of mass frame}
Amplitudes for the photon-electron scattering in this frame of reference with specified polarisations are given by
\begin{align}
\frac{1}{2}\sum_{\mathrm{spins}}|\mathcal{M}(e\gamma_- \to e\gamma_-)|^2=&\frac{1}{2}\sum_{\mathrm{S_e}}|\mathcal{M}(e\gamma_+ \to e\gamma_+)|^2= (1+\cos\theta) \frac{ \left(E_\gamma^2 (1-\cos \theta)^2+(E_\gamma+E_e)^2 (\cos \theta+1)\right)}{(E_e+E_\gamma \cos \theta)^2}  \,, \nonumber\\ 
\frac{1}{2}\sum_{\mathrm{spins}}|\mathcal{M}(e\gamma_+ \to e\gamma_-)|^2=&\frac{1}{2}\sum_{\mathrm{S_e}}|\mathcal{M}(e\gamma_- \to e\gamma_+)|^2 = (1-\cos\theta)^2 \frac{ (E_e-E_\gamma)  \left(E_e^2-E_\gamma^2 \cos \theta\right)}{(E_e+E_\gamma) (E_e+E_\gamma \cos \theta)^2} \,.
\label{eq:com_frame_amp}
\end{align}
In the high energy limit, i.e., the ultra-relativistic limit, they approximate to 
\begin{align}
\frac{1}{2}\sum_{\mathrm{spins}}|\mathcal{M}(e\gamma_\pm \to e\gamma_\pm)|^2=& 1+ \cos \theta +\frac{4}{1 + \cos \theta } \,,\nonumber\\ 
\frac{1}{2}\sum_{\mathrm{spins}}|\mathcal{M}(e\gamma_\pm \to e\gamma_\mp)|^2=& \frac{m_e^2 (1-\cos\theta)^3 }{4 E_\gamma^2 (1+\cos\theta)^2 } \,.
\label{eq:com_frame_amp_high}
\end{align}
Here, we have dropped the small mass in the denominator. These formulas are not valid if $\theta$ is close to $\pi$. 
\subsubsection{Rest frame \label{app:rest_frame}}
Working in the rest frame of reference, the corresponding squared amplitudes for processes where the photon helicity is preserved  and  changed are, respectively
\begin{align}
\frac{1}{2}\sum_{\mathrm{spins}}|\mathcal{M}(e\gamma_\pm \to e\gamma_\pm)|^2=&\frac{(2E_{\gamma,1}E_{\gamma,2}-E_{\gamma,1}m_e+E_{\gamma,2}m_e)(E_{\gamma,1}^2+E_{\gamma,2}^2-E_{\gamma,1}m_e+E_{\gamma,2}m_e)}{E_{\gamma,1}^2E^2_{\gamma,2}} \,, \nonumber\\ 
\frac{1}{2}\sum_{\mathrm{spins}}|\mathcal{M}(e\gamma_\pm \to e\gamma_\mp)|^2=&\frac{m_e(E_{\gamma,1}-E_{\gamma,2})^2(E_{\gamma,1}-E_{\gamma,2}+m_e)}{E_{\gamma,1}^2E^2_{\gamma,2}} \,.
\label{eq:rest_frame_amp}
\end{align}
In the ultra relativistic limit ($E_{\gamma,1}\rightarrow \infty$), we get

\begin{align}
\frac{1}{2}\sum_{\mathrm{spins}}|\mathcal{M}(e\gamma_\pm \to e\gamma_\pm)|^2=& 2 \frac{E_{\gamma,1}^2+ E_{\gamma,2}^2}{E_{\gamma,1}E_{\gamma,2}} \,, \nonumber\\ 
\frac{1}{2}\sum_{\mathrm{spins}}|\mathcal{M}(e\gamma_\pm \to e\gamma_\mp)|^2=& \frac{m_e (E_{\gamma,1}- E_{\gamma,2})^3}{E_{\gamma,1}^2E_{\gamma,2}^2} \,.
\label{eq:rest_frame_amp_high}
\end{align}

\subsubsection{Spin frame}
 Based on this frame, the amplitudes for the different polarisation transitions are
\begin{align}
\frac{1}{2}\sum_{\mathrm{spins}}|\mathcal{M}(e\gamma_\pm \to e\gamma_\pm)|^2=& \frac{4E^2_e \sin^2{\theta}\left(E^2_\gamma\cos^2{\theta}\left(1+\cos^2{\theta}\right)\, +\, m^2_e\sin^2{\theta} \right)}{\left( E_e^2\,-\, E^2_\gamma\cos^4{\theta}\right)^2} \,,\nonumber\\ 
\frac{1}{2}\sum_{\mathrm{spins}}|\mathcal{M}(e\gamma_\pm \to e\gamma_\mp)|^2=&\frac{4m^2_e\cos^4{\theta} \left(E_e^2+E^2_\gamma \cos^2{\theta}\right)}{\left( E_e^2\,-\, E^2_\gamma\cos^4{\theta}\right)^2
 } \,.
\end{align}

\subsubsection{Fixed frame}
When we sum/average over final/initial spins, the different polarisation transitions in the fixed frame are:
\begin{align}
\frac{1}{2}\sum_{\mathrm{spins}}|\mathcal{M}(e\gamma_\pm \to e\gamma_\pm)|^2
&= \frac{\Big{(}m^2_eE_{\gamma,2}(E_{e,1}-p_{e,z}\mu_2)-E_{\gamma,1}(E_{e,1}-p_{e,z}\mu_1)(m^2_e-2E_{\gamma,2}(E_{e,1}-p_{e,z}\mu_2))\Big{)}}{E_{\gamma,1}^2E^2_{k_2}(E_{e,1}-p_{e,z}\mu_1)^2(E_{e,1}-p_{e,z}\mu_2)^2} \nonumber \\
&\hspace{-2cm}\times\Big{(}E^2_{k_1}(E_{e,1}-p_{e,z}\mu_1)^2-m^2_eE_{\gamma,1}(E_{e,1}-p_{e,z}\mu_1)+E_{\gamma,2}(E_{e,1}-p_{e,z}\mu_2)(m_e^2 + E_{\gamma,2}(E_{e,1}-p_{e,z}\mu_2))\Big{)} \,, \nonumber\\[1mm]
\frac{1}{2}\sum_{\mathrm{spins}}|\mathcal{M}(e\gamma_\pm \to e\gamma_\mp)|^2
&= \frac{m^2_e(\Delta-1)^2\Big{(}m^2_e+E_{\gamma,1}E_{\gamma,2}(1-\Delta)\Big{)}}{(E_{e,1}-p_{e,z}\mu_1)^2(E_{e,1}-p_{e,z}\mu_2)^2} \,,
\label{eq:fixed_frame_amp}
\end{align}
where $\Delta\equiv \cos\theta =\sqrt{1-\mu_1^2}\sqrt{1-\mu_2^2}\cos(\theta_1-\theta_2)+ \mu_1 \mu_2$, $\mu_1=\cos\phi_1$ and  $\mu_2=\cos\phi_2$. In the ultra ($E_{\gamma,1} \rightarrow \infty$) relativistic limit we obtain

\begin{align}
\frac{1}{2}\sum_{\mathrm{spins}}|\mathcal{M}(e\gamma_\pm \to e\gamma_\pm)|^2=& \frac{\big{(}E_{e,1}-\mu_1 p_{e,z}\big{)}\big{(}1-\Delta^2\big{)}}{m^2_e}E_{\gamma,1} \,, \nonumber\\
\frac{1}{2}\sum_{\mathrm{spins}}|\mathcal{M}(e\gamma_\pm \to e\gamma_\mp)|^2=& \frac{\big{(}E_{e,1}-\mu_1 p_{e,z}\big{)}\big{(}1-\Delta\big{)}^2}{m^2_e}E_{\gamma,1} \,.
\label{eq:fixed_frame_amp_high}
\end{align}

\section{Second Approach \label{App:SecondAppr}} 
In section \ref{sec:compton} we have calculated the $\mathbf{A}$--matrix elements given in Eq.~\eqref{eq:M_matrix_elements}, by  directly substituting the explicit form of the electron spinors $u(p)_{\pm \frac{1}{2}}$, into the matrix element $\mathcal{M}$ accordingly to the chosen frame of reference. However, for the case of unpolarised electrons there is another way to get explicit form of these elements. Nonetheless the form of the general expression is not simple until applying the kinematics. 

In the second approach instead of replacing all the elements in the amplitude, we sum/average over the electron spin as usual and use trace technology. Therefore, this approach can only be applied to the case of unpolarised electrons.  Normally, if we do not care about the polarisation of the photons during the scattering process, we also sum/average over their helicity states. However, in order to find the polarised amplitude of polarised photon scattering,  we assume definite helicity states for each polarisation vector in the initial and final states.  

The calculation of the squared polarised amplitude using this approach starts from Eq.~\eqref{eq:General_amplitude}, which can be written as
\begin{equation}
\frac{1}{2}\sum_{\alpha,\beta=\pm}\mathcal{M}(e_\alpha\gamma_i\rightarrow e_\beta\gamma_{i'})\mathcal{M}^*(e_\alpha\gamma_j\rightarrow e_\beta\gamma_{j'})\equiv \left(M_{i'i}\,M_{j'j}^*\right)_{ s}+\left(M_{i'i}M_{j'j}^{*}\right)_{ t}+2\,\mathrm{Re}[M_{i'i}M_{j'j}^{*}]_{st} \,,
\label{eq:genAmp}
\end{equation}
where we have summed/averaged over electron spin, but not photon helicity. The sub indices correspond to the s and u--channel contributions and $i'$, $i$, $j'$, $j$ indicates the polarisation.  More explicitly, each term in the equation above is given by,
\begin{align}
\left(M_{i'i}M_{j'j}^*\right)_{ s}=&\frac{e^4}{(s-m_e^2)^2}\left(M_{\mu \nu}M_{\mu' \nu'}^*\right)_{s} \ \epsilon_i^{(1) \mu } \epsilon_{i'}^{*(2)  \nu} \epsilon_{j}^{*(1)  \mu'} \epsilon_{j'}^{(2)\ \nu' } \,, \nonumber\\
\left(M_{i'i}M_{j'j}^{*}\right)_{ t}=&\frac{e^4}{(t-m_e^2)^2}\left(M_{\mu \nu}M_{\mu' \nu'}^*\right)_{t} \ \epsilon_i^{(1) \mu } \epsilon_{i'}^{*(2)  \nu} \epsilon_{j}^{*(1)  \mu'} \epsilon_{j'}^{(2)\ \nu' } \,, \nonumber\\
\left(M_{i'i}M_{j'j}^{*}\right)_{st}=&\frac{e^4}{(s-m_e^2)(t-m^2_e)} \left(M_{\mu \nu} M_{\mu' \nu'}^*\right)_{st} \ \epsilon_i^{(1) \mu } \epsilon_{i'}^{*(2)  \nu} \epsilon_{j}^{*(1)  \mu'} \epsilon_{j'}^{(2)\ \nu'} \,.
\end{align}
 $ \left(M_{\mu \nu} M_{\mu' \nu'}^*\right)_{s,t,st}$ are the squared polarisation vector--independent amplitude corresponding to each channel. This is the most general expression for the polarised squared amplitude of photon--scattering.
 
For simplicity, and comparison with Eq.~\eqref{eq:Rprime_matrix} in section \ref{sec:compton}, we apply  the rest frame kinematics. Therefore, the polarised squared amplitude takes the form,

\begin{align}
\frac{1}{2}\sum_{\alpha,\beta}&\mathcal{M}(e_\alpha \gamma_i \to e_\beta \gamma_{i'})\mathcal{M}(e_\alpha \gamma_j \to e_\beta \gamma_{j'})^\ast=\frac{1}{2 m_e E_{\gamma,1}^2 E_{\gamma,2}}\times\nonumber\\
\times&\left[2 E_{\gamma,1}
k_1\cdot\epsilon_i^{*(2)} \left[(E_{\gamma,2}-E_{\gamma,1})   \epsilon^{(1)}_{i'}\cdot \epsilon_{j}^{(2)} k_2\cdot \epsilon^{*(1)}_{j'}+(E_{\gamma,1}+E_{\gamma,2}) k_2 \cdot\epsilon^{(1)}_{i'} \epsilon_{j'}^{(1)}\cdot \epsilon_{j}^{(2)}\right]\right.\nonumber\\
+&2 E_{\gamma,1}k_1\cdot\epsilon_{j}^{(2)} \left[(E_{\gamma,1}-E_{\gamma,2}) k_2 \cdot\epsilon_{i'}^{(1)}\epsilon_{j'}^{*(1)}\cdot \epsilon_{i}^{*(2)}-(E_{\gamma,1}+E_{\gamma,2}) k_2\cdot \epsilon_{j'}^{*(1)} \epsilon_{i'}^{(1)} \cdot\epsilon_{i}^{*(2)}\right]\nonumber\\
+&E_{\gamma,2} \left[ \epsilon_{j'}^{*(1)}\cdot \epsilon_{j}^{(2)} \epsilon_{i'}^{(1)} \cdot\epsilon_{i}^{*(2)}\left(2 E_{\gamma,1}^2 \left[(E_{\gamma,2}-E_{\gamma,1}) (\cos \theta -1)+4 m_e\right]+m^3_e\right)\right.\nonumber\\
-&\left.\left.\left(\epsilon_{j'}^{*(1)} \cdot\epsilon_{j}^{*(2)}\epsilon_{i'}^{(1)} \cdot \epsilon^{i}(k_2)-\epsilon_{j'}^{*(1)}\cdot\epsilon_{i'}^{(1)} \epsilon_{i}^{*(2)} \cdot\epsilon_{j}^{(2)}\right)(2 E_{\gamma,1}^2 (\cos\theta-1) (E_{\gamma,2}-E_{\gamma,1})+m^3_e)\right]\right] \,,
\label{SecAppRestFrame}
\end{align}
where we have neither specified the photon polarisation in the initial nor final state.

Considering the case where polarisation is conserved (changed), i.e., $e\gamma_{\pm}\rightarrow e\gamma_{\pm}$ ($e\gamma_{\pm}\rightarrow e\gamma_{\mp}$), the expression for the polarised squared amplitude simplifies even more and leads to the same result as in \eqref{eq:rest_frame_amp}.
Furthermore,  the result in Eq.~\eqref{SecAppRestFrame} allows to calculate the $\mathbf{A}$--matrix by just doing every combination of $M_{i'i}M_{j'j}^*$ required in Eq.~\eqref{eq:General_amplitude}.
In here, we have done it for the rest frame. However, is possible to do all this process for any frame of reference in Appendix \ref{Ap:kinematics}. 

\section{Polarised photons scattering off polarised electrons\label{App:C}}  

We can study the transitions between different polarisation states for the incoming and outgoing particles by expressing this amplitude as a function of the incoming and outgoing electron and photon polarisation states ($\alpha$, $\beta$ and $k$, $l$ respectively):
\begin{equation}
\mathcal{M}(e_\alpha \gamma_k \to e_\beta \gamma_j)=-ie^2\bar{u}_\beta(p_2) \left(  \frac{\slashed{\epsilon}^\ast_j(k_2) \left(\slashed{p}_1 +\slashed{k}_1 + m_e \right)\slashed{\epsilon}_k(k_1)}{p_1 \cdot k_1} -   \frac{ \slashed{\epsilon}_k(k_1)\left(\slashed{p}_1 -\slashed{k}_2 + m_e \right)\slashed{\epsilon}^\ast_j(k_2) }{p_1 \cdot k_2} \right) u_\alpha(p_1) \, , \label{eq:pol_amp}
\end{equation}
where $p_{1,2}$ and $k_{1,2}$ are the incoming/outgoing electron and photon 4--momenta respectively.

In order to calculate the amplitude  for different photon polarisation transitions, one needs to express polarised vectors in the 4 dimensional Lorentz covariant form. In the Lorentz gauge, they are defined as $\epsilon_{l,r}^\mu = (0,\epsv_{l,r})$ and $\epsilon_{\pm}^\mu = (0,\epsv_{\pm})$, where $\epsv_{l,r}$ and $\epsv_{\pm}$ are listed in equations (\ref{eq:pol_vec_def}) and (\ref{eq:circ_pol_vec_def}).

Similarly, we can express the spinors $u(p)_{\pm }$,$v(p)_{\pm}$ for the two electron polarisation as a function of its 4--momentum $p^\mu=(E_e, p_x, p_y, p_z)$ in any frame by: 
\begin{eqnarray}
u_{\alpha}(p) &=& \begin{pmatrix}
+ \sqrt{p\cdot\sigma}~ \xi_{\alpha}  \\
+ \sqrt{p\cdot\bar{\sigma}}~ \xi_{\alpha}
\end{pmatrix} =
\frac{1}{\sqrt{2(E+m_e)}}
\begin{pmatrix}
+ (p\cdot\sigma + m_e) ~ \xi_{\alpha}  \\
+ (p\cdot\bar{\sigma} + m_e) ~ \xi_{\alpha}
\end{pmatrix} \,, \nonumber\\
v_{\alpha}(p) &=& \begin{pmatrix}
+\sqrt{p\cdot\sigma}~ \eta_{\alpha}  \\
-\sqrt{p\cdot\bar{\sigma}}~ \eta_{\alpha}
\end{pmatrix} =
\frac{1}{\sqrt{2(E+m_e)}}
\begin{pmatrix}
+(p\cdot\sigma + m_e) ~ \eta_{\alpha}  \\
-(p\cdot\bar{\sigma} + m_e) ~ \eta_{\alpha}
\end{pmatrix} \,, 
\label{eq:spinor}
\end{eqnarray}
where the electron polarisation states are given by
\begin{eqnarray}
\xi_{+}=\eta_{-}=\begin{pmatrix}
1\\0
\end{pmatrix}\,,\hspace{1cm}
\xi_{-\frac{1}{2}}=\eta_{+\frac{1}{2}}=\begin{pmatrix}
0\\1
\end{pmatrix} \, ,
\label{eq:2_component_spinor}
\end{eqnarray}
and 
\begin{eqnarray}
\sigma^\mu &=& \{\mathbf{1}, \sigma^1, \sigma^2, \sigma^3\} \,, \nonumber \\
\bar{\sigma}^\mu &=& \{\mathbf{1}, -\sigma^1, -\sigma^2, -\sigma^3\}
\end{eqnarray}

The right hand side of Eq.~\eqref{eq:spinor}  can be found by diagonalizing the matrix $p\cdot \sigma = o \Lambda o^\dagger$ where $o$ is the normalized matrix of eigenvectors of $p\cdot \sigma$. In this way $\sqrt{p \cdot \sigma}=o \sqrt{\Lambda} o^\dagger$, which simplifies the calculations. Note that these are just two independent solutions of the Dirac equation and can only be understood as eigenstates of the momentum operator when the electron is at rest or moving along de $z$ direction. 

For $2 \to 2$ scattering processes, the amplitude, which is Lorentz invariant, is only dependent upon 2 Lorentz-invariant invariants. Here, we choose them to be $p_1\cdot k_1$ and $p_1 \cdot k_2$. We also choose spinors ($u_{+}, u_{-}$) in Eqs.~\eqref{eq:spinor} and \eqref{eq:2_component_spinor} as the basis of electron polarisation. Note that for massive fermions, this basis is not Lorentz-invariant so we are constrain to  work in a specific frame of reference. Here we are working in the electron rest frame and these amplitudes are obtained, ignoring the coefficient $-ie^2$, as
\begin{eqnarray}
\mathcal{M}(e_+\gamma_+ \to e_+\gamma_+) = \mathcal{M}(e_-\gamma_- \to e_-\gamma_-) &=&
 \frac{m_e (\piki+\piko) \left(2 \piki \ \piko-m_e^2 (\piki-\piko)\right)}{(\piki)^2 \ \piko \sqrt{4 m_e^2+2 \piki-2 \piko}} \,,  \nonumber\\[1mm] 
\mathcal{M}(e_+\gamma_+ \to e_+\gamma_-) = \mathcal{M}(e_-\gamma_- \to e_-\gamma_+) &=&
 -\frac{m_e (\piki-\piko) \left(m_e^2 (3 \piki-\piko)+2 \piki (\piki-\piko)\right)}{(\piki)^2 \ \piko \sqrt{4 m_e^2+2 \piki-2 \piko}}\,,  \nonumber\\[1mm]   
\mathcal{M}(e_+\gamma_+ \to e_-\gamma_+) = - \mathcal{M}(e_-\gamma_- \to e_+\gamma_-) &=&
 \frac{\sqrt{\piki-\piko} \left(m_e^2 (\piko-\piki)+2 \piki \ \piko\right)^{3/2}}{(\piki)^2 \ \piko \sqrt{4 m_e^2+2 \piki-2 \piko}} \,,  \nonumber\\[1mm] 
\mathcal{M}(e_+\gamma_+ \to e_-\gamma_-) = - \mathcal{M}(e_-\gamma_- \to e_+\gamma_+) &=&-\mathcal{M}(e_+\gamma_- \to e_-\gamma_+) =  \mathcal{M}(e_-\gamma_+ \to e_+\gamma_-) \nonumber\\[1mm]
&=&
 \frac{m_e^2 (\piki-\piko)^{3/2} \sqrt{m_e^2 (\piko-\piki)+2 \piki \ \piko}}{(\piki)^2 \ \piko \sqrt{4 m_e^2+2 \piki-2 \piko}} \,,  \nonumber\\[1mm] 
\mathcal{M}(e_+\gamma_- \to e_+\gamma_+) = \mathcal{M}(e_-\gamma_+ \to e_-\gamma_-)  &=& 
 -\frac{m_e^3 (\piki-\piko) (\piki+\piko)}{(\piki)^2 \ \piko \sqrt{4 m_e^2+2 \piki-2 \piko}} \nonumber\\[1mm] 
\mathcal{M}(e_+\gamma_- \to e_+\gamma_-) = \mathcal{M}(e_-\gamma_+ \to e_-\gamma_+) &=&
 -\frac{m_e (3 \piki-\piko) \left(m_e^2 (\piki-\piko)-2 \piki \ \piko\right)}{(\piki)^2 \ \piko \sqrt{4 m_e^2+2 \piki-2 \piko}} \,,  \nonumber\\[1mm] 
\mathcal{M}(e_+\gamma_- \to e_-\gamma_-) = - \mathcal{M}(e_-\gamma_+ \to e_+\gamma_+) &&\nonumber\\[1mm]
&&\hspace{-4cm}= \frac{\left(m_e^2 (\piki-\piko)+2 (\piki)^2\right) \sqrt{(\piki-\piko) \left(m_e^2 (\piko-\piki)+2 \piki \ \piko\right)}}{(\piki)^2 \ \piko \sqrt{4 m_e^2+2 \piki-2 \piko}}  \,.     
\label{eq:lorentz_invariant_ampTotal}
\end{eqnarray}

Although the basis of electron polarisation specify the frame, we still use Lorentz-invariant products ($\piki$, $\piko$) to express the amplitude here. In the rest frame, it is simple to replace Lorentz invariants to particular variables, i.e., incoming photon energy $E_{\gamma,1}$ and outgoing photon energy $E_{\gamma,2}$ by doing the replacements $\piki=m_e E_{\gamma,1}$ and $\piko=m_e E_{\gamma,2}$ to re-express these formulas. 

\section{Cross section calculations \label{App:xsec_frames}}
The cross section for the process preserving circular polarisation ($e \gamma_\pm \to e \gamma_\pm$) and the process changing it ($e \gamma_\pm \to e \gamma_\mp$) can be calculated by following the procedure described in Appendix B of \cite{Campo:2017nwh}, by which we express the amplitude in terms of the variables $\chi$ and $\eta$, where $\chi=(p_1 \cdot k_1)/m_e^2$ and $\omega=(p_1\cdot k_2)/m_e^2$ and  perform the integral over $\omega$:
\begin{equation}
\sigma = \int^{2\chi}_\frac{2\chi}{2\chi+1}\frac{1}{32\pi m_e^2\chi^2}|\mathcal{M}(\chi,\omega)|^2d\omega \, .
\end{equation}

We can then use the Lorentz invariant amplitudes in Eq.~\eqref{eq:lorentz_invariant_amp} to find the corresponding cross sections
\begin{eqnarray}
&&\sigma(e\gamma_\pm \to e\gamma_\pm) = \frac{3 \sigma_{\text{T}}}{8} \left(\frac{2+3 \chi-\chi ^2}{\chi ^2 ( 1+2 \chi)}-\frac{2+\chi-2 \chi ^2 }{2 \chi ^3} \log (1+ 2 \chi) \right) \, ,\nonumber\\
&&\sigma(e\gamma_\pm \to e\gamma_\mp) =
\frac{3 \sigma_{\text{T}}}{8} \left(\frac{2+9 \chi+13 \chi ^2+4 \chi ^3}{\chi ^2 (1+2 \chi)^2}-\frac{2+3 \chi}{2 \chi ^3} \log (1+2 \chi)\right)\,,
\end{eqnarray}
where $\sigma_{\text{T}}=\frac{1}{6\pi m_e^2}$. Note that the same method can be used to calculate the cross sections for the photon scattering off polarised electrons discussed in Appendix \ref{App:C}.

\begin{figure}[ht!]
\includegraphics[width=.95 \textwidth, angle = 0]{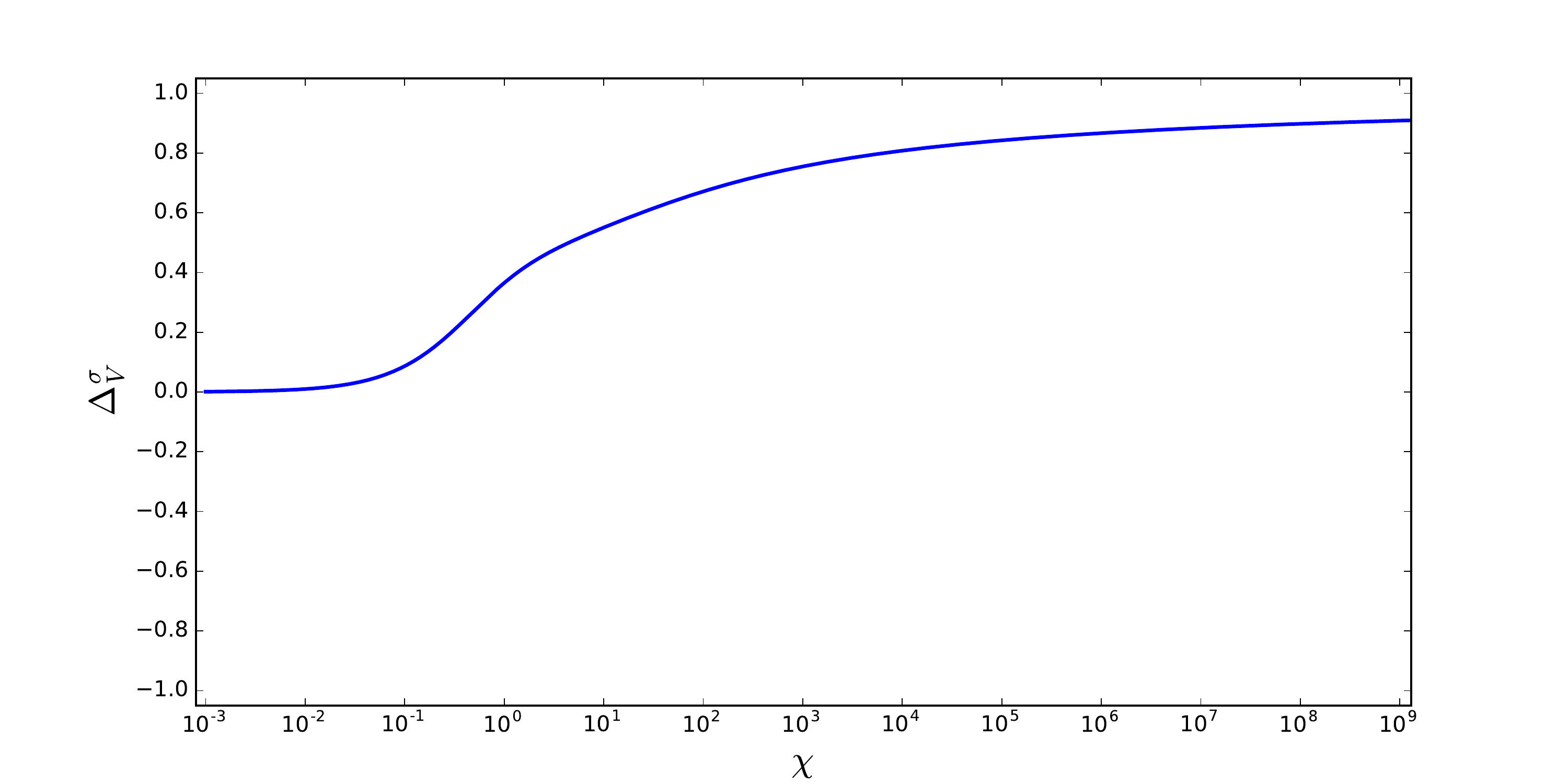}
\caption{The change of the net circular polarisation at the cross section level, $\Delta_V^\sigma$, as a function of the Lorentz invariant parameter $\chi = \piki/m_e^2$.}\label{fig:PolChi}
\end{figure}

The summation of both photon and electron polarisations gives the standard Compton scattering amplitude 
\begin{eqnarray}
|\mathcal{M}(e\gamma \to e\gamma)|^2 &=& \frac{1}{4}\sum_{k,l=\pm} \sum_{\alpha,\beta=\pm}| \mathcal{M}(e_\alpha\gamma_k \to e_\beta\gamma_l)|^2 \nonumber\\ 
&=& 2 \left(\frac{\piki}{\piko}+\frac{\piko}{\piki}\right) + 4 m_e^2 \left(\frac{1}{\piki}-\frac{1}{\piko}\right) + 2 m_e^4 \left(\frac{1}{\piki}-\frac{1}{\piko}\right)^2 \, 
\label{eq:sum_amp}
\end{eqnarray}
and  the total Compton scattering cross section is
\begin{eqnarray}
\sigma_{\text{C}} = \frac{3 \sigma_{\text{T}}}{4} \left(\frac{ 2+8 \chi+9 \chi ^2+\chi ^3}{\chi ^2 (1+2 \chi)^2}+\frac{-2-2 \chi+\chi ^2}{2 \chi ^3} \log (1+2 \chi) \right) \, ,
\end{eqnarray}
which is consistent with former result in \cite{Boehm:2008nj}. 
It is useful to define the ``summed'' asymmetry between the two photon helicity states at the cross section level, 
\begin{eqnarray}
\Delta_V^\sigma 
&=& \frac{\sigma(e \gamma_+ \to e \gamma_+) + \sigma(e \gamma_- \to e \gamma_-) - \sigma(e \gamma_+ \to e \gamma_-) - \sigma(e \gamma_- \to e \gamma_+)}{\sigma(e \gamma_+ \to e \gamma_+) + \sigma(e \gamma_- \to e \gamma_-) + \sigma(e \gamma_+ \to e \gamma_-) + \sigma(e \gamma_- \to e \gamma_+)} \nonumber\\
&=& \frac{\chi  (1+ \chi) \left(-2 \chi  (1+3 \chi)+(1+2 \chi)^2 \log (1+2 \chi)\right)}
{2 \chi  \left(2+8 \chi+9 \chi ^2+\chi ^3\right)+ \left(-2 -2 \chi+\chi ^2\right) (1+2 \chi)^2 \log (1+2 \chi)} \, , \label{eq:rcp} 
\end{eqnarray}
which is the ratio of total conserved circular polarisation after integrating over phase space. Its behaviour as a function of $\chi$ is numerically shown in Fig. \ref{fig:PolChi}. The Thomson scattering refers to the limit $\chi\to 0$. In this case, only $\Delta_V^\sigma = 0$, which is well-known \cite{chandrasekhar1960radiative}. The circular polarisation is likely to be preserved with larger $\chi$, which corresponds to larger energy momentum transfer between photon and electron. For $\chi \gtrsim 10^4$ (corresponding to $E_{\gamma,1} \gtrsim 0.361$ GeV in the COM frame, or $E_{\gamma,1} \gtrsim 511$ GeV in the rest frame), $\Delta_V^\sigma$ can reach 0.8. 
 
\newpage

\section{More plots in the fixed frame \label{App:FixedFramePlots}}
\begin{figure}[t!]
\centering
\includegraphics[width=.97\linewidth]{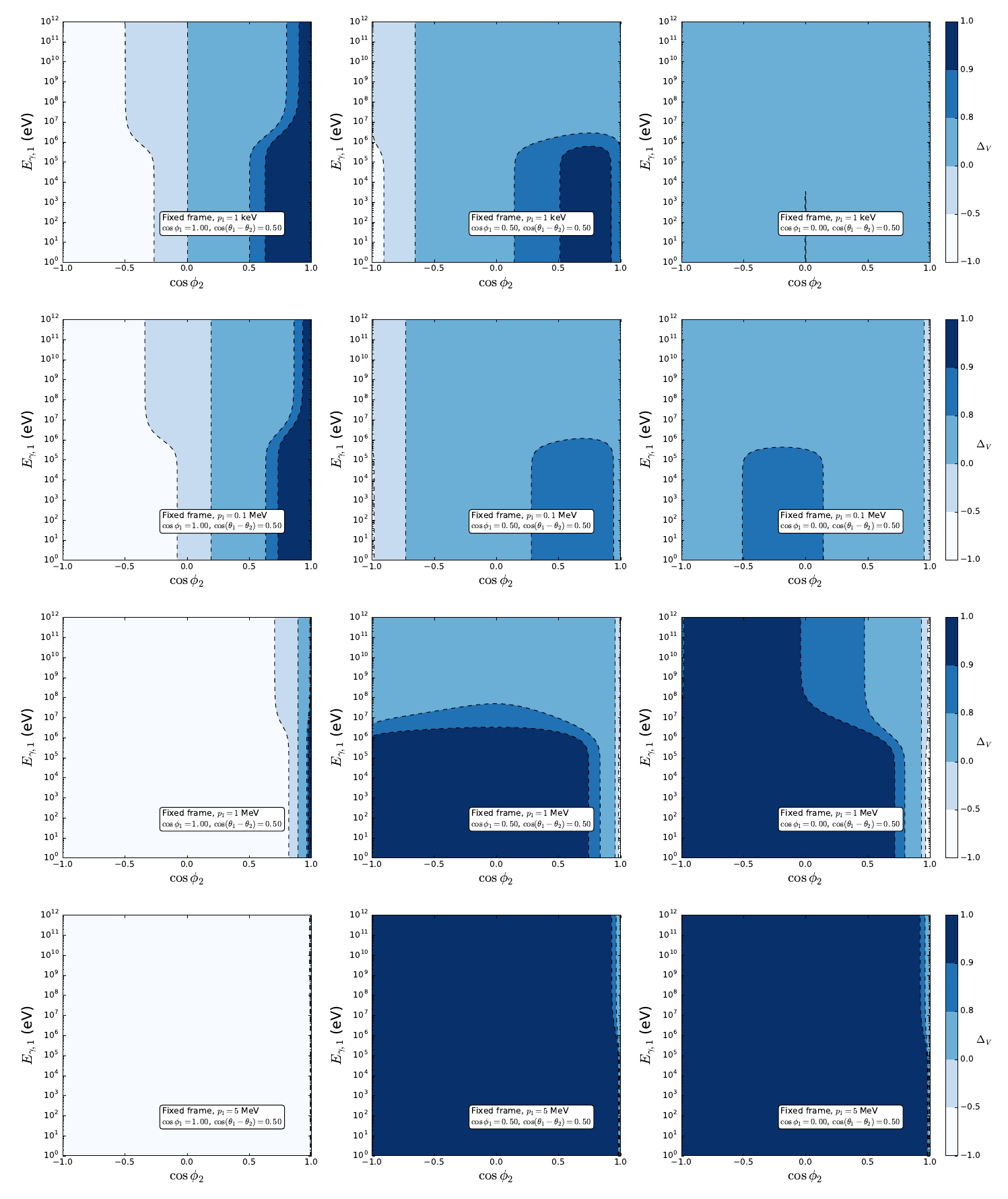}
\caption{The asymmetry of photon helicity states $\Delta_V$ in fixed frames with respect to different direction and energy information. For each row, we fix incoming electron kinetic energy $p_1=$ 1 keV, 0.1 MeV, 1 MeV and 5 MeV, respectively and for each column, we fix the angle $\phi_1$ between incoming photon and electron with $\cos\phi_1=$ 1, 0.5 and 0, respectively. The angle difference between incoming and outgoing photons projected on the $x-y$ plane is fixed $\cos(\theta_1-\theta_2)=0.5$. } 
\label{fig:FixedFramePercResults1}
\end{figure}

\begin{figure}[ht!]
\centering
\includegraphics[width=.95\linewidth]{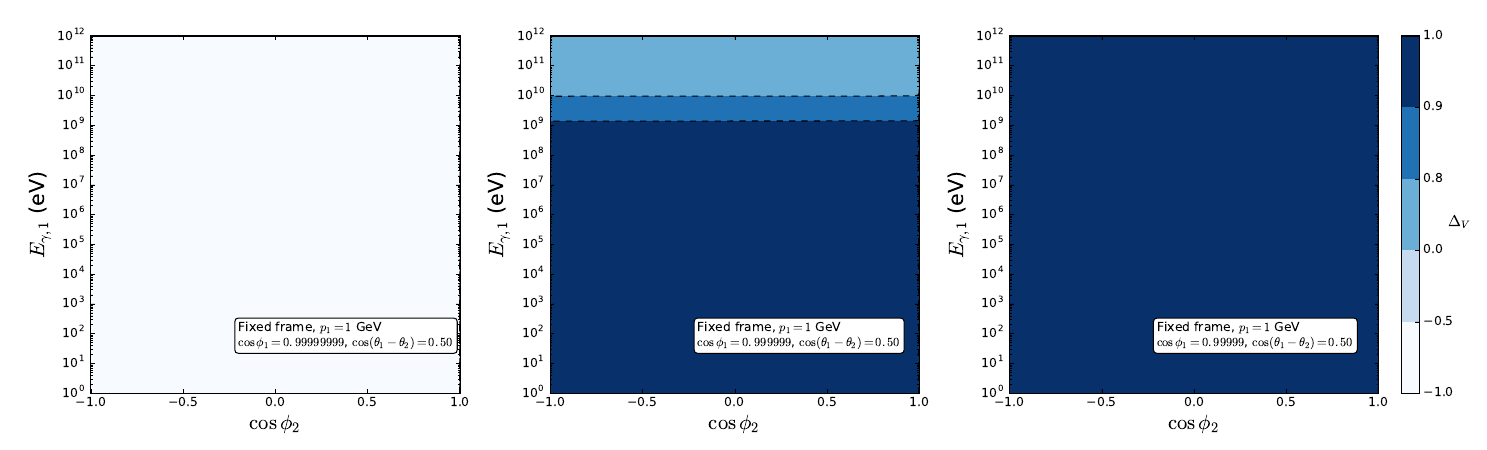}
\caption{The asymmetry of photon helicity states $\Delta_V$ for photon scattering with energetic electron in the fixed frame. The kinetic energy of incoming electron is fixed at 1 GeV, $\cos(\theta_1-\theta_2)=0.5$ is used as in Fig.~\ref{fig:FixedFramePercResults1}, and $\cos\phi_1 =$ 0.99999999, 0.999999, 0.99999 respectively in each subfigure. } 
\label{fig:FixedFramePercResults2}
\end{figure}

The fixed frame is the most generic frame of reference we consider and consequently, has more kinematic parameters that can be changed. For the scattering with high energy incoming electron, circular polarisation is conserved for any angular distribution and any energy of the incoming photon, which agrees with the results in the spin and COM frames. However, for low energy incoming electrons, taking different kinematic configurations from the one shown in the main text (Fig. \ref{fig:d}), the plot in the fixed frame changes dramatically. Here, we show two sets of plots (Fig. \ref{fig:FixedFramePercResults1} and \ref{fig:FixedFramePercResults2}) where we vary the energy of the incoming electron  and the angular configuration of the scattering process. In every set of plots, the vertical axis corresponds to the energy of the photon in the initial state and the horizontal axis corresponds to $\cos\phi_2$ which is the angle between outgoing photon and incoming electron as discussed in section \ref{sec:compton}.

In Fig. \ref{fig:FixedFramePercResults1}, we fixed the incident angle of the photon, $\phi_1$, i.e., the angle between the incoming photon and electron, with its cosine value fixed at $\cos\phi_1=$ 1, 0.5 and 0 in the left, middle and right panels, respectively. For each row, we fix the kinetic energy of the incoming electron $p_1 = $ 1 keV, 0.1 MeV, 1 MeV, 5 MeV, respectively. $\theta_1-\theta_2$ is the difference between incoming and outgoing photons projected on the $x-y$ plane at $\pi/3$. The influence of $\theta_1-\theta_2$ is less important, so we fix its cosine value at 0.5 for all plots. As shown in the left-top corner, by setting the incident angle $\phi_1=0$, the result for photon scattering with low energy electron is almost the same as that in the rest frame. But this behaviour changes largely once the incident angle increases. Increasing the energy of the incoming electron in general lead to larger $\Delta_V$, i.e., less change of the net circular polarisation after scattering. Once $p_1 \gg m_e$, the circular polarisation is more likely to be preserved for large incident angles. In Fig.~\ref{fig:FixedFramePercResults2}, we fixed $p_1= 1$ GeV, and see that the $\Delta_V<0.9$ only for very small $\phi_1$, noting that $\cos\phi_1=$ 0.99999999 and 0.999999 correspond to $\phi_1 \approx  0.0081^\circ$ and $0.081^\circ$, respectively. Only for $\phi_1 \gtrsim 0.1^\circ$, $\Delta_V>0.9$ is satisfied.

\clearpage
\bibliographystyle{ieeetr} 
\bibliography{Circular_Polarisation}

\end{document}